\documentclass[aps,prb,reprint,showpacs,superscriptaddress]{revtex4-1}
\usepackage[english]{babel}
\usepackage{amsmath,amssymb,bbm,graphicx,color,comment,txfonts}
\usepackage[bookmarks=true,colorlinks,citecolor=blue,urlcolor=blue]{hyperref}
\usepackage{dsfont}
\usepackage[normalem]{ulem}
\usepackage{physics}
\usepackage{svg}
\usepackage{mwe}
\newcommand{\cre}[1]{c^{\dagger}_{#1}}
\newcommand{\ann}[1]{c^{\phantom{\dagger}}_{#1}}

\begin{document}

\date{\today}

\title{Spectral Transitions of the Entanglement Hamiltonian in Monitored Free Fermions}

\author{K. Chahine}
\affiliation{Institut f\"ur Theoretische Physik, Universit\"at zu K\"oln, D-50937 Cologne, Germany}
\author{M. Buchhold}
\affiliation{Institut f\"ur Theoretische Physik, Universit\"at zu K\"oln, D-50937 Cologne, Germany}
\affiliation{Universität Innsbruck, Institut für Theoretische Physik, Technikerstraße 21a, 6020 Innsbruck, Austria}

\begin{abstract}
We numerically investigate measurement-induced phase transitions in monitored free fermions through the spectral and eigenstate properties of the entanglement Hamiltonian. By analyzing entanglement scaling, we identify three non-trivial fixed points, for which the entanglement follows exact analytical scaling forms: chaotic unitary dynamics at infinitesimal monitoring, characterized by a Gaussian Page law; a Fermi-liquid fixed point at moderate monitoring, defining a metallic phase with logarithmic entanglement growth and emergent space-time invariance; and a quantum Lifshitz fixed point marking the measurement-induced phase transition into a localized area-law phase. Adopting a random-matrix perspective on the entanglement Hamiltonian, we show that short-range spectral correlations, such as the adjacent gap ratio $\langle \tilde r\rangle$ and the Kullback–Leibler divergence $KL_1$, sharply detect an ergodic to non-ergodic phase transition at the quantum Lifshitz fixed point, and yield precise estimates of the critical point and correlation length exponent. Long-range probes, including the spectral form factor and the associated Thouless time, corroborate this picture, while the variant $KL_2$ uncovers signatures of a possible non-ergodic extended (multifractal) regime at intermediate monitoring strengths. Together, these results establish the entanglement Hamiltonian as a powerful framework for diagnosing metallic, localized, and multifractal regimes in monitored quantum dynamics, and highlight unexpected scaling structures, Fermi-liquid and Lifshitz criticality, that lie beyond current field-theoretic approaches.  
\end{abstract}

\maketitle
\section{Introduction}

Recent advances in controllable quantum platforms have opened new avenues for the study of non-equilibrium quantum critical phenomena. In particular, the combination of coherent unitary evolution with mid-circuit measurements has motivated investigations of a new class of \emph{monitored quantum systems}, whose dynamics result from the interplay between unitary evolution and measurements. In these systems, entanglement and quantum information spread through coherent dynamics but can be suppressed through measurement-induced collapse. The competition between these tendencies can give rise to a measurement-induced phase transition (MIPT), separating distinct dynamical phases characterized by different entanglement structures in space and time.

While much of the theoretical and experimental work on MIPTs has focused on qubit-based hybrid circuits~\cite{Skinner2019,Fisher2018,Li2019b,gullans2019,choi2020prl,Jian2020,fan2020selforganized,lifisher2021,nahum2021prxq, jian2021syk,  bao2021symmetry,turkeshi2021measurementinduced, Ivanov2020,Buonaiuto2021,  Zabalo2020,Zabalo2022,ippoliti2021,circuitreview,Romito2020, Schomerus2019, IadecolaControl, Iadecola2025,Agraval2021,turkeshi2022measurement,Popperl,alba2022,KhemaniKitaev,Chakraborty2024,Zhu2022,Zhu2023,ippoliti2020,Klocke2022,Klocke2023,SagarKitaev,Morral2022,lirasolanilla2025}, fermionic monitored systems, particularly those with particle number conservation, exhibit distinctive features. In free-fermion models, the many-body wavefunction is fully determined by the single-particle correlation matrix, enabling exact computation of entanglement measures. This structure allows for efficient analytical and numerical study of entanglement dynamics, providing access to properties often difficult to probe in qubit circuits~\cite{Cao2019,alberton2021enttrans,Bernard_2018,buchhold2021effective,poboiko,poboiko2023measurementinduced,Fava2023,Fava2024, chahine2024,klocke2025,Ladewig2022,minato2021fate,szyniszewski2023disordered, kells2023, jin2023measurementinduced,poboiko2025, poboiko2025interacting, poboiko2025levyflights,doggen2021generalized, tsitsishvili2023measurement, turkeshi2022entanglement,piccitto2022, Mueller2022,oshima2023,paviglianiti2023multipartite, minoguchi2021continuous, tirrito2023, xing2023interactions, turkeshi2023entanglement,fuji2020,Altland2022, turkeshi2023density,Yang2023Keldysh}.

Analytical progress in fermionic monitored systems has been achieved using field-theoretic approaches, which map a $D$-dimensional $R$-replica model to a nonlinear sigma model (NL$\sigma$M) in $D+1$ dimensions~\cite{poboiko,chahine2024, poboiko2023measurementinduced,Fava2023,Fava2024,klocke2025}. This mapping establishes a connection between monitored free fermions in $D$ dimensions and disordered fermion Hamiltonian ground states in $D+1$ dimensions, suggesting that a measurement-induced phase transition for number-conserving dynamics occurs only for $D\ge 2$. Despite these advances, several universal features of the phase diagram remain elusive, including the detailed structure of entanglement and the nature of fixed points in higher dimensions. Recent numerical studies have reported behavior deviating from NL$\sigma$M predictions~\cite{chahine2024,fan2025}, for example a scale-invariant metallic fixed point at intermediate measurement strength~\cite{chahine2024} and critical exponents that differ from perturbative renormalization group predictions~\cite{fan2025}.

In disordered systems, random matrix theory (RMT), applied to the random Hamiltonian energy levels or eigenstates, provides a complementary perspective on localization-delocalization transitions~\cite{brody1981,mehta1967}. In the context of monitored fermions, the dynamics is described by stochastic wave function updates, and no static energy functional or Hamiltonian is available. Nevertheless, the entanglement Hamiltonian $\mathcal{H}_A$, defined via the reduced density matrix of a subsystem $A$ as $\rho_A = \mathcal{Z}_A^{-1}\exp(-\mathcal{H}_A)$, serves a similar role. Even in the absence of energy conservation, $\mathcal{H}_A$ defines an effective single-particle spectrum and wavefunctions. We show that both can be analyzed using RMT techniques to probe ergodicity, localization, and spectral correlations in monitored systems. This approach complements entanglement scaling, offering a more detailed view of the internal structure of quantum states.

In this work, we combine these two complementary analyses to characterize monitored free fermions. First, we determine the entanglement phase diagram through scaling analysis of the entanglement entropy, identifying three nontrivial fixed points: (i) a Gaussian Page-law fixed point at $\gamma\to 0^+$, (ii) a Fermi-liquid fixed point at finite monitoring $\gamma_\text{Fl}$, characterized by log-law entanglement scaling and space-time invariance, and (iii) a quantum Lifshitz fixed point at the critical monitoring strength $\gamma_c$, separating the metallic and area-law phases. Second, we perform a detailed RMT analysis of the entanglement Hamiltonian $\mathcal{H}_A$, which provides complementary insights into ergodicity, localization, and universal spectral properties not captured by entanglement scaling alone. Both the short-range and long-range spectral correlations, e.g., the spectral gap statistics and the spectral form factor, display a clear ergodic to non-ergodic transition at $\gamma=\gamma_c$, i.e., at the quantum Lifshitz fixed point. This is further confirmed by the behavior of the eigenstates of $\mathcal{H}_A$, which we probe through the Kullback-Leibler divergence. All signatures display a clear, sharp crossing point at $\gamma=\gamma_c$. Finite size scaling reveals a correlation length exponent $\nu=0.83\pm0.03$ ($\nu=0.88\pm0.03$ for $\langle\tilde r\rangle$) within the range of critical, three-dimensional percolation. 

The primary focus is on a two-dimensional model~\cite{chahine2024, poboiko2023measurementinduced,jin2023measurementinduced,poboiko2025, fan2025}, while the one- and three-dimensional cases are also considered to illustrate the generality of the approach. Together, these analyses provide a comprehensive characterization of the universal features and dynamical phases of monitored free fermion systems. While we focus on the case of U$(1)$-symmetric, particle number conserving dynamics, the approach is readily applicable to the case of $\mathbb{Z}_2$-symmetric Majorana circuits~\cite{Bernard_2018, paviglianiti2023multipartite, turkeshi2021measurementinduced, Malakar2024, zerba2023, turkeshi2023entanglement, Nahum20a, SagarKitaev, Klocke2022, Klocke2023, klocke2025, merritt2023, sang2021, sang2021b, Fava2023}.

\section{Monitored Free Fermions }

\begin{figure*}[th!]
\includegraphics[width=\linewidth]{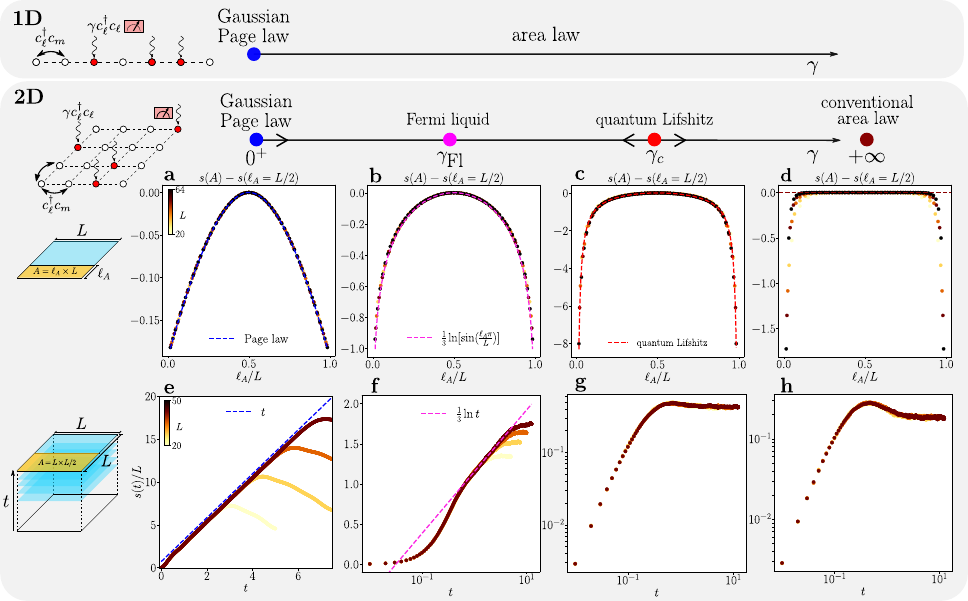}
\caption{\label{fig:cover} \textbf{Fixed points and entanglement properties of monitored free fermions.} The phase diagram of monitored free fermions in $1D$ displays a repulsive, volume-law entanglement fixed point for $\gamma\to0^+$ and an attractive, area-law fixed point for $\gamma\to+\infty$. Correspondingly, the entanglement entropy displays an area law at asymptotic scales for any $\gamma>0$~\cite{poboiko,fan2025}. For $D\geq2$ the phase diagram is enriched by a repulsive critical point at finite critical measurement rate $\gamma_c$, which is repulsive in the sense of the renormalization group (RG) and corresponds to the measurement-induced localization transition. The presence of two repulsive fixed points $\gamma=0^+, \gamma_c$ calls for a third, attractive fixed point $0^+<\gamma_\text{Fl}<\gamma_c$ in the RG sense. (a)-(d) Entanglement density $s(A)=S(A)/L$ for a cut $A=\ell_A\times L$ (modulo its value at $\ell_A=L/2$). We identify in total four distinct fixed points, described in the main text, at which the entanglement entropy follows exact analytical expressions. Here $\gamma_\text{Fl}=2.15$ and $\gamma_c=5.1$. We simulate systems of size $L^2$ with linear dimension $L=20, 30, 40, 50, 64$. (e)-(h) Half-system entanglement density $s(t)$ with $\ell_A=L/2$ as a function of time, starting from an initial Néel state for the same parameters and system sizes up to $L=50$. }
\end{figure*}

\subsection{Setup and implementation}

We consider number-conserving free fermions on a half-filled, hypercubic lattice in $d$-dimensions. Setting $L$ the linear dimension we consider creation and annihilation operators obeying $\{{c}_\ell^{\phantom{^\dagger}}, {c}_{\ell'}^\dagger\}=\delta_{\ell,\ell'}$ on sites $\ell,\ell'\in\{1,\dots, L^d\}$. At each time step $t\to t+dt$, the fermions undergo coherent nearest-neighbour hopping governed by the Hamiltonian $  H = -\sum_{\langle \ell,m\rangle} {c}_m^\dagger {c}_\ell^{\phantom{^\dagger}} + {c}_\ell^\dagger {c}_m^{\phantom{^\dagger}},$
where $\langle \ell,m\rangle$ denotes nearest-neighbouring sites. 

In addition, the local particle number ${n}_\ell$ at each site is continuously monitored with rate $\gamma$. Monitoring site $\ell$ at time step $t\to t+dt$ is implemented by the generalized projector~\cite{book_QO_Walls,book_QControl_Wiseman}
\begin{align}\label{eq:proj}
 M_\ell(J_\ell)=\big[\tfrac{2\gamma dt}{\pi}\big]^{\frac{1}{4}}
\exp[-\gamma dt (J_\ell- n_\ell)^2],
\end{align}
with $J_\ell\in\mathbb{R}$ being the local measurement outcome and normalization $\int dJ_\ell\,  M^2_\ell(J_\ell)=\mathds{1}$. This can be implemented by the infinitesimal coupling of site $\ell$ to an ancilla, followed by a projective ancilla-measurement.  

For each measurement trajectory, the normalized wave function evolves as $|\psi_{t}\rangle\to |\psi_{t+dt}\rangle$ in two steps. First, the measurement outcome $J_\ell$ at each site is drawn from the Born probabilities $p(J_\ell)=\langle\psi_t| M^2_\ell(J_\ell)|\psi_t\rangle.$ 
Second, measurement and unitary dynamics are applied, yielding
\begin{equation}\label{eq:stateupdate}
     |\psi_{t+dt}\rangle=\exp(-i H dt)\left[\prod_\ell (p(J_\ell))^{-\frac12}  M_{\ell}(J_{\ell})\right]|\psi_{t}\rangle.
\end{equation}

For the numerical implementation, it is convenient to take the limit $dt\to 0^+$, for which the variables $J_{\ell}$ approach a Gaussian distribution with mean $\mathbb{E}(J_{\ell})=\langle n_\ell\rangle$ and variance $\mathrm{var}(J_{\ell})=\tfrac{1}{2\gamma dt}$. To see this, consider the $m$-th moment
\begin{align}
    \mathbb{E}[J_{\ell}^m]=\int_{-\infty}^{+\infty} dJ_{\ell}\ J_{\ell}^m  \times p(J_{\ell}).
\end{align} 
Using the operator identity ${n}_\ell^2={n}_\ell$, one readily obtains the relation $\mathbb{E}[{J_{\ell,t}^m}]=\mathbb{E}_0[{J_{\ell,t}^m}]
    +\langle{n}_\ell\rangle\,\mathbb{E}_0[(J_{\ell,t}+1)^m-J_{\ell,t}^m].$
Here, the expression $\mathbb{E}_0[\cdot]$ denotes the average over a Gaussian distribution with zero mean and standard deviation $\sigma=(2\gamma dt)^{-\frac12}$. In the limit $dt\to 0^+$, only the highest non-vanishing moment contributes, yielding
\begin{equation}
    \mathbb{E}[{J_{\ell}^m}]= \sigma^{m-1}(m-1)!!\times
    \begin{cases}
        \sigma & m \ \text{even},\\
        \langle{n}_\ell\rangle\,  & m \ \text{odd}.
    \end{cases} 
\end{equation}
Hence, the distribution of $J_\ell$ is  Gaussian  with mean $\mu=\langle{n}_\ell\rangle$ and variance $\sigma^2$ and at each time $t$ we may replace
\begin{equation}\label{eq:replaceJ}
    J_{\ell} \to \langle n_\ell\rangle_t+\xi_{\ell,t}(2\gamma dt)^{-\frac12},
\end{equation}
with Gaussian white noise $\xi_{\ell,t}$. It fulfills  $\mathbb{E}(\xi_{\ell,t})=0$ and $\mathbb{E}(\xi_{\ell,t}\xi_{\ell',t'})=\delta_{\ell,\ell'}\delta_{t,t'}$. This allows in the limit $dt\to0^+$ to shift the state dependence of the Born probabilities into the mean of $J_\ell$, while its variance is state independent. 


Since both $ H,  M_\ell$ are quadratic in fermionic operators, any initial Gaussian state $|\psi_t\rangle$ remains Gaussian under the update in Eq.~\eqref{eq:stateupdate}~\cite{Cao2019, alberton2021enttrans, chahine2024}. With $N$ fermion states occupied (at half-filling $N=L^d/2$), it can hence be written as
\begin{equation}\label{eq:gstate}
    \ket{\psi_t}=\prod_{1\le s\le N} c^\dagger_s\ket{0}, 
    \quad 
    c^\dagger_s=\sum_{1\le\ell\le L^d} \psi^s_{\ell,t}\, c^\dagger_\ell.
\end{equation}
Here $\psi^s_{\ell,t}\in\mathds{C}$ are the single-particle wave functions, which are stochastic variables that depend on the full noise history $\{\xi_{\ell,t'<t}\}$. Normalization requires 
\begin{equation}
    \sum_\ell (\psi_{\ell,t}^s)^*\psi_{\ell,t}^{s'}=\delta_{s,s'}
     \Leftrightarrow 
    \psi_{t}^\dagger \psi_{t}=\mathds{1},
\end{equation}
where the second identity is the matrix notation with $\psi_t=\{\psi_{\ell,t}^{s}\}\in \mathbb{C}^{L^d\times N}$.

We implement the trajectory evolution by updating $\psi_t$ directly via Eq.~\eqref{eq:stateupdate} and Eq.~\eqref{eq:replaceJ}. This yields
\begin{equation}\nonumber
    \psi_{t+dt}=\text{diag}\!\left( e^{\xi_{1,t}+\frac{\gamma dt}{2} (2\langle {n}_1\rangle_t-1)},\dots,
    e^{\xi_{N,t}+\frac{\gamma dt}{2} (2\langle {n}_N\rangle_t-1)} \right)
    e^{-ihdt}\psi_{t},
\end{equation}
where $h_{\ell,\ell'}=-\delta_{\ell\text{ n.n.}\ell'}$ is the nearest-neighbour hopping matrix. Normalization is enforced via a QR (Gram–Schmidt) decomposition, i.e., by first expressing $\psi=QR$ and then redefining $\psi\equiv Q$~\cite{Cao2019}.  

Unless otherwise stated, trajectories are initialized in random Gaussian states at half-filling and evolved until observables and statistics of $\psi_{\ell,t}^s$ reach stationarity. Observables are then obtained by averaging over sufficiently many trajectories.

Due to Gaussianity, and Wick's theorem, any observable can be computed from the system's correlation matrix $\langle \psi_t|c^
\dagger_{\ell}c^{\phantom{\dagger}}_m|\psi_t\rangle=(\psi_t^{\phantom{\dagger}}\psi_t^\dagger)_{\ell,m}$. In particular, in order to compute the entanglement entropy for a subsystem $A$, one considers the correlation matrix $(\psi_t^{\phantom{\dagger}}\psi_t^\dagger)|_A$ restricted to that subsystem and computes its eigenvalues $\{\lambda_\alpha\}|_A$. The von Neumann entanglement entropy for the subsystem is then
\begin{equation}\nonumber
S_A=-\sum_\alpha\lambda_\alpha\log\lambda_\alpha+(1-\lambda_\alpha)\log(1-\lambda_\alpha).
\end{equation}

\subsection{Phase diagram of monitored free fermions in 2D}

In order to prepare the analysis of the entanglement Hamiltonian, we first discuss the phase diagram of monitored free fermions in two spatial dimensions based on subsystem entanglement scaling. Our results here reconcile and advance previously reported findings~\cite{chahine2024,poboiko2023measurementinduced,fan2025} and establish the existence of four characteristic ``fixed points'' as a function of the measurement strength $\gamma$, illustrated in Fig.~\ref{fig:cover}:  
\begin{itemize}
    \item[(i)] a Gaussian Page-law fixed point at $\gamma=0^+$,  
    \item[(ii)] a Fermi-liquid fixed point at $\gamma=\gamma_{\text{Fl}}\approx 2.15$,  
    \item[(iii)] a quantum Lifshitz fixed point at $\gamma=\gamma_c\approx 5.16$,  
    \item[(iv)] an area-law fixed point at $\gamma\to+\infty$.  
\end{itemize}
To probe the entanglement characteristics at these fixed points, we study the dependence of the \emph{linear entanglement entropy density} $s_A\equiv S_A/L$ on subsystem geometry, focusing on strips of size $A=\ell_A\times L$ with linear length $L$ in one dimension and varying length $\ell_A$ in the other dimension. For all fixed points and finite system sizes, we find matching analytic expressions describing the subsystem scaling behavior exactly. The analytical dependence contains no free parameters at (i), (ii), and (iv), while point (iii) involves a single fitting parameter. We refer to them as fixed points in the sense of the renormalization group (RG): (i) and (iii) are repulsive, corresponding respectively to (i) a Gaussian page law, which is unstable to infinitesimal monitoring and representing the limit of fast scrambling, chaotic unitary dynamics~\cite{Cao2019} and (iii) to the measurement-induced localization-delocalization transition. The points (ii) and (iv) as a consequence need to be attractive and define robust phases, respectively a metallic (Fermi-liquid-type) phase and a localized phase.

In addition, we characterize the dynamical properties of entanglement by considering a half-system cut $A=L/2\times L$. Starting from a Néel initial state $|\psi_{t=0}\rangle$, we analyze the entanglement growth as a function of time due to the competition between unitary dynamics and monitoring.

We emphasize that, while aspects of this phase diagram have been analyzed previously~\cite{chahine2024,poboiko2023measurementinduced,fan2025}, our work is the first to establish the coexistence of all four fixed points within a single unified phase diagram and to demonstrate their exact correspondence with analytical expressions.

\subsubsection*{Fixed point I: Gaussian Page law}

For $\gamma\to0^+$, the dynamics is governed by the integrable free-fermion Hamiltonian $ H$. Infinitesimal monitoring breaks integrability and drives the system toward a well-defined stationary entanglement entropy. The resulting state is maximally chaotic while remaining within the Gaussian manifold. Consequently, one expects a Page-law–type scaling modified by Gaussian constraints. Indeed, a Page curve for pure fermionic Gaussian states in a one-dimensional geometry was recently derived using random matrix theory~\cite{bianchi2021}. For a perfectly random state, the results are expected to apply independently of the dimensionality. Hence, we expect the subsystem entanglement entropy density to be described by
\begin{align}
     s_A=&\left(L-\tfrac{1}{2}\right)\Psi(2L)
    +\left(\tfrac{1}{2}+\ell_A-L\right)\Psi(2L-2\ell_A) \nonumber \\
    &+\left(\tfrac{1}{4}-\ell_A\right)\Psi(L)
    -\tfrac{1}{4}\Psi(L-\ell_A)-\ell_A,
\end{align}
where $\Psi(z)=\Gamma'(z)/\Gamma(z)$ is the digamma function. As shown in Fig.~\ref{fig:cover}(a), this expression exactly describes the entanglement entropy for arbitrary finite $L$ and $\ell_A$, yielding a perfect match with our numerical simulations.  

Turning to dynamics, we initialize the system in a highly excited Néel state. Entanglement spreads ballistically and exhibits linear growth in time, $s_A(t)\sim t$, consistent with analytical predictions~\cite{Alba2018, Calabrese_2005}. Our numerical results confirm this behavior within the monitored free-fermion setup, see Fig.~\ref{fig:cover}(e). 

\subsubsection*{Fixed point II: Fermi-liquid scaling}

At intermediate monitoring strength, $\gamma=\gamma_{\text{Fl}}=2.15$, Ref.~\cite{chahine2024} reported a fixed point characterized by scale-invariant behavior of the entanglement entropy. We confirm this result by explicitly computing the subsystem entanglement entropy, finding an exact match with
\begin{equation}
    s_A = \tfrac{1}{3}\ln\!\left[L\sin\!\left(\tfrac{\pi \ell_A}{L}\right)\right] + s_0,
\end{equation}
as shown in Fig.~\ref{fig:cover}(b). Remarkably, this is precisely the scaling form of the ground state of $ H$ at half filling, i.e., of a perfect Fermi liquid. Thus, the interplay between unitary dynamics and continuous monitoring dynamically realizes a state with the same  entanglement properties as the ground-state Fermi liquid. In analogy with conventional ground-state physics in the presence of moderate disorder or interactions, we therefore refer to this regime as the \emph{Fermi-liquid fixed point}.  

Further, we analyze the real-time build-up of entanglement starting from a Néel initial state. Here we find that the entanglement entropy grows logarithmically in time,
\begin{equation}
    s_A(t)= \tfrac{1}{3}\ln(t),
\end{equation}
with the same prefactor as in the static scaling law; see Fig.~\ref{fig:cover}(f) and Fig.~\ref{fig:MI}(a) in Appendix~\ref{appendixA} for an independent estimate of the prefactor via static law scaling with $L$. This logarithmic growth highlights the emergence of space–time invariance at the fixed point and is consistent with a dynamical critical exponent $z=1$~\cite{chahine2024}. Together, these results reinforce the interpretation of $\gamma=\gamma_{\text{Fl}}$ as a genuine scale-invariant fixed point of the monitored free-fermions.

\subsubsection*{Fixed point III: Quantum Lifshitz scaling}

A further fixed point, previously identified from a nonlinear sigma model analysis and from numerical simulations of projectively measured free fermions in $D=2$~\cite{poboiko2023measurementinduced}, emerges at stronger monitoring, $\gamma=\gamma_c\gtrsim 5.16$. This point separates an entangled phase with logarithmic growth of entanglement from a localized area-law phase, and its nature has been linked to Anderson localization transitions in $D+1$ dimensions. For continuously monitored fermions, we locate this transition at $\gamma_c=5.1$ (see Appendix~\ref{appendixA}).  

At this fixed point, we find that the subsystem entanglement entropy obeys an area law with a distinctive correction of quantum Lifshitz type; see Fig.~\ref{fig:cover}(c). Specifically, the entropy takes the form (with $u=\ell_A/L$)
\begin{align}
    s_A &= a\,J(u)/L + b,\nonumber\\
    J(u) &= \log\!\left(\frac{\theta_3(i\lambda u)\,\theta_3(i\lambda (1-u))}{\eta(2iu)\,\eta(2i(1-u))}\right),
\end{align}
with $a\approx-1$ and $b\approx0.3$. Here, $J(u)$ is the Lifshitz scaling function with $\theta_3$ the Jacobi theta function, $\eta$ the Dedekind eta function, and $\lambda$ a free parameter. Numerically we find $\lambda=1$. This scaling form was first derived in the quantum Lifshitz model (QLM)~\cite{ardonne2004} and subsequently observed in several $(2\!+\!1)$-dimensional CFTs~\cite{inglis2013, chen2015}. More recently, it has also been reported at the phase transition of monitored Kitaev circuits~\cite{klocke2025}. In the QLM, $\lambda$ corresponds to the exponent of the two-point dimer correlation function~\cite{Fradkin2013}, with $\lambda=2$ at the Rokhsar–Kivelson critical point~\cite{chen2015}, while no physical interpretation of $\lambda$ is reported for fermionic models~\cite{chen2015}. The appearance of quantum Lifshitz scaling at the monitored free-fermion transition is therefore unprecedented as it has not been linked to Anderson-like localization transitions.  

Turning to dynamics, we find that the entanglement entropy grows rapidly and saturates on time scales $t\sim\mathcal{O}(1)$, consistent with area-law behavior. We do not observe substructures in the growth resembling the Lifshitz correction.

\subsubsection*{Fixed point IV: Conventional area law}

In the limit $\gamma\to+\infty$, measurements dominate and drive the steady state into a featureless area-law regime. The subsystem entanglement entropy becomes flat for $\ell_A\gg 1$, with no fine structure, as only short-range correlations across the subsystem boundary contribute. The entanglement dynamics in this regime closely resembles that at the critical point $\gamma_c$, but with saturation to the area-law value.

\subsubsection*{Phases and phase transitions}

The fixed points define a consistent phase diagram of monitored fermions in two dimensions, in agreement with Refs.~\cite{chahine2024,poboiko2023measurementinduced}. For $0<\gamma<\gamma_c$ the system is metallic, showing a logarithmic violation of the area law. Within this regime, the Fermi-liquid point at $\gamma=\gamma_{\text{Fl}}$ is attractive and displays space-time scale invariance. In contrast, the Gaussian Page point at $\gamma\to 0^+$ is repulsive and represents the unstable limit of chaotic unitary dynamics. For $\gamma>\gamma_c$ the system becomes localized, with area-law entanglement and an attractive fixed point at $\gamma\to+\infty$. Finally, at $\gamma=\gamma_c$ a measurement-induced phase transition occurs, governed by a repulsive quantum Lifshitz fixed point with characteristic scaling corrections to the entanglement.

\begin{figure}[th!]
\includegraphics[width=\linewidth]{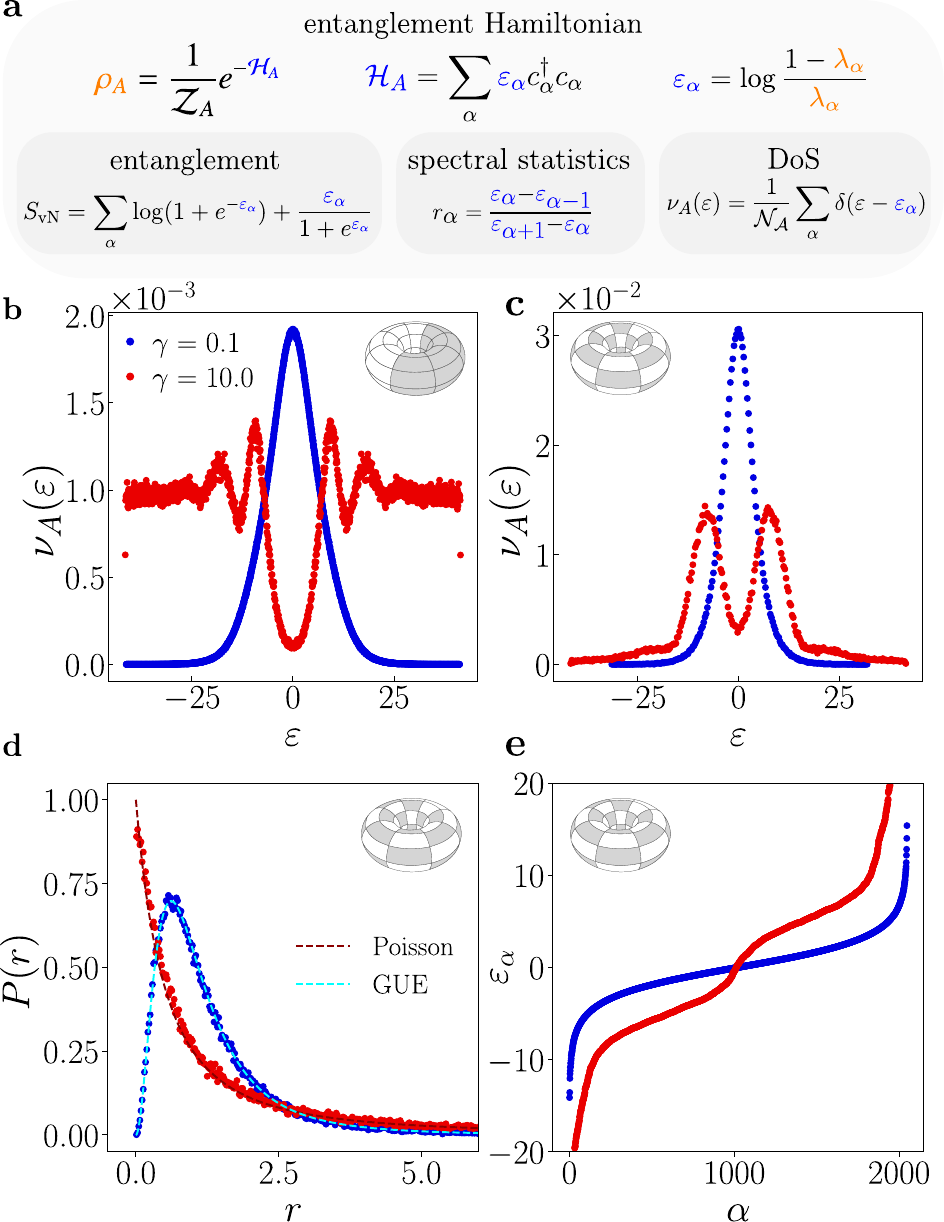}
\caption{\label{fig:DoS}\textbf{Subsystem entanglement Hamiltonian.} (a) Definition of the entanglement Hamiltonian $\mathcal{H}_A$ from the subsystem density matrix $\rho_A$ on some region $A$. $\mathcal{H}_A$ gives access to the entanglement spectrum $\{\epsilon_\alpha\}$ and a set of unambiguous single-particle wave functions $\{\psi^A_\alpha(\ell)\}$, which in turn can be used to compute entanglement entropies, spectral statistics and spectral and wave function correlations. (b) and (c) show the density of states (DoS) at $L=50$, $\gamma=0.1$ and $\gamma=10$ for $A$ being a half-cut geometry or a checkerboard pattern, respectively. Here the DoS is averaged over several trajectories. The half-cut geometry (b) features fat tails of the DoS, which correspond to an over-representation of localized wave functions in the bulk of the region $A$ at high $\gamma$. These tails do not provide any relevant information on the structure of the eigenstates and are absent on the checkerboard geometry (c). (d) Probability distibution of ratios of next-level spacings as defined in (a) for $\gamma=0.1$ (blue) and $\gamma=10$ (red) on the checkerboard geometry. This is for $r$ values computed over the full ensemble of eigenvalues, i.e. bunched up across trajectories. Our numerical data perfectly matches the theoretical predictions for a localized phase (Poissonian distribution) or an ergodic one (Wigner-Dyson GUE distribution). (e) Example spectrum for a single trajectory on the checkerboard subsystem for $\gamma=0.1$ (blue) and $\gamma=10$ (red). The spectrum becomes less dense around $0$ as the measurement rate increases, as it can be seen in the DoS in (a), (b). }
\end{figure}

\section{Entanglement Hamiltonian of monitored free fermions}

To complement the entanglement analysis presented above, we now turn to the entanglement Hamiltonian, analyzed from a random matrix theory perspective. We show that this provides a powerful diagnostic: it clearly identifies an ergodic phase for $\gamma<\gamma_c$ and a non-ergodic phase for $\gamma>\gamma_c$, consistent with the metallic and localized phase determined from the entanglement entropy. It further displays a sharp transition between them at $\gamma=\gamma_c$. In particular, the entanglement Hamiltonian not only reproduces the  phase structure qualitatively and quantitatively but it also enables highly accurate predictions for the critical point and critical exponents already for moderate system sizes.

\subsection{Definition of the entanglement Hamiltonian}
For any subsystem density matrix $\rho_A=\text{tr}_{\tilde A}\rho$, with $A$ the subsystem and $\tilde A$ its complement, the corresponding entanglement Hamiltonian $\mathcal{H}_A$ is defined as $\rho_A=\mathcal{Z}_A^{-1}\exp(-\mathcal{H}_A)$ with $\mathcal{Z}_A=\text{tr}(\exp(-\mathcal{H}_A))$. This definition is well-posed since $\rho_A$ is Hermitian and positive semi-definite and has unit trace. For a general system, the normalization, or partition function, $\mathcal{Z}_A$ can, in principle, be absorbed in the definition of $\mathcal{H}_A$. For Gaussian states, however, it will be useful to keep $\mathcal{Z}_A$ explicit. If the original state $\rho$ is a Gaussian state, then $\rho_A$ is also Gaussian~\footnote{This can be easily verified by confirming Wick's theorem for $\rho_A$.}. As a consequence, $\mathcal{H}_A$ has a free fermion structure
\begin{align}
    \mathcal{H}_A=\sum_{\ell,m\in A}\cre{\ell}\Theta_{\ell,m}^A\ann{m}.
\end{align}
This defines the $|A|\times|A|$ Hermitian subsystem entanglement matrix $\Theta^A$, which is in one-to-one correspondence with the subsystem correlation functions 
\begin{align}
    \langle \cre{\ell}\ann{m}\rangle_A=\tfrac12[\mathds{1}-\tanh(\tfrac12\Theta^A)]_{\ell,m}
\end{align}

The matrix $\Theta^A$, and hence the  entanglement Hamiltonian $\mathcal{H}_A$ and the correlation matrix $\langle \cre{\ell}\ann{m}\rangle_A$ can be diagonalized by a unitary transformation $V^A$ on the subsystem. This yields $\mathcal{H}_A=\sum_\alpha\varepsilon_\alpha \tilde c^\dagger_\alpha \tilde c_\alpha$ and $\langle \tilde c^\dagger_\alpha \tilde c_\beta \rangle=\delta_{\alpha\beta}\lambda_\beta$ with occupations 
\begin{align}
\lambda_\alpha=\tfrac12[1-\tanh(\tfrac{\varepsilon_\alpha}{2})]=(e^{\varepsilon_\alpha}+1)^{-1}.
\end{align}
The operators  $\tilde{c}^\dagger_\alpha=V^A_{\ell,\alpha}c^\dagger_\ell$ represent the single-particle eigenmodes of subsystem $A$ and $V^A_{\alpha,\ell}=\psi_\alpha(\ell)$ are the corresponding single-particle wave functions. 

Hence, the entanglement Hamiltonian provides an effective Hamiltonian for any partition $A$, with an unambiguous set of eigenvalues $\{\varepsilon_\alpha\}$ and eigenfunctions $\{\psi_\alpha\}$~\footnote{The original wave functions $\psi^s_{\ell,t}$ get renormalized at every time step. This changes the basis at each time step and provides an ambiguous set of wave functions, hindering numerical probes applied to them~\cite{Szyniszewski2024}.}. In the following, we analyze these spectral and wave-function properties in detail using tools from random matrix theory (RMT). The definition of the entanglement Hamiltonian and the tools we use to analyze it are summarized in Fig.~\ref{fig:DoS}(a). 

\subsection{Spectrum of the entanglement Hamiltonian and entanglement geometries}

The entanglement Hamiltonian fully encodes the subsystem entanglement entropy. For example, the von Neumann entropy is obtained as the global average
\begin{align}
    S_A=-\mathrm{tr}(\rho_A\ln\rho_A)=\ln \mathcal{Z}_A + \langle \mathcal{H}_A\rangle .
\end{align}

A particularly useful diagnostic is the density of states (DoS) of $\mathcal{H}_A$, defined in analogy to conventional Hamiltonians as
\begin{align}
    \nu_A(\varepsilon)=\frac{1}{\mathcal{N}_A}\sum_\alpha \delta(\varepsilon-\varepsilon_\alpha),
\end{align}
where $\{\varepsilon_\alpha\}$ are the eigenvalues of $\mathcal{H}_A$, $\mathcal{N}_A$ is the number of levels and the Dirac $\delta$-function is suitably broadened for finite systems. $\nu_A(\varepsilon)$ counts the density of entanglement modes in a small window around $\varepsilon$.

In Fig.~\ref{fig:DoS}(b) we show $\nu_A(\varepsilon)$ for weak and strong monitoring rates $\gamma$ in the half-cut geometry $A=L\times L/2$, illustrated by the colored torus in Fig.~\ref{fig:DoS}(b). For weak monitoring, single-particle wave functions are delocalized and the cut removes roughly half of each mode. This produces a peak in the DoS at $\varepsilon=0$, corresponding to $\lambda_\alpha=\tfrac12$ occupation. For strong monitoring, wave functions are localized and are either fully included in or excluded from the subsystem. Consequently, $\lambda_\alpha\to0,1$, the DoS develops a gap around $\varepsilon=0$, and weight is shifted towards $\varepsilon\to\pm\infty$.

Since $\mathcal{H}_A$ is defined from $\rho_A=\mathrm{tr}_{\bar A}\rho$, its spectrum depends both on the underlying state $\rho$ and the subsystem geometry $A$. For example, switching from the half-cut to a checkerboard geometry $A=\{\text{even columns} \text{ in odd rows and odd columns in even rows }\}$, illustrated in Fig.~\ref{fig:DoS}(c), drives the entanglement entropy towards a volume law irrespective of the underlying state (except for the trivial limit $\gamma\to+\infty$). By contrast, the DoS -- a local quantity in entanglement energy -- remains qualitatively unchanged: as shown in Fig.~\ref{fig:DoS}(c), it retains the same weak- and strong-monitoring features as for the half-cut.

In fact, the checkerboard geometry is even advantageous for spectral properties of $\mathcal{H}_A$. As seen in Figs.~\ref{fig:DoS}(b,c), the half-cut carries substantial spectral weight in the tails of the distribution, especially in the localized regime, while the checkerboard geometry confines the spectral density to the center of the spectrum. This originates from the fact that the half-cut strongly perturbs wave functions near the boundary but leaves those deep in the bulk nearly unaffected. The checkerboard, in contrast, samples all single-particle states more evenly, making it better suited for detecting correlations across the entire spectrum. We stress, however, that both geometries yield consistent results for the measurement-induced phase transition; the checkerboard is simply more efficient for spectral analysis.

Our goal is to analyze the spectrum of $\mathcal{H}_A$ using tools of random matrix theory (RMT). A central perspective in RMT is that correlations between eigenstates manifest in the statistics of their eigenvalues: extended eigenstates $\{\psi_\alpha\}$ have overlapping densities $|\psi_\alpha|^2$, which induces level repulsion between nearby eigenvalues. The same reasoning applies to the entanglement Hamiltonian, viewing $\Theta^A$ as a random matrix in real space. Figure~\ref{fig:DoS}(d) confirms this picture: for weak monitoring, the ratios of consecutive level spacings 
\begin{align}
    r_\alpha=\frac{\varepsilon_\alpha-\varepsilon_{\alpha-1}}{\varepsilon_{\alpha+1}-\varepsilon_\alpha}
\end{align}
follow a Wigner--Dyson distribution, with the probability $P(r)$ of finding level spacings $r$ being strongly suppressed for $r\to0$. This signals level repulsion and thus correlations between eigenstates. By contrast, strong monitoring yields a Poisson distribution peaked at $r\to0$, consistent with the absence of correlations. This establishes that the entanglement Hamiltonian faithfully captures the metallic phase through Wigner--Dyson statistics and the localized phase through Poisson statistics.

\subsection{Characterizing monitored fermion states from the entanglement Hamiltonian}

We now extend the above analysis to obtain quantitative signatures of the metallic and localized phases of monitored fermions directly from the spectrum and eigenstates of the entanglement Hamiltonian $\mathcal{H}_A$ (or equivalently, $\Theta^A$). This framework will allow us to accurately characterize the measurement-induced phase transition in $d=2$ dimensions, and later in $d=1$ and $d=3$. Our approach distinguishes two complementary probes: (a) short-range correlations in the spectrum and eigenstates, and (b) long-range correlations.

Spectral analysis provides a powerful and universal toolbox, as many physical systems fall into one of two limiting regimes: the Poissonian distribution, characteristic of integrable systems with localized eigenstates and no level repulsion, or the Wigner–Dyson distribution, characteristic of chaotic systems with extended eigenstates and strong level repulsion. Depending on symmetry, the latter is described by one of the three Gaussian ensembles (GOE, GUE, GSE)~\cite{mehta1967}. Interpolations between these limits play a central role in problems ranging from Anderson localization to many-body localization~\cite{Zharekeshev1997, Serbyn2016, Sierant2019, Pal2010, suntajs2020}. In what follows, we adapt this perspective to the entanglement Hamiltonian to distinguish metallic from localized behavior in monitored fermions.

\subsubsection*{Short-range spectral correlations}

\begin{figure}[t]
\includegraphics[width=\linewidth]{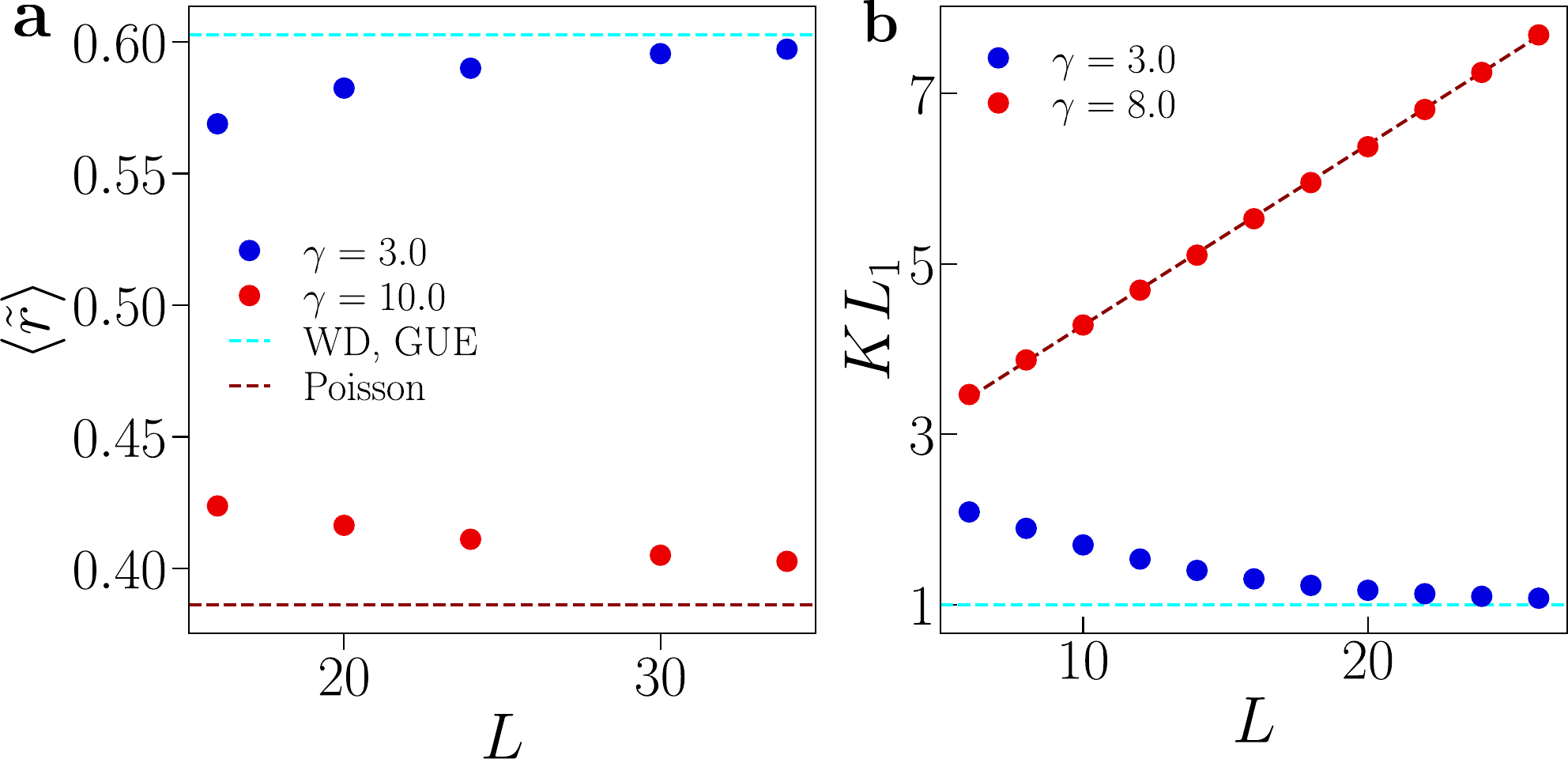}
\caption{\label{fig:short_range_limits}\textbf{Short-range spectral correlations of monitored fermions.} 
Average gap ratio (a) and Kullback–Leibler divergence $KL_1$ (b) for a two-dimensional system as a function of the linear system size $L$ for weak (blue) and strong (red) measurement rates, corresponding to the ergodic and non-ergodic phases, respectively. (a) In the large-$L$ limit, $\langle\tilde r\rangle$ converges to the predictions of the Wigner–Dyson and Poisson distributions. 
(b) For weak monitoring $KL_1=O(1)$, consistent with delocalized wave functions, while for strong monitoring $KL_1\propto L$, as expected for localized states.}
\end{figure}

We begin with short-range spectral correlations~\cite{brody1981, berry1977, bohigas1984, atas2013}, i.e., correlations between neighboring eigenvalues and their corresponding eigenstates. In ergodic phases, extended eigenstates overlap, producing correlations between adjacent levels and hence Wigner–Dyson (WD) statistics. In contrast, localized states do not overlap, leading to uncorrelated Poisson statistics.

A convenient probe of these correlations is the ratio of consecutive spacings, 
\begin{align}
        \tilde{r}_\alpha = \min[r_\alpha, r_\alpha^{-1}].
\end{align}
The average $\langle \tilde{r}\rangle=|A|^{-1}\sum_{\alpha =1}^{|A|}\tilde r_\alpha$ takes universal values~\cite{atas2013}: $\langle \tilde{r}\rangle_{\text{P}}\approx0.38629$ for the Poissonian limit and $\langle \tilde{r}\rangle_{\text{GUE}}\approx0.60266$ for the Gaussian unitary ensemble (GUE). As a single scalar quantity, $\langle \tilde{r}\rangle$ is particularly useful, effectively acting as an order parameter distinguishing localized from ergodic behavior without requiring the full distribution $P(r)$.

Short-range correlations can also be accessed directly through eigenfunctions. Localized states are spatially disjoint, while extended states overlap and hybridize, producing both spectral level repulsion and enhanced wave-function overlap. A convenient measure of this overlap is the Kullback–Leibler (KL) divergence~\cite{Kullback1951}
\begin{equation}\label{eq:KL1}
    KL_1 = \frac{2}{L}\sum_\alpha\sum_i|\psi_\alpha(i)|^2
    \ln\frac{|\psi_\alpha(i)|^2}{|\psi_{\alpha+1}(i)|^2},
\end{equation}
where $\psi_\alpha$ is the $\alpha$-th eigenvector ordered by increasing eigenvalue, and the sums run over components $i$ and indices $\alpha$. For delocalized wave functions $KL_1=\mathcal{O}(1)$, while for localized ones $KL_1$ grows with system size~\cite{Luitz2015,Colbois2024, Khaymovich2020, kravtsov2020, Pino2019}. For the entanglement Hamiltonian, the interpretation differs slightly: consecutive eigenmodes correspond to states with similar total weight in the subsystem, but the same logic applies: small $KL_1$ signals extended entanglement modes, while large $KL_1$ indicates localization.

 We now numerically analyze the eigenvalues and eigenvectors of the entanglement Hamiltonian $\mathcal{H}_A$, obtained from the reduced density matrix $\rho_A$ on the checkerboard subsystem. Since $\mathcal{H}_A$ is a Hermitian Hamiltonian, its spectrum is invariant under unitary transformations and should follow GUE statistics in the delocalized phase. From the entanglement spectrum of each trajectory we compute the average spacing ratio $\langle\tilde{r}\rangle$, first averaging over levels of a single spectrum and then over trajectories. As shown in Fig.~\ref{fig:short_range_limits}(a), $\langle\tilde{r}\rangle$ converges with system size $L$ to the RMT values for WD and Poisson statistics, respectively confirming the ergodic and localized limits.  

The same dichotomy appears in the eigenstates of $\mathcal{H}_A$. For each trajectory we compute $KL_1$ [Eq.~\eqref{eq:KL1}] and average over trajectories. Fig.~\ref{fig:short_range_limits}(b) shows $KL_1\sim \mathcal{O}(1)$ under weak monitoring, consistent with extended states, and $KL_1\sim L$ under strong monitoring, as expected for localized states.

\begin{figure}[t] \includegraphics[width=\linewidth]{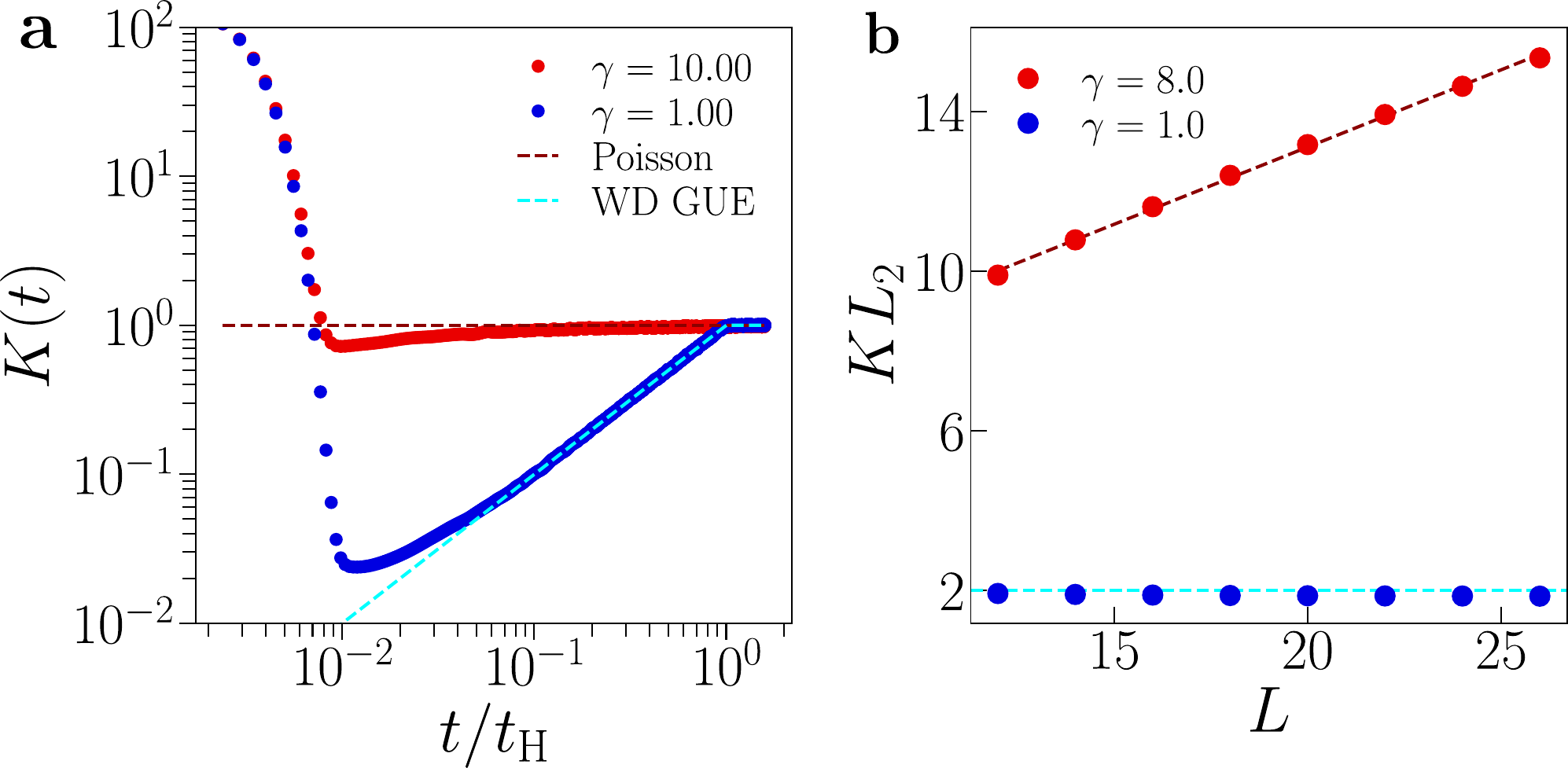} \caption{\label{fig:long_range_limits}\textbf{Long-range spectral correlations of monitored fermions.} Spectral form factor (SFF) (a) and Kullback–Leibler divergence $KL_2$ (b) for a two-dimensional system as a function of the linear system size $L$ for weak (blue) and strong (red) measurement rates, corresponding to the ergodic and non-ergodic phases, respectively.
(a) For weak monitoring, the SFF approaches the universal WD form of Eq.~\eqref{eq:SFF_GUE}. For strong monitoring, the SFF matches that of  a Poisson distribution after an initial non-universal dip. (b)  For weak (strong) monitoring $KL_2\sim O(1)$ $(KL_2\sim L)$, as predicted for delocalized (localized) wave functions.} \end{figure}

\subsubsection*{Long-range spectral correlations}

Short-range level statistics such as $\langle \tilde{r}\rangle$ and $KL_1$ probe correlations between adjacent levels or states. To reveal long-range spectral correlations, one uses the spectral form factor (SFF)~\cite{suntajs2020,edwards1972,Hopjan2023,Kos2018},  
\begin{equation}
    K(t) = \frac{1}{Z}\Big|\sum_\alpha e^{i\varepsilon_\alpha t}\Big|^2,
\end{equation}
the Fourier transform of the two-point spectral correlator. Here $t$ plays the role of time, and the normalization $Z$ is chosen such that $K(t\!\to\!\infty)=1$. For an uncorrelated (Poisson) spectrum, $K(t)=1$ at all times. For a GUE spectrum, rescaling time as $\tau=t/T_\text{H}$ with $T_\text{H}$ the Heisenberg time (inverse mean level spacing), one finds
\begin{equation}\label{eq:SFF_GUE}
   K(\tau) = \begin{cases}
        \tau, & \tau<1,\\
        1, & \tau\ge 1.
    \end{cases}
\end{equation}
Physically, $T_\text{H}$ marks the scale beyond which the discreteness of the spectrum becomes relevant. Since the mean level spacing typically decays exponentially with system size (i.e., linear in the Hilbert space size), $T_\text{H}$ grows exponentially.  

In realistic ergodic systems, the SFF deviates from this universal form at early times $t<T_\text{Th}$, defining the Thouless time $T_\text{Th}$. In diffusive systems $T_\text{Th}$ grows polynomially with system size, so that $\tau_\text{Th}=T_\text{Th}/T_\text{H}\to0$ in the thermodynamic limit and the RMT prediction becomes exact~\footnote{In systems for which the Hilbert space is not exponentially growing with system size, the Heisenberg time also scales polynomially. Crucially, it still grows faster than $T_\text{Th}$ so that the ratio $\tau_\text{Th}=T_\text{Th}/T_\text{H}$ still vanishes. For instance, in the $3D$ Anderson model, the Thouless time in the diffusive regime is found to scale quadratically in $L$, while the Heisenberg time scales as $T_\text{H}\sim L^3$~\cite{sierant2020thouless}.}. In localized or integrable systems, by contrast, the system never fully explores its Hilbert space: the SFF remains flat, and $\tau_\text{Th}=1$. The ratio $\tau_\text{Th}=T_\text{Th}/T_\text{H}$, often interpreted as a dimensionless conductance, thus quantifies ergodicity.  

Intermediate regimes have also been proposed, in which correlations persist only on short energy scales but vanish on larger ones, making such regimes invisible under short-range analysis. This ``non-ergodic extended'' (NEE) or multifractal phase is characterized by a Thouless time $\tau_\text{Th}\in(0,1)$ in the thermodynamic limit, i.e., by long-range correlations that are invisible to short-range probes of $\langle \tilde{r}\rangle$ or $KL_1$. An analytically tractable example is provided by the Rosenzweig-Porter model, a single-parameter random matrix ensemble displaying an ergodic, an NEE and a localized phase~\cite{Rosenzweig1960,Kravtsov2015rp}.

To probe eigenstate correlations in such NEE phases directly, one can use a second variant of the Kullback--Leibler divergence~\cite{kravtsov2020,Khaymovich2020},  
\begin{equation}\label{eq:KL2}
    KL_2 = \frac{2}{L}\sum_\alpha \sum_i |\psi_\alpha(i)|^2
    \ln \frac{|\psi_\alpha(i)|^2}{|\tilde{\psi}_{\alpha+1}(i)|^2},
\end{equation}
where $\psi_\alpha$ and $\tilde{\psi}_{\alpha+1}$ are eigenstates from different realizations but close in energy. As with $KL_1$, one finds $KL_2=\mathcal{O}(1)$ for extended states and $KL_2\sim L$ for localized ones. Crucially, in an NEE phase eigenstates occupy only a vanishing fraction of the system, yet remain uncorrelated across realizations. This yields a divergent $KL_2\sim L$, while $KL_1\sim \mathcal{O}(1)$ in the same regime. Hence, while $KL_1$ and $KL_2$ exhibit the same scaling in ergodic and localized phases, they differ in a NEE phase, where $KL_1\neq KL_2$.

\subsubsection*{Numerical computation of the spectral form factor}
Accurate numerical computation of the SFF requires unfolding the spectrum to remove the influence of the local density of states. We adopt the following procedure~\cite{suntajs2020}: for each trajectory, we define the cumulative spectral function  
\begin{equation}
    C(\varepsilon) = \sum_\alpha \Theta(\varepsilon-\varepsilon_\alpha).
\end{equation}
We then smoothen it across all trajectories with a spline fit $\overline{C}_s(\varepsilon)$, defining unfolded eigenvalues as $\tilde{\varepsilon}_\alpha = \overline{C}_s(\varepsilon_\alpha)$. To suppress contributions from spectral edges, we apply a Gaussian filter  
\begin{equation}
    g(\varepsilon) = \exp\Big[-\frac{(\varepsilon-\overline{\varepsilon})^2}{2 (\eta \Gamma)^2}\Big],
\end{equation}
where $\overline{\varepsilon}$ and $\Gamma^2$ are the mean and variance of the spectrum for a given trajectory, and $\eta=0.5$ controls the fraction of states included. The SFF is then computed as  
\begin{equation}
    K(t) = \Big\langle \frac{1}{Z} \Big|\sum_\alpha g(\tilde{\varepsilon}_\alpha) e^{i \tilde{\varepsilon}_\alpha t}\Big|^2 \Big\rangle,
\end{equation}
with $Z=\sum_\alpha |g(\tilde{\varepsilon}_\alpha)|^2$.

Fig.~\ref{fig:long_range_limits} shows the long-range correlations for weak and strong monitoring. Under weak monitoring, the SFF follows the GUE prediction from early times, while under strong monitoring $K(t)\approx 1$, consistent with a Poisson spectrum. Similarly, $KL_2$ confirms the ergodic vs. localized dichotomy: $KL_2=\mathcal{O}(1)$ for weak monitoring and $KL_2\sim L$ for strong monitoring, mirroring the behavior of $KL_1$ for single-trajectory eigenstates.

\section{Measurement-induced spectral phase transition in the entanglement Hamiltonian}
In the previous section, we described the phases of monitored free fermions in two dimensions, i.e., an ergodic, metallic phase and a non-ergodic, localized phase, using the spectrum and eigenstates of the  entanglement Hamiltonian. Here, we show that this framework is also powerful in characterizing the measurement-induced phase transition and its critical properties. We start with the two-dimensional setup and then analyze three and one spatial dimensions. We also explore the possibility of a non-ergodic extended phase in the spectrum, which provides inconclusive results.

\subsection{Spectral transition in two dimensions}

\subsubsection*{Short-range correlations}

\begin{figure}[t]
\includegraphics[width=\linewidth]{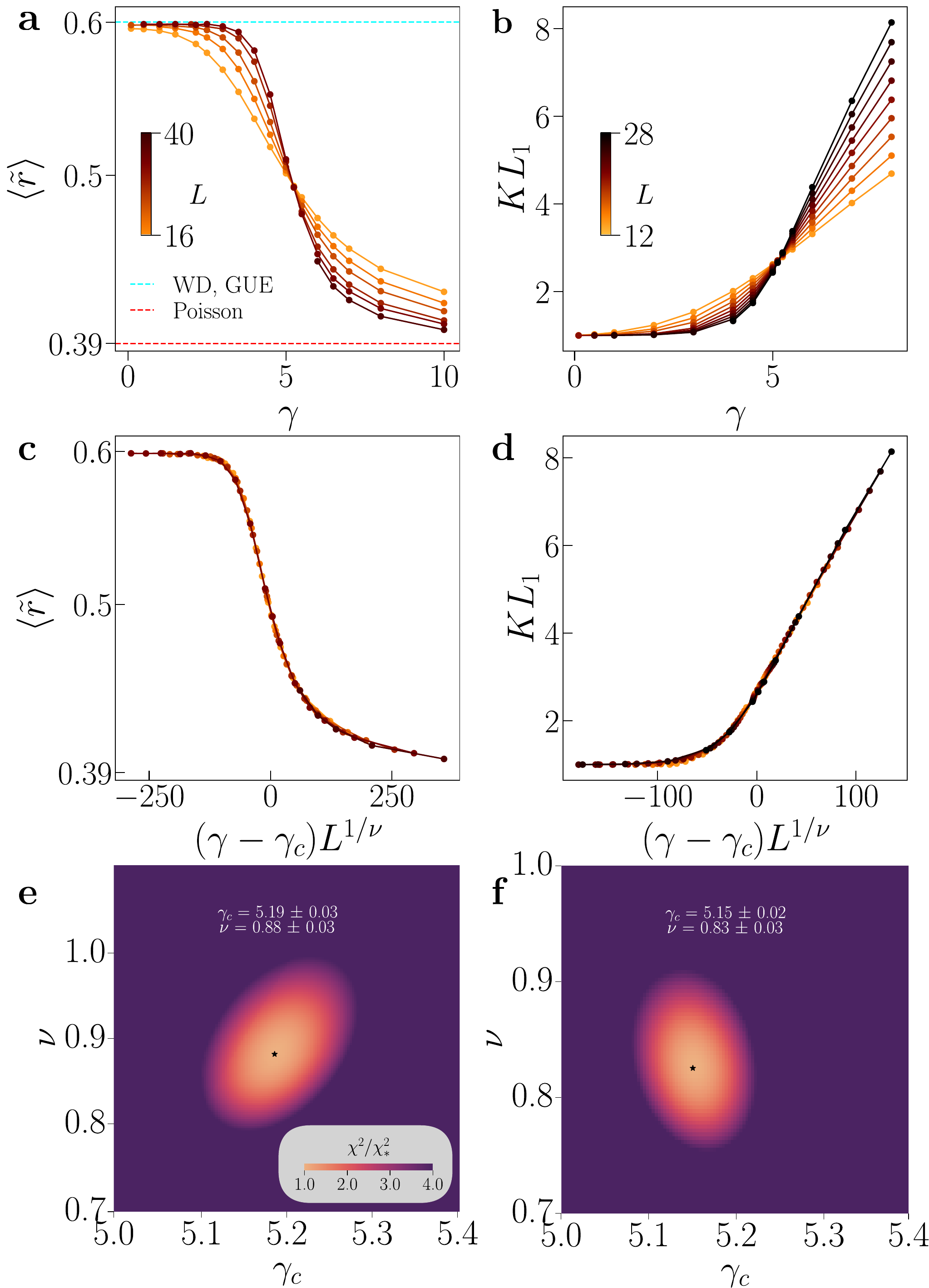}
\caption{\label{fig:r_tilde_KL1} \textbf{Spectral transition in the entanglement Hamiltonian in two dimensions.} 
(a) Average gap ratio $\langle \tilde r\rangle$ for system sizes $L\in[16,40]$ as a function of $\gamma$, showing a sharp crossing and converging to the ergodic (localized) limits at weak (strong) monitoring. 
(b) $KL_1$ for $L\in[12,28]$ as a function of $\gamma$, likewise displaying a sharp crossing. 
(c,d) Finite-size scaling collapse of $\langle \tilde r\rangle$ and $KL_1$ using the optimal critical parameters $(\gamma_c,\nu)$ extracted from the cost-function analysis. 
(e,f) Heatmaps of the normalized cost function for $\langle \tilde r\rangle$ and $KL_1$, truncated at $4\chi_*$ to visualize the uncertainty of the critical parameters. The optimal values minimize the cost function, with error bars estimated from the contour $\chi_*+4$. Estimates from the two observables are mutually consistent.}
\end{figure}
We analyze the average ratio $\langle\tilde{r}\rangle$ and the KL$_1$ as a function of $\gamma$ for different system sizes in Fig.~\ref{fig:r_tilde_KL1}. Both parameters display a transition from their characteristic values at small $\gamma$ and in the ergodic, metallic phase towards the localized phase at large $\gamma$. Both regimes are separated by a sharp crossing point, detecting the critical value $\gamma_c$ of the transition with large accuracy. 

We perform a scaling collapse analysis to extract the values of the critical point $\gamma_c$ and critical exponent $\nu$. In particular, we assume a scaling ansatz of the form $F(\gamma, L)=F((\gamma-\gamma_c)L^{\frac{1}{\nu}})$ and determine $(\gamma_c,\nu)$ through a minimization procedure of a cost function $F_\text{cost}(\gamma_c, \nu)$. The cost function is a measure of the weighted residual sum of squares between the collapsed ensemble of data for a given $(\gamma_c, \nu)$ and a spline fitting the data. We find an excellent scaling collapse and extract 
\begin{align}
    \begin{array}{|l|l|l|}
    \hline
     & \gamma_c & \nu\\
     \hline
     \langle \tilde r\rangle: & 5.19\pm0.03 & 0.88\pm 0.03\\
     \text{KL}_1: & 5.15\pm 0.02 & 0.83\pm 0.03\\
     \hline
    \end{array}
\end{align}
for the values of the critical point and correlation length exponent. A weighted average yields $\nu=0.86\pm0.03$ and $\gamma_c=5.16\pm0.03$ . 

We note that both the location of the critical point $\gamma_c$ and the value of the critical exponent $\nu$ differ significantly from the prediction of a perturbative analysis of the nonlinear sigma model of monitored fermions~\cite{chahine2024,poboiko2023measurementinduced} and from previously obtained numerical results~\cite{poboiko2023measurementinduced,fan2025}, which both predict $\nu>1$. Instead, the exponent $\nu$ is rather close to the value for critical percolation in $d+1=3$ spatial dimensions, which is $\nu_{3DP}\sim 0.87$. We also note that the data computed from the entanglement Hamiltonian features much less noise and smaller error bars compared to similar system sizes in previous numerical approaches.

\begin{figure}[t]
\includegraphics[width=\linewidth]{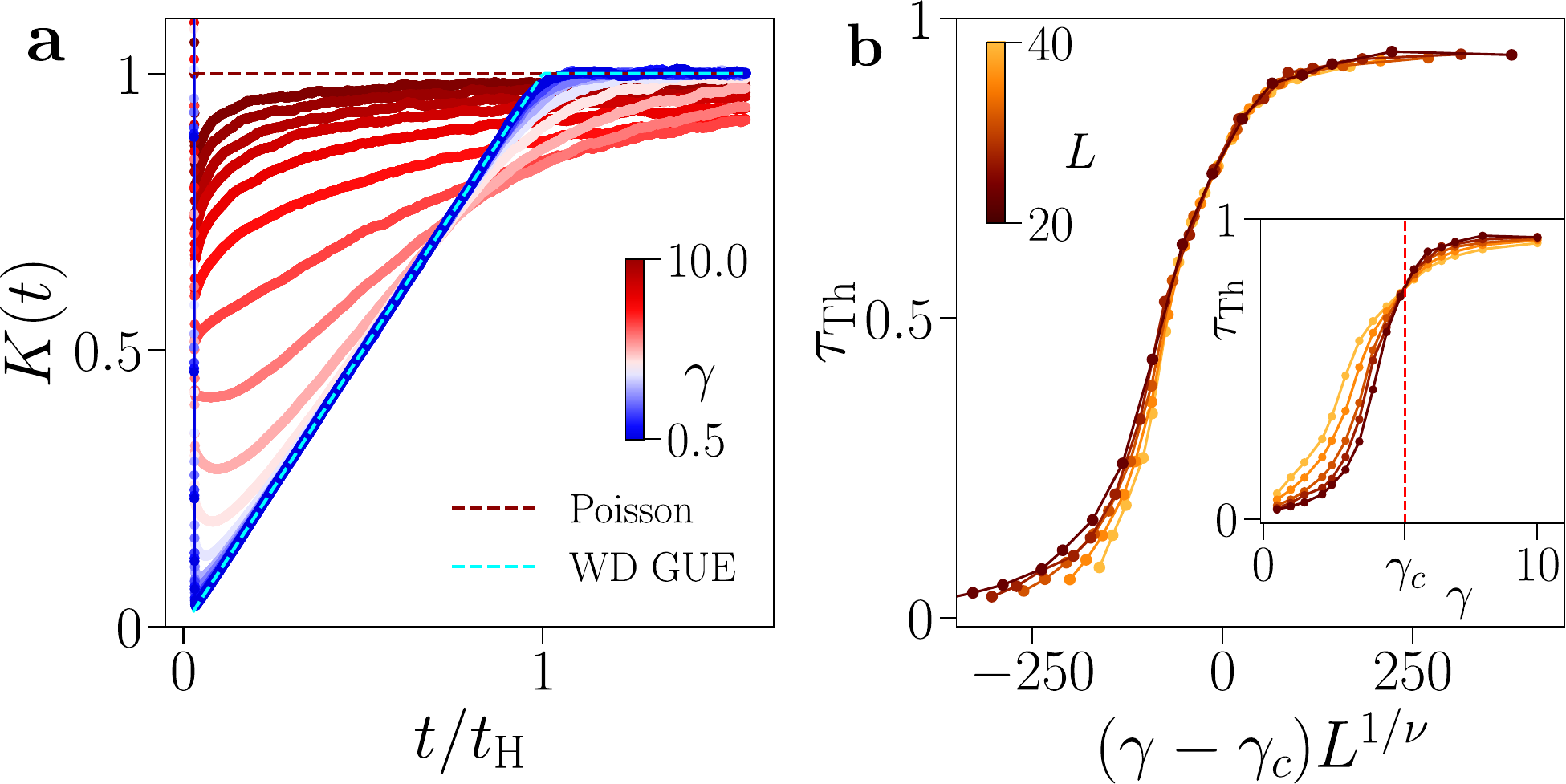}
\caption{\label{fig:thouless}\textbf{Spectral form factor transition.} (a) Spectral form factor (SFF) for $L=30$ and $\gamma\in[0.5,10]$. At weak monitoring, the SFF exhibits the typical chaotic ramp and plateau. As $\gamma$ increases, $K(t)$ flattens. (b) Scaling collapse of the Thouless time $\tau_\text{Th}$ as a function of $\gamma$, using $\gamma_c = 5.16$ and $\nu = 0.86$. Inset: unrescaled data.}
\end{figure}

To probe long-range correlations, we compute the spectral form factor (SFF) across the phase diagram (Fig.~\ref{fig:thouless}(a)). For weak monitoring, the SFF shows the characteristic ramp and plateau of chaotic systems. As $\gamma$ increases, the SFF flattens toward the integrable limit $K(t)=1$.  

We quantify the transition using the Thouless time $\tau_\text{Th}$, defined as the earliest time satisfying  
\[
|\log K(t) - \log K_\text{GUE}(t)| < \epsilon,
\]  
with $\epsilon = 0.05$. Figure \ref{fig:thouless}(b) shows that $\tau_\text{Th}$ captures the localization transition, exhibiting a crossing point and saturation at large $\gamma$. A reasonable scaling collapse is obtained using the values $\gamma_c=5.16$ and $\nu=0.86$, i.e. the optimal results from the short-range correlation analysis. For weak measurements, current system sizes are insufficient to conclusively determine the full extent of the ergodic regime or whether a non-ergodic extended (NEE) phase may exist.

\begin{figure}[t]
\includegraphics[width=\linewidth]{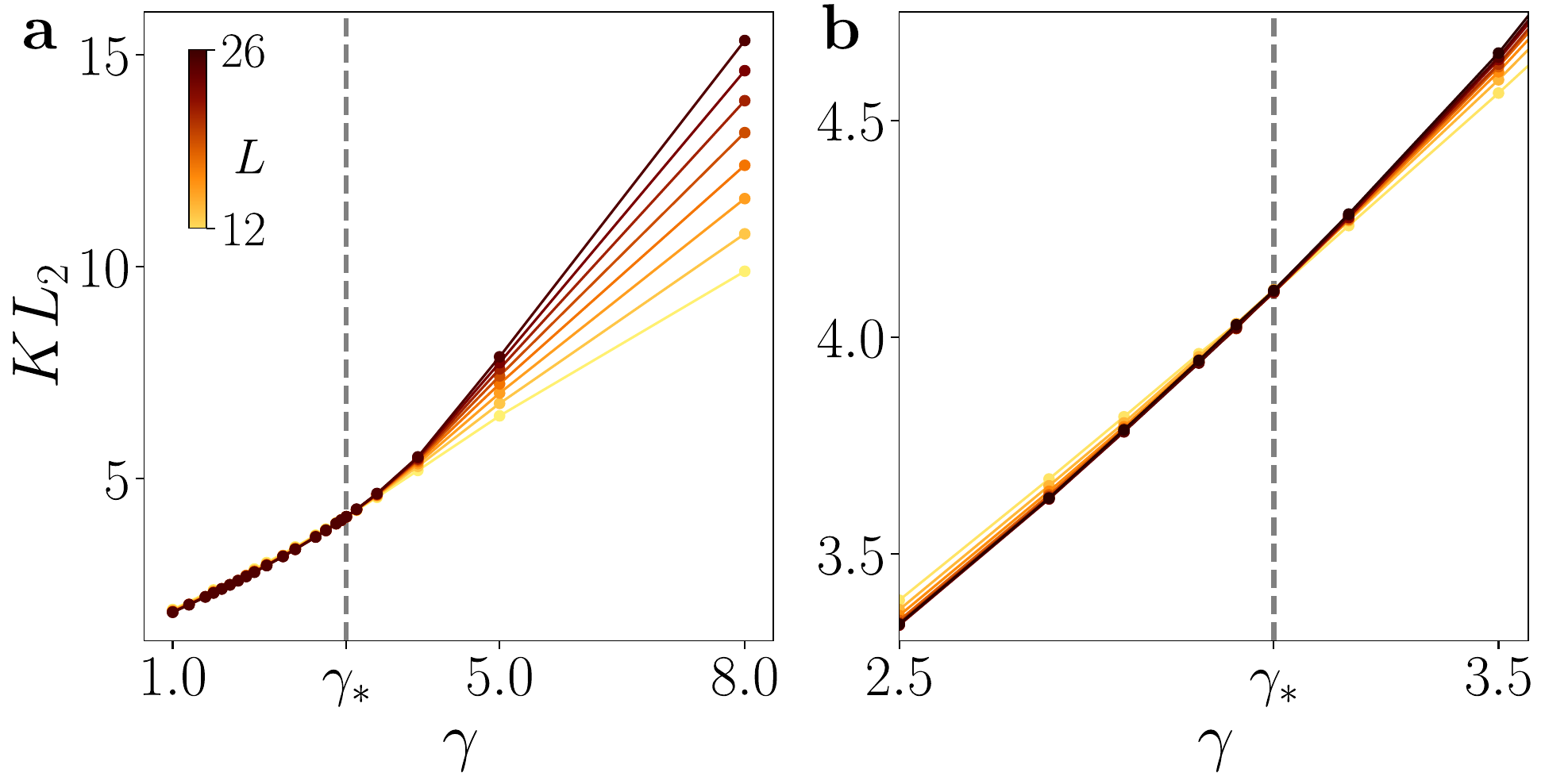}
\caption{\label{fig:KL2} \textbf{Long-range Kullback-Leibler divergence.} (a) $KL_2$ as a function of $\gamma$ for system sizes $L\in[12,26]$, confirming the weak- and strong-measurement limits. (b) Zoom on the crossing at $\gamma_*<\gamma_c$.}
\end{figure}

Finally, we evaluate the long-range $KL_2$ divergence (Fig. \ref{fig:KL2}). Both $KL_1$ and $KL_2$ are expected to display the same behavior in an ergodic phase and in a localized phase, while they may differ from each other inside a NEE phase. Interestingly, $KL_2$ features a crossing point at an intermediate measurement rate $\gamma_*$ with $\gamma_\text{Fl} < \gamma_* < \gamma_c$ and hence at a different location compared to the previous data. 

Hence, while the spectral form factor yields results that are consistent with the short range correlations in the spectrum, the KL$_2$ data suggests the possible existence of an intermediate regime where a NEE or multifractal phase emerges at intermediate $\gamma$. However, for the system sizes accessible here, the data do not allow us to conclusively confirm or rule out such a regime, as predicted in Ref.~\cite{chahine2024}.

\subsection{Spectral transition in three dimensions}
We extend the entanglement Hamiltonian analysis to monitored free fermions in $D=3$ dimensions, both to confirm the robustness of the method and to explore possible universal connections with the lower-dimensional case.  

\begin{figure}[t]
\includegraphics[width=\linewidth]{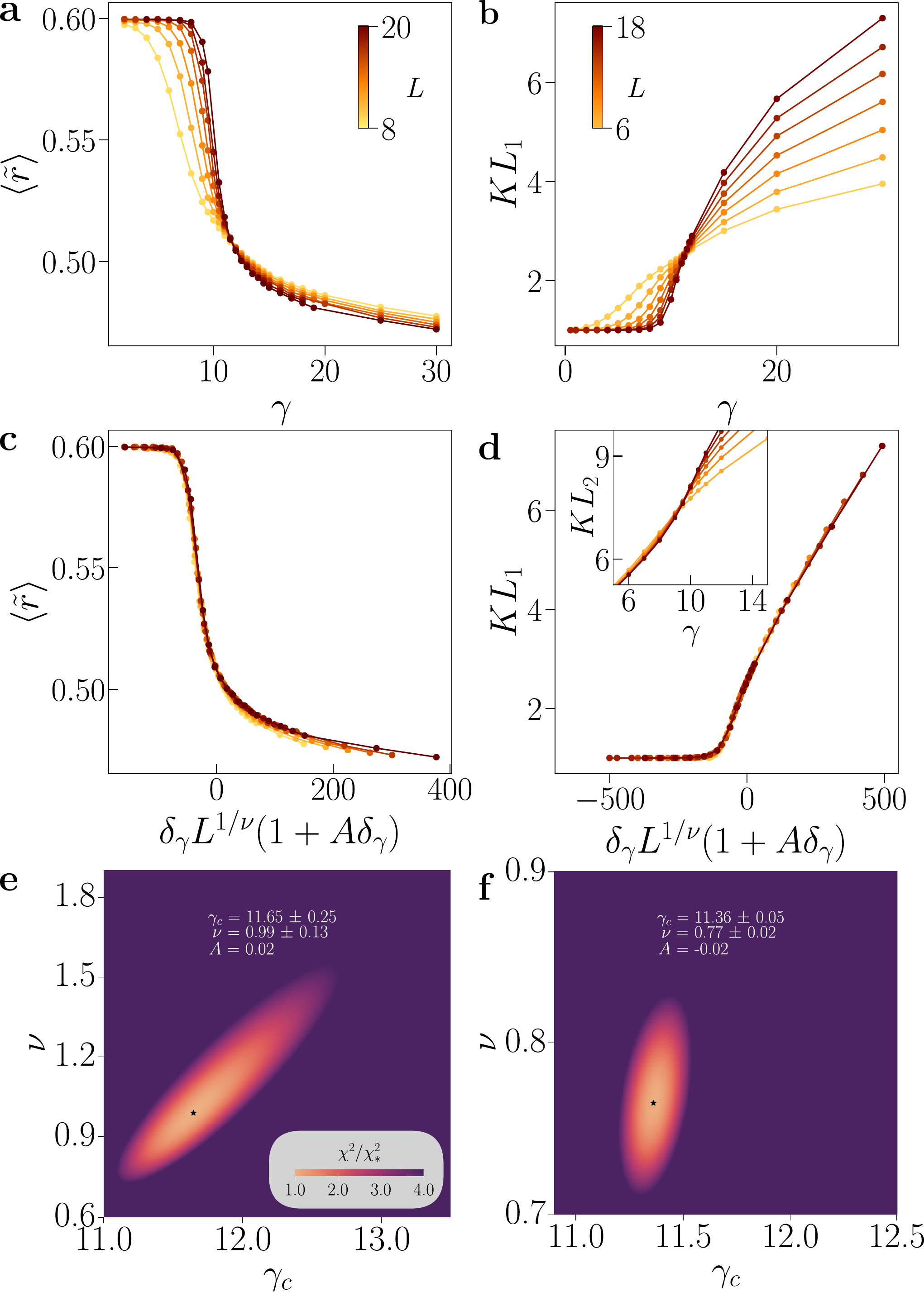}
\caption{\label{fig:3D}\textbf{Spectral transition in the entanglement Hamiltonian in three dimensions.} 
(a) Average gap ratio $\langle \tilde r\rangle$ for system sizes $L\in[8,20]$ as a function of $\gamma$, showing a sharp crossing and approaching the ergodic limit at weak monitoring, while the localized limit is reached only slowly for accessible sizes. 
(b) $KL_1$ for $L\in[6,18]$ likewise displays a sharp crossing. 
(c,d) Finite-size scaling collapse of $\langle \tilde r\rangle$ and $KL_1$ using the optimal critical parameters $(\gamma_c,\nu)$ obtained from the cost-function analysis. The $KL_1$ data collapse holds across the full range of $\gamma$, while $\langle \tilde r\rangle$ collapses well at weak monitoring but shows finite-size deviations at strong monitoring. The inset in (d) shows $KL_2$ for $L\in[8,18]$ as a function of $\gamma$. 
(e,f) Heatmaps of the normalized cost function for $\langle \tilde r\rangle$ and $KL_1$, truncated at $4\chi_*$ to indicate the uncertainty in the critical parameters. Optimal values minimize the cost function, and error bars are estimated from the contour $\chi_*+4$ at fixed optimal $A$.
 }
\end{figure}

As in $D=2$, we find a phase transition from ergodic, metallic behavior into a non-ergodic, localized area-law phase. This transition is clearly signaled by short-range correlations in the entanglement spectrum and eigenfunctions. In particular, we compute the average gap ratio $\langle\tilde r\rangle$ and $KL_1$, see Fig.~\ref{fig:3D}. The scaling behavior is best captured by a non-linear ansatz,
\begin{equation}
    F(\gamma,L)=F\!\big(\delta_\gamma L^{1/\nu}(1+A\delta_\gamma)\big), \qquad 
    \delta_\gamma=\gamma-\gamma_c,
\end{equation}
with parameters $(\gamma_c,\nu,A)$ extracted by minimizing the cost function $F_\text{cost}(\gamma_c,\nu,A)$ as outlined above. A conventional ansatz, i.e., with $(A=0)$ is unable to produce a scaling collapse in three dimensions~\footnote{Similar behavior was observed for the monitored Kitaev model in $D=2$~\cite{klocke2025}.}. The resulting estimates are
\begin{align}
    \begin{array}{|l|l|l|l|}
    \hline
     & \gamma_c & \nu & A \\
     \hline
     \langle \tilde r\rangle: & 11.65\pm0.25 & 0.99\pm0.13 & 0.02 \\
     \text{KL}_1: & 11.36\pm0.05 & 0.77\pm0.02 & -0.02 \\
     \hline
    \end{array}
\end{align}
for the critical point and correlation-length exponent. While $\gamma_c$ is consistent across observables, the exponents differ drastically from each other. We attribute the inconsistency to $\langle \tilde r\rangle$, which appears to suffer from strong finite-size effects: it converges rapidly to the GUE limit on the ergodic side but very slowly to the Poisson limit under strong monitoring. By contrast, $KL_1$, based on single-particle wave functions, is less affected by this issue and provides more precise estimates with smaller error bars.   

The transition is also detected by the $KL_2$ diagnostic, shown in the inset of Fig.~\ref{fig:3D}(d). It yields a critical measurement rate consistent with that from short-range correlations. This indicates the absence of any multifractal (NEE) phase in $D=3$. 

\subsection{Absence of a spectral transition in one dimension}
The case of monitored free fermions in $D=1$ dimension has attracted considerable attention, owing to its intrinsic difficulties. Early analyses predicted either a weak-localization picture or a Berezinskii–Kosterlitz–Thouless (BKT) transition~\cite{alberton2021enttrans,szyniszewski2023disordered,poboiko2023measurementinduced}, but numerical studies were strongly hampered by the exponentially divergent correlation length. Consensus has now emerged, based largely on NL$\sigma$M arguments~\cite{poboiko,Starchl2025} and numerical simulations for large system sizes~\cite{fan2025}, that no transition occurs at finite monitoring.  

\begin{figure}[t]
\includegraphics[width=\linewidth]{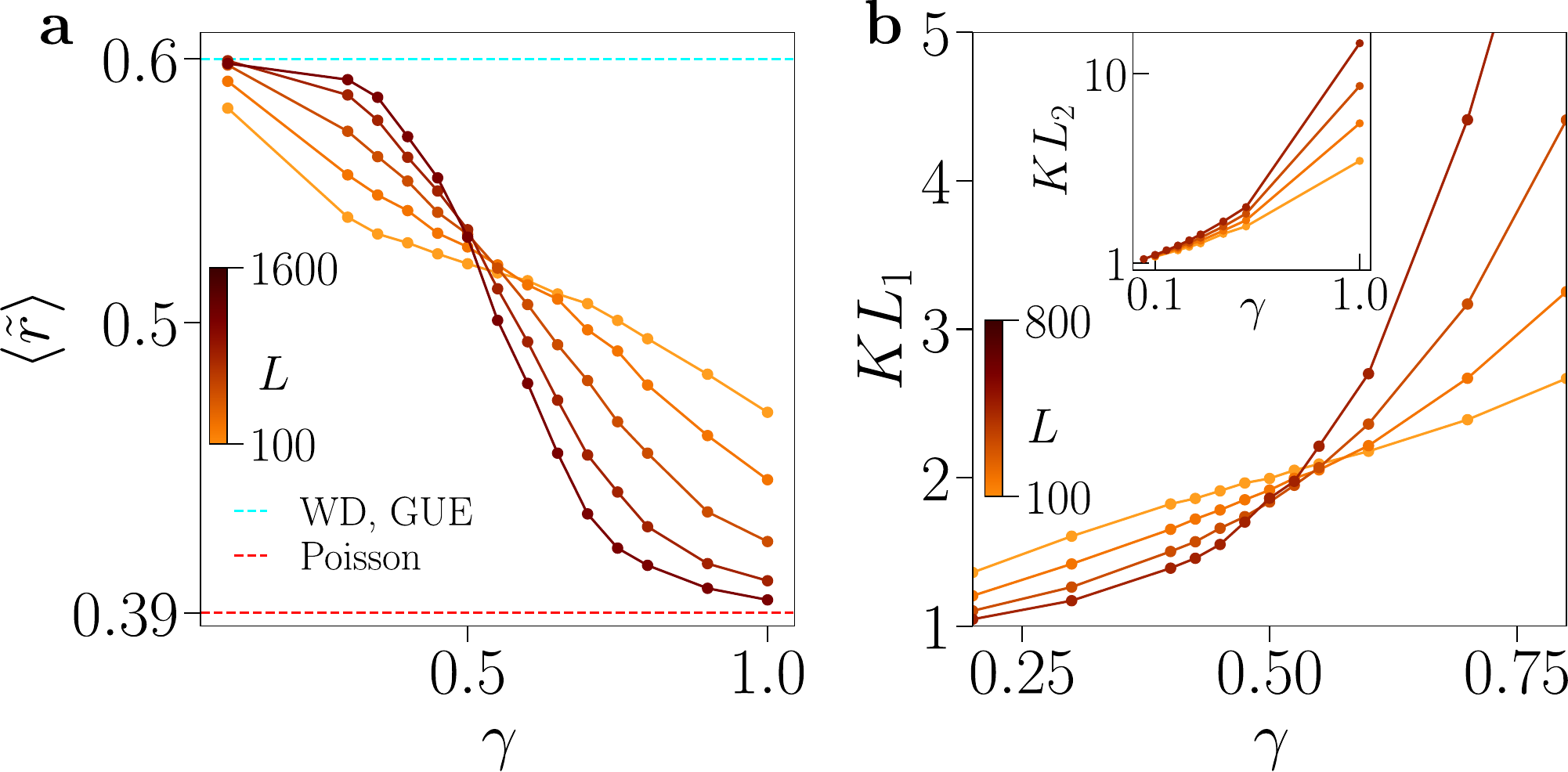}
\caption{\label{fig:1D} \textbf{Entanglement Hamiltonian analysis in 1D.} (a) $\langle \tilde r\rangle $ as a function of $\gamma$ and increasing $L\in[100,1600]$. (b) $KL_1$ and $KL_2$ (inset) as a function of $\gamma$ and increasing $L\in[100,800]$. Both $\langle \tilde r\rangle$ and $KL_1$ show a drift of the crossing point with increasing $L$, while the $KL_2$ displays no crossings at all, consistent with the absence of a critical point for finite $\gamma$.}
\end{figure}

We briefly revisit the problem and show that the entanglement Hamiltonian spectrum is consistent with the absence of a transition in one dimension. Figure~\ref{fig:1D} shows results for $\langle \tilde r\rangle$ and $KL_1$ for system sizes up to $L=1200$ and $L=800$, respectively. Both observables exhibit crossing points that systematically drift toward $\gamma=0$ with increasing system size, consistent with the absence of a transition. This is further reinforced by the $KL_2$, which shows no indication of a crossing point at all. Given the extremely sharp crossing points in $D=2,3$, this shows the consistency of the behavior of the entanglement Hamiltonian spectrum for monitored fermion phases in arbitrary dimensions.

\section{Conclusion}

Monitored free fermions host a rich phase diagram with three non-trivial fixed points controlled by the measurement rate $\gamma$. At $\gamma=0^+$, the system exhibits chaotic unitary dynamics, characterized by a generalized Page law for Gaussian states. At $\gamma=\gamma_\text{Fl}$, an attractive Fermi-liquid fixed point stabilizes a metallic phase with logarithmic entanglement scaling and emergent scale- and space–time invariance. Finally, at $\gamma=\gamma_c$, the system undergoes a measurement-induced entanglement transition governed by a repulsive quantum Lifshitz fixed point.  

At the Lifshitz critical point $\gamma=\gamma_c$, the entanglement Hamiltonian displays a spectral transition from an ergodic to a localized, non-ergodic phase. Short-range spectral correlations, such as the adjacent gap ratio $\langle\tilde r\rangle$ and the Kullback–Leibler divergence $KL_1$, yield precise estimates of the critical point and correlation length exponent. Long-range probes corroborate these findings, while the variant $KL_2$ reveals signatures of a non-ergodic extended (multifractal) regime at intermediate monitoring strengths $\gamma_*<\gamma<\gamma_c$, though this requires further clarification. A striking open puzzle is the discrepancy between our estimate of the critical exponent in $D=2$, $\nu\approx0.86$, and values reported in other numerical studies, $\nu\approx1.3$~\cite{poboiko2023measurementinduced,fan2025}, which calls for deeper investigation.  

Several features uncovered here lie beyond the reach of current theoretical approaches such as the NL$\sigma$M. In particular, the emergence of Lifshitz-type critical scaling at the transition and Fermi-liquid scaling deep in the log-law phase are unexpected. Together with indications of a possible non-ergodic extended regime, these results highlight the need for new analytic frameworks. A deeper theoretical understanding of the entanglement Hamiltonian in monitored systems could provide the missing bridge between numerical probes and field-theoretic descriptions.  

The RMT-based approach developed here also opens new directions. Natural extensions include probing the robustness of the phase diagram to quenched disorder or interactions~\cite{szyniszewski2023disordered, poboiko2025interacting}, and exploring alternative geometries such as the Bethe lattice or random regular graphs~\cite{deluca2014,Sierant2023,biroli2022}, which have played a central role in Anderson and many-body localization. Since multifractality is often enhanced in higher dimensions, our observation that it appears absent in $3D$ suggests important constraints on its universality. Finally, rapid progress in programmable quantum simulators and hybrid unitary–measurement protocols provides promising experimental platforms to test these predictions.

\begin{acknowledgments}
 
  We acknowledge support from the Deutsche Forschungsgemeinschaft (DFG, German Research Foundation) under Germany's Excellence Strategy Cluster of Excellence Matter and Light for Quantum Computing (ML4Q) EXC 2004/1 390534769, and by the DFG Collaborative Research Center (CRC) 183 Project No. 277101999 - project B02. M.B. acknowledges support from the Heisenberg programme of the Deutsche Forschungsgemeinschaft (DFG, German Research Foundation), project no. 549109008. The code for our numerical computations was implemented in Julia~\cite{bezanson17}. 

\end{acknowledgments}

\bibliography{EntEnt}

\begin{thebibliography}{120}%
\makeatletter
\providecommand \@ifxundefined [1]{%
 \@ifx{#1\undefined}
}%
\providecommand \@ifnum [1]{%
 \ifnum #1\expandafter \@firstoftwo
 \else \expandafter \@secondoftwo
 \fi
}%
\providecommand \@ifx [1]{%
 \ifx #1\expandafter \@firstoftwo
 \else \expandafter \@secondoftwo
 \fi
}%
\providecommand \natexlab [1]{#1}%
\providecommand \enquote  [1]{``#1''}%
\providecommand \bibnamefont  [1]{#1}%
\providecommand \bibfnamefont [1]{#1}%
\providecommand \citenamefont [1]{#1}%
\providecommand \href@noop [0]{\@secondoftwo}%
\providecommand \href [0]{\begingroup \@sanitize@url \@href}%
\providecommand \@href[1]{\@@startlink{#1}\@@href}%
\providecommand \@@href[1]{\endgroup#1\@@endlink}%
\providecommand \@sanitize@url [0]{\catcode `\\12\catcode `\$12\catcode `\&12\catcode `\#12\catcode `\^12\catcode `\_12\catcode `\%12\relax}%
\providecommand \@@startlink[1]{}%
\providecommand \@@endlink[0]{}%
\providecommand \url  [0]{\begingroup\@sanitize@url \@url }%
\providecommand \@url [1]{\endgroup\@href {#1}{\urlprefix }}%
\providecommand \urlprefix  [0]{URL }%
\providecommand \Eprint [0]{\href }%
\providecommand \doibase [0]{http://dx.doi.org/}%
\providecommand \selectlanguage [0]{\@gobble}%
\providecommand \bibinfo  [0]{\@secondoftwo}%
\providecommand \bibfield  [0]{\@secondoftwo}%
\providecommand \translation [1]{[#1]}%
\providecommand \BibitemOpen [0]{}%
\providecommand \bibitemStop [0]{}%
\providecommand \bibitemNoStop [0]{.\EOS\space}%
\providecommand \EOS [0]{\spacefactor3000\relax}%
\providecommand \BibitemShut  [1]{\csname bibitem#1\endcsname}%
\let\auto@bib@innerbib\@empty
\bibitem [{\citenamefont {Skinner}\ \emph {et~al.}(2019)\citenamefont {Skinner}, \citenamefont {Ruhman},\ and\ \citenamefont {Nahum}}]{Skinner2019}%
  \BibitemOpen
  \bibfield  {author} {\bibinfo {author} {\bibfnamefont {B.}~\bibnamefont {Skinner}}, \bibinfo {author} {\bibfnamefont {J.}~\bibnamefont {Ruhman}}, \ and\ \bibinfo {author} {\bibfnamefont {A.}~\bibnamefont {Nahum}},\ }\href {\doibase 10.1103/PhysRevX.9.031009} {\bibfield  {journal} {\bibinfo  {journal} {Phys. Rev. X}\ }\textbf {\bibinfo {volume} {9}},\ \bibinfo {pages} {031009} (\bibinfo {year} {2019})}\BibitemShut {NoStop}%
\bibitem [{\citenamefont {Li}\ \emph {et~al.}(2018)\citenamefont {Li}, \citenamefont {Chen},\ and\ \citenamefont {Fisher}}]{Fisher2018}%
  \BibitemOpen
  \bibfield  {author} {\bibinfo {author} {\bibfnamefont {Y.}~\bibnamefont {Li}}, \bibinfo {author} {\bibfnamefont {X.}~\bibnamefont {Chen}}, \ and\ \bibinfo {author} {\bibfnamefont {M.~P.~A.}\ \bibnamefont {Fisher}},\ }\href {\doibase 10.1103/PhysRevB.98.205136} {\bibfield  {journal} {\bibinfo  {journal} {Phys. Rev. B}\ }\textbf {\bibinfo {volume} {98}},\ \bibinfo {pages} {205136} (\bibinfo {year} {2018})}\BibitemShut {NoStop}%
\bibitem [{\citenamefont {Li}\ \emph {et~al.}(2019)\citenamefont {Li}, \citenamefont {Chen},\ and\ \citenamefont {Fisher}}]{Li2019b}%
  \BibitemOpen
  \bibfield  {author} {\bibinfo {author} {\bibfnamefont {Y.}~\bibnamefont {Li}}, \bibinfo {author} {\bibfnamefont {X.}~\bibnamefont {Chen}}, \ and\ \bibinfo {author} {\bibfnamefont {M.~P.~A.}\ \bibnamefont {Fisher}},\ }\href {\doibase 10.1103/PhysRevB.100.134306} {\bibfield  {journal} {\bibinfo  {journal} {Phys. Rev. B}\ }\textbf {\bibinfo {volume} {100}},\ \bibinfo {pages} {134306} (\bibinfo {year} {2019})}\BibitemShut {NoStop}%
\bibitem [{\citenamefont {Gullans}\ and\ \citenamefont {Huse}(2020)}]{gullans2019}%
  \BibitemOpen
  \bibfield  {author} {\bibinfo {author} {\bibfnamefont {M.~J.}\ \bibnamefont {Gullans}}\ and\ \bibinfo {author} {\bibfnamefont {D.~A.}\ \bibnamefont {Huse}},\ }\href {\doibase 10.1103/PhysRevX.10.041020} {\bibfield  {journal} {\bibinfo  {journal} {Phys. Rev. X}\ }\textbf {\bibinfo {volume} {10}},\ \bibinfo {pages} {041020} (\bibinfo {year} {2020})}\BibitemShut {NoStop}%
\bibitem [{\citenamefont {Choi}\ \emph {et~al.}(2020)\citenamefont {Choi}, \citenamefont {Bao}, \citenamefont {Qi},\ and\ \citenamefont {Altman}}]{choi2020prl}%
  \BibitemOpen
  \bibfield  {author} {\bibinfo {author} {\bibfnamefont {S.}~\bibnamefont {Choi}}, \bibinfo {author} {\bibfnamefont {Y.}~\bibnamefont {Bao}}, \bibinfo {author} {\bibfnamefont {X.-L.}\ \bibnamefont {Qi}}, \ and\ \bibinfo {author} {\bibfnamefont {E.}~\bibnamefont {Altman}},\ }\href {\doibase 10.1103/PhysRevLett.125.030505} {\bibfield  {journal} {\bibinfo  {journal} {Phys. Rev. Lett.}\ }\textbf {\bibinfo {volume} {125}},\ \bibinfo {pages} {030505} (\bibinfo {year} {2020})}\BibitemShut {NoStop}%
\bibitem [{\citenamefont {Jian}\ \emph {et~al.}(2020)\citenamefont {Jian}, \citenamefont {You}, \citenamefont {Vasseur},\ and\ \citenamefont {Ludwig}}]{Jian2020}%
  \BibitemOpen
  \bibfield  {author} {\bibinfo {author} {\bibfnamefont {C.-M.}\ \bibnamefont {Jian}}, \bibinfo {author} {\bibfnamefont {Y.-Z.}\ \bibnamefont {You}}, \bibinfo {author} {\bibfnamefont {R.}~\bibnamefont {Vasseur}}, \ and\ \bibinfo {author} {\bibfnamefont {A.~W.~W.}\ \bibnamefont {Ludwig}},\ }\href {\doibase 10.1103/PhysRevB.101.104302} {\bibfield  {journal} {\bibinfo  {journal} {Phys. Rev. B}\ }\textbf {\bibinfo {volume} {101}},\ \bibinfo {pages} {104302} (\bibinfo {year} {2020})}\BibitemShut {NoStop}%
\bibitem [{\citenamefont {Fan}\ \emph {et~al.}(2021)\citenamefont {Fan}, \citenamefont {Vijay}, \citenamefont {Vishwanath},\ and\ \citenamefont {You}}]{fan2020selforganized}%
  \BibitemOpen
  \bibfield  {author} {\bibinfo {author} {\bibfnamefont {R.}~\bibnamefont {Fan}}, \bibinfo {author} {\bibfnamefont {S.}~\bibnamefont {Vijay}}, \bibinfo {author} {\bibfnamefont {A.}~\bibnamefont {Vishwanath}}, \ and\ \bibinfo {author} {\bibfnamefont {Y.-Z.}\ \bibnamefont {You}},\ }\href {\doibase 10.1103/PhysRevB.103.174309} {\bibfield  {journal} {\bibinfo  {journal} {Phys. Rev. B}\ }\textbf {\bibinfo {volume} {103}},\ \bibinfo {pages} {174309} (\bibinfo {year} {2021})}\BibitemShut {NoStop}%
\bibitem [{\citenamefont {Li}\ and\ \citenamefont {Fisher}(2021)}]{lifisher2021}%
  \BibitemOpen
  \bibfield  {author} {\bibinfo {author} {\bibfnamefont {Y.}~\bibnamefont {Li}}\ and\ \bibinfo {author} {\bibfnamefont {M.~P.~A.}\ \bibnamefont {Fisher}},\ }\href {\doibase 10.1103/PhysRevB.103.104306} {\bibfield  {journal} {\bibinfo  {journal} {Phys. Rev. B}\ }\textbf {\bibinfo {volume} {103}},\ \bibinfo {pages} {104306} (\bibinfo {year} {2021})}\BibitemShut {NoStop}%
\bibitem [{\citenamefont {Nahum}\ \emph {et~al.}(2021)\citenamefont {Nahum}, \citenamefont {Roy}, \citenamefont {Skinner},\ and\ \citenamefont {Ruhman}}]{nahum2021prxq}%
  \BibitemOpen
  \bibfield  {author} {\bibinfo {author} {\bibfnamefont {A.}~\bibnamefont {Nahum}}, \bibinfo {author} {\bibfnamefont {S.}~\bibnamefont {Roy}}, \bibinfo {author} {\bibfnamefont {B.}~\bibnamefont {Skinner}}, \ and\ \bibinfo {author} {\bibfnamefont {J.}~\bibnamefont {Ruhman}},\ }\href {\doibase 10.1103/PRXQuantum.2.010352} {\bibfield  {journal} {\bibinfo  {journal} {PRX Quantum}\ }\textbf {\bibinfo {volume} {2}},\ \bibinfo {pages} {010352} (\bibinfo {year} {2021})}\BibitemShut {NoStop}%
\bibitem [{\citenamefont {Jian}\ \emph {et~al.}(2021)\citenamefont {Jian}, \citenamefont {Liu}, \citenamefont {Chen}, \citenamefont {Swingle},\ and\ \citenamefont {Zhang}}]{jian2021syk}%
  \BibitemOpen
  \bibfield  {author} {\bibinfo {author} {\bibfnamefont {S.-K.}\ \bibnamefont {Jian}}, \bibinfo {author} {\bibfnamefont {C.}~\bibnamefont {Liu}}, \bibinfo {author} {\bibfnamefont {X.}~\bibnamefont {Chen}}, \bibinfo {author} {\bibfnamefont {B.}~\bibnamefont {Swingle}}, \ and\ \bibinfo {author} {\bibfnamefont {P.}~\bibnamefont {Zhang}},\ }\href {\doibase 10.1103/PhysRevLett.127.140601} {\bibfield  {journal} {\bibinfo  {journal} {Phys. Rev. Lett.}\ }\textbf {\bibinfo {volume} {127}},\ \bibinfo {pages} {140601} (\bibinfo {year} {2021})}\BibitemShut {NoStop}%
\bibitem [{\citenamefont {Bao}\ \emph {et~al.}(2021)\citenamefont {Bao}, \citenamefont {Choi},\ and\ \citenamefont {Altman}}]{bao2021symmetry}%
  \BibitemOpen
  \bibfield  {author} {\bibinfo {author} {\bibfnamefont {Y.}~\bibnamefont {Bao}}, \bibinfo {author} {\bibfnamefont {S.}~\bibnamefont {Choi}}, \ and\ \bibinfo {author} {\bibfnamefont {E.}~\bibnamefont {Altman}},\ }\href {\doibase https://doi.org/10.1016/j.aop.2021.168618} {\bibfield  {journal} {\bibinfo  {journal} {Annals of Physics}\ ,\ \bibinfo {pages} {168618}} (\bibinfo {year} {2021})}\BibitemShut {NoStop}%
\bibitem [{\citenamefont {Turkeshi}\ \emph {et~al.}(2021)\citenamefont {Turkeshi}, \citenamefont {Biella}, \citenamefont {Fazio}, \citenamefont {Dalmonte},\ and\ \citenamefont {Schir\'o}}]{turkeshi2021measurementinduced}%
  \BibitemOpen
  \bibfield  {author} {\bibinfo {author} {\bibfnamefont {X.}~\bibnamefont {Turkeshi}}, \bibinfo {author} {\bibfnamefont {A.}~\bibnamefont {Biella}}, \bibinfo {author} {\bibfnamefont {R.}~\bibnamefont {Fazio}}, \bibinfo {author} {\bibfnamefont {M.}~\bibnamefont {Dalmonte}}, \ and\ \bibinfo {author} {\bibfnamefont {M.}~\bibnamefont {Schir\'o}},\ }\href {\doibase 10.1103/PhysRevB.103.224210} {\bibfield  {journal} {\bibinfo  {journal} {Phys. Rev. B}\ }\textbf {\bibinfo {volume} {103}},\ \bibinfo {pages} {224210} (\bibinfo {year} {2021})}\BibitemShut {NoStop}%
\bibitem [{\citenamefont {Ivanov}\ \emph {et~al.}(2020)\citenamefont {Ivanov}, \citenamefont {Ivanova}, \citenamefont {Caballero-Benitez},\ and\ \citenamefont {Mekhov}}]{Ivanov2020}%
  \BibitemOpen
  \bibfield  {author} {\bibinfo {author} {\bibfnamefont {D.~A.}\ \bibnamefont {Ivanov}}, \bibinfo {author} {\bibfnamefont {T.~Y.}\ \bibnamefont {Ivanova}}, \bibinfo {author} {\bibfnamefont {S.~F.}\ \bibnamefont {Caballero-Benitez}}, \ and\ \bibinfo {author} {\bibfnamefont {I.~B.}\ \bibnamefont {Mekhov}},\ }\href {\doibase 10.1103/PhysRevLett.124.010603} {\bibfield  {journal} {\bibinfo  {journal} {Phys. Rev. Lett.}\ }\textbf {\bibinfo {volume} {124}},\ \bibinfo {pages} {010603} (\bibinfo {year} {2020})}\BibitemShut {NoStop}%
\bibitem [{\citenamefont {Buonaiuto}\ \emph {et~al.}(2021)\citenamefont {Buonaiuto}, \citenamefont {Carollo}, \citenamefont {Olmos},\ and\ \citenamefont {Lesanovsky}}]{Buonaiuto2021}%
  \BibitemOpen
  \bibfield  {author} {\bibinfo {author} {\bibfnamefont {G.}~\bibnamefont {Buonaiuto}}, \bibinfo {author} {\bibfnamefont {F.}~\bibnamefont {Carollo}}, \bibinfo {author} {\bibfnamefont {B.}~\bibnamefont {Olmos}}, \ and\ \bibinfo {author} {\bibfnamefont {I.}~\bibnamefont {Lesanovsky}},\ }\href {\doibase 10.1103/PhysRevLett.127.133601} {\bibfield  {journal} {\bibinfo  {journal} {Phys. Rev. Lett.}\ }\textbf {\bibinfo {volume} {127}},\ \bibinfo {pages} {133601} (\bibinfo {year} {2021})}\BibitemShut {NoStop}%
\bibitem [{\citenamefont {Zabalo}\ \emph {et~al.}(2020)\citenamefont {Zabalo}, \citenamefont {Gullans}, \citenamefont {Wilson}, \citenamefont {Gopalakrishnan}, \citenamefont {Huse},\ and\ \citenamefont {Pixley}}]{Zabalo2020}%
  \BibitemOpen
  \bibfield  {author} {\bibinfo {author} {\bibfnamefont {A.}~\bibnamefont {Zabalo}}, \bibinfo {author} {\bibfnamefont {M.~J.}\ \bibnamefont {Gullans}}, \bibinfo {author} {\bibfnamefont {J.~H.}\ \bibnamefont {Wilson}}, \bibinfo {author} {\bibfnamefont {S.}~\bibnamefont {Gopalakrishnan}}, \bibinfo {author} {\bibfnamefont {D.~A.}\ \bibnamefont {Huse}}, \ and\ \bibinfo {author} {\bibfnamefont {J.~H.}\ \bibnamefont {Pixley}},\ }\href {\doibase 10.1103/PhysRevB.101.060301} {\bibfield  {journal} {\bibinfo  {journal} {Phys. Rev. B}\ }\textbf {\bibinfo {volume} {101}},\ \bibinfo {pages} {060301} (\bibinfo {year} {2020})}\BibitemShut {NoStop}%
\bibitem [{\citenamefont {Zabalo}\ \emph {et~al.}(2022)\citenamefont {Zabalo}, \citenamefont {Gullans}, \citenamefont {Wilson}, \citenamefont {Vasseur}, \citenamefont {Ludwig}, \citenamefont {Gopalakrishnan}, \citenamefont {Huse},\ and\ \citenamefont {Pixley}}]{Zabalo2022}%
  \BibitemOpen
  \bibfield  {author} {\bibinfo {author} {\bibfnamefont {A.}~\bibnamefont {Zabalo}}, \bibinfo {author} {\bibfnamefont {M.~J.}\ \bibnamefont {Gullans}}, \bibinfo {author} {\bibfnamefont {J.~H.}\ \bibnamefont {Wilson}}, \bibinfo {author} {\bibfnamefont {R.}~\bibnamefont {Vasseur}}, \bibinfo {author} {\bibfnamefont {A.~W.~W.}\ \bibnamefont {Ludwig}}, \bibinfo {author} {\bibfnamefont {S.}~\bibnamefont {Gopalakrishnan}}, \bibinfo {author} {\bibfnamefont {D.~A.}\ \bibnamefont {Huse}}, \ and\ \bibinfo {author} {\bibfnamefont {J.~H.}\ \bibnamefont {Pixley}},\ }\href {\doibase 10.1103/PhysRevLett.128.050602} {\bibfield  {journal} {\bibinfo  {journal} {Phys. Rev. Lett.}\ }\textbf {\bibinfo {volume} {128}},\ \bibinfo {pages} {050602} (\bibinfo {year} {2022})}\BibitemShut {NoStop}%
\bibitem [{\citenamefont {Ippoliti}\ \emph {et~al.}(2022)\citenamefont {Ippoliti}, \citenamefont {Rakovszky},\ and\ \citenamefont {Khemani}}]{ippoliti2021}%
  \BibitemOpen
  \bibfield  {author} {\bibinfo {author} {\bibfnamefont {M.}~\bibnamefont {Ippoliti}}, \bibinfo {author} {\bibfnamefont {T.}~\bibnamefont {Rakovszky}}, \ and\ \bibinfo {author} {\bibfnamefont {V.}~\bibnamefont {Khemani}},\ }\href {\doibase 10.1103/PhysRevX.12.011045} {\bibfield  {journal} {\bibinfo  {journal} {Phys. Rev. X}\ }\textbf {\bibinfo {volume} {12}},\ \bibinfo {pages} {011045} (\bibinfo {year} {2022})}\BibitemShut {NoStop}%
\bibitem [{\citenamefont {Fisher}\ \emph {et~al.}(2023)\citenamefont {Fisher}, \citenamefont {Khemani}, \citenamefont {Nahum},\ and\ \citenamefont {Vijay}}]{circuitreview}%
  \BibitemOpen
  \bibfield  {author} {\bibinfo {author} {\bibfnamefont {M.~P.}\ \bibnamefont {Fisher}}, \bibinfo {author} {\bibfnamefont {V.}~\bibnamefont {Khemani}}, \bibinfo {author} {\bibfnamefont {A.}~\bibnamefont {Nahum}}, \ and\ \bibinfo {author} {\bibfnamefont {S.}~\bibnamefont {Vijay}},\ }\href {\doibase 10.1146/annurev-conmatphys-031720-030658} {\bibfield  {journal} {\bibinfo  {journal} {Annual Review of Condensed Matter Physics}\ }\textbf {\bibinfo {volume} {14}},\ \bibinfo {pages} {335} (\bibinfo {year} {2023})},\ \Eprint {http://arxiv.org/abs/https://doi.org/10.1146/annurev-conmatphys-031720-030658} {https://doi.org/10.1146/annurev-conmatphys-031720-030658} \BibitemShut {NoStop}%
\bibitem [{\citenamefont {Szyniszewski}\ \emph {et~al.}(2020)\citenamefont {Szyniszewski}, \citenamefont {Romito},\ and\ \citenamefont {Schomerus}}]{Romito2020}%
  \BibitemOpen
  \bibfield  {author} {\bibinfo {author} {\bibfnamefont {M.}~\bibnamefont {Szyniszewski}}, \bibinfo {author} {\bibfnamefont {A.}~\bibnamefont {Romito}}, \ and\ \bibinfo {author} {\bibfnamefont {H.}~\bibnamefont {Schomerus}},\ }\href {\doibase 10.1103/PhysRevLett.125.210602} {\bibfield  {journal} {\bibinfo  {journal} {Phys. Rev. Lett.}\ }\textbf {\bibinfo {volume} {125}},\ \bibinfo {pages} {210602} (\bibinfo {year} {2020})}\BibitemShut {NoStop}%
\bibitem [{\citenamefont {Szyniszewski}\ \emph {et~al.}(2019)\citenamefont {Szyniszewski}, \citenamefont {Romito},\ and\ \citenamefont {Schomerus}}]{Schomerus2019}%
  \BibitemOpen
  \bibfield  {author} {\bibinfo {author} {\bibfnamefont {M.}~\bibnamefont {Szyniszewski}}, \bibinfo {author} {\bibfnamefont {A.}~\bibnamefont {Romito}}, \ and\ \bibinfo {author} {\bibfnamefont {H.}~\bibnamefont {Schomerus}},\ }\href {\doibase 10.1103/PhysRevB.100.064204} {\bibfield  {journal} {\bibinfo  {journal} {Phys. Rev. B}\ }\textbf {\bibinfo {volume} {100}},\ \bibinfo {pages} {064204} (\bibinfo {year} {2019})}\BibitemShut {NoStop}%
\bibitem [{\citenamefont {Iadecola}\ \emph {et~al.}(2023)\citenamefont {Iadecola}, \citenamefont {Ganeshan}, \citenamefont {Pixley},\ and\ \citenamefont {Wilson}}]{IadecolaControl}%
  \BibitemOpen
  \bibfield  {author} {\bibinfo {author} {\bibfnamefont {T.}~\bibnamefont {Iadecola}}, \bibinfo {author} {\bibfnamefont {S.}~\bibnamefont {Ganeshan}}, \bibinfo {author} {\bibfnamefont {J.~H.}\ \bibnamefont {Pixley}}, \ and\ \bibinfo {author} {\bibfnamefont {J.~H.}\ \bibnamefont {Wilson}},\ }\href {\doibase 10.1103/PhysRevLett.131.060403} {\bibfield  {journal} {\bibinfo  {journal} {Phys. Rev. Lett.}\ }\textbf {\bibinfo {volume} {131}},\ \bibinfo {pages} {060403} (\bibinfo {year} {2023})}\BibitemShut {NoStop}%
\bibitem [{\citenamefont {Iadecola}\ \emph {et~al.}(2025)\citenamefont {Iadecola}, \citenamefont {Wilson},\ and\ \citenamefont {Pixley}}]{Iadecola2025}%
  \BibitemOpen
  \bibfield  {author} {\bibinfo {author} {\bibfnamefont {T.}~\bibnamefont {Iadecola}}, \bibinfo {author} {\bibfnamefont {J.~H.}\ \bibnamefont {Wilson}}, \ and\ \bibinfo {author} {\bibfnamefont {J.}~\bibnamefont {Pixley}},\ }\href {\doibase 10.1103/PRXQuantum.6.010351} {\bibfield  {journal} {\bibinfo  {journal} {PRX Quantum}\ }\textbf {\bibinfo {volume} {6}},\ \bibinfo {pages} {010351} (\bibinfo {year} {2025})}\BibitemShut {NoStop}%
\bibitem [{\citenamefont {Agrawal}\ \emph {et~al.}(2022)\citenamefont {Agrawal}, \citenamefont {Zabalo}, \citenamefont {Chen}, \citenamefont {Wilson}, \citenamefont {Potter}, \citenamefont {Pixley}, \citenamefont {Gopalakrishnan},\ and\ \citenamefont {Vasseur}}]{Agraval2021}%
  \BibitemOpen
  \bibfield  {author} {\bibinfo {author} {\bibfnamefont {U.}~\bibnamefont {Agrawal}}, \bibinfo {author} {\bibfnamefont {A.}~\bibnamefont {Zabalo}}, \bibinfo {author} {\bibfnamefont {K.}~\bibnamefont {Chen}}, \bibinfo {author} {\bibfnamefont {J.~H.}\ \bibnamefont {Wilson}}, \bibinfo {author} {\bibfnamefont {A.~C.}\ \bibnamefont {Potter}}, \bibinfo {author} {\bibfnamefont {J.~H.}\ \bibnamefont {Pixley}}, \bibinfo {author} {\bibfnamefont {S.}~\bibnamefont {Gopalakrishnan}}, \ and\ \bibinfo {author} {\bibfnamefont {R.}~\bibnamefont {Vasseur}},\ }\href {\doibase 10.1103/PhysRevX.12.041002} {\bibfield  {journal} {\bibinfo  {journal} {Phys. Rev. X}\ }\textbf {\bibinfo {volume} {12}},\ \bibinfo {pages} {041002} (\bibinfo {year} {2022})}\BibitemShut {NoStop}%
\bibitem [{\citenamefont {Turkeshi}(2022)}]{turkeshi2022measurement}%
  \BibitemOpen
  \bibfield  {author} {\bibinfo {author} {\bibfnamefont {X.}~\bibnamefont {Turkeshi}},\ }\href {\doibase 10.1103/PhysRevB.106.144313} {\bibfield  {journal} {\bibinfo  {journal} {Phys. Rev. B}\ }\textbf {\bibinfo {volume} {106}},\ \bibinfo {pages} {144313} (\bibinfo {year} {2022})}\BibitemShut {NoStop}%
\bibitem [{\citenamefont {P\"opperl}\ \emph {et~al.}(2023)\citenamefont {P\"opperl}, \citenamefont {Gornyi},\ and\ \citenamefont {Gefen}}]{Popperl}%
  \BibitemOpen
  \bibfield  {author} {\bibinfo {author} {\bibfnamefont {P.}~\bibnamefont {P\"opperl}}, \bibinfo {author} {\bibfnamefont {I.~V.}\ \bibnamefont {Gornyi}}, \ and\ \bibinfo {author} {\bibfnamefont {Y.}~\bibnamefont {Gefen}},\ }\href {\doibase 10.1103/PhysRevB.107.174203} {\bibfield  {journal} {\bibinfo  {journal} {Phys. Rev. B}\ }\textbf {\bibinfo {volume} {107}},\ \bibinfo {pages} {174203} (\bibinfo {year} {2023})}\BibitemShut {NoStop}%
\bibitem [{\citenamefont {Carollo}\ and\ \citenamefont {Alba}(2022)}]{alba2022}%
  \BibitemOpen
  \bibfield  {author} {\bibinfo {author} {\bibfnamefont {F.}~\bibnamefont {Carollo}}\ and\ \bibinfo {author} {\bibfnamefont {V.}~\bibnamefont {Alba}},\ }\href {\doibase 10.1103/PhysRevB.106.L220304} {\bibfield  {journal} {\bibinfo  {journal} {Phys. Rev. B}\ }\textbf {\bibinfo {volume} {106}},\ \bibinfo {pages} {L220304} (\bibinfo {year} {2022})}\BibitemShut {NoStop}%
\bibitem [{\citenamefont {Sriram}\ \emph {et~al.}(2023)\citenamefont {Sriram}, \citenamefont {Rakovszky}, \citenamefont {Khemani},\ and\ \citenamefont {Ippoliti}}]{KhemaniKitaev}%
  \BibitemOpen
  \bibfield  {author} {\bibinfo {author} {\bibfnamefont {A.}~\bibnamefont {Sriram}}, \bibinfo {author} {\bibfnamefont {T.}~\bibnamefont {Rakovszky}}, \bibinfo {author} {\bibfnamefont {V.}~\bibnamefont {Khemani}}, \ and\ \bibinfo {author} {\bibfnamefont {M.}~\bibnamefont {Ippoliti}},\ }\href {\doibase 10.1103/PhysRevB.108.094304} {\bibfield  {journal} {\bibinfo  {journal} {Phys. Rev. B}\ }\textbf {\bibinfo {volume} {108}},\ \bibinfo {pages} {094304} (\bibinfo {year} {2023})}\BibitemShut {NoStop}%
\bibitem [{\citenamefont {Chakraborty}\ \emph {et~al.}(2024)\citenamefont {Chakraborty}, \citenamefont {Chen}, \citenamefont {Zabalo}, \citenamefont {Wilson},\ and\ \citenamefont {Pixley}}]{Chakraborty2024}%
  \BibitemOpen
  \bibfield  {author} {\bibinfo {author} {\bibfnamefont {A.}~\bibnamefont {Chakraborty}}, \bibinfo {author} {\bibfnamefont {K.}~\bibnamefont {Chen}}, \bibinfo {author} {\bibfnamefont {A.}~\bibnamefont {Zabalo}}, \bibinfo {author} {\bibfnamefont {J.~H.}\ \bibnamefont {Wilson}}, \ and\ \bibinfo {author} {\bibfnamefont {J.~H.}\ \bibnamefont {Pixley}},\ }\href {\doibase 10.1103/PhysRevB.110.045135} {\bibfield  {journal} {\bibinfo  {journal} {Phys. Rev. B}\ }\textbf {\bibinfo {volume} {110}},\ \bibinfo {pages} {045135} (\bibinfo {year} {2024})}\BibitemShut {NoStop}%
\bibitem [{\citenamefont {Zhu}\ \emph {et~al.}(2023)\citenamefont {Zhu}, \citenamefont {Tantivasadakarn}, \citenamefont {Vishwanath}, \citenamefont {Trebst},\ and\ \citenamefont {Verresen}}]{Zhu2022}%
  \BibitemOpen
  \bibfield  {author} {\bibinfo {author} {\bibfnamefont {G.-Y.}\ \bibnamefont {Zhu}}, \bibinfo {author} {\bibfnamefont {N.}~\bibnamefont {Tantivasadakarn}}, \bibinfo {author} {\bibfnamefont {A.}~\bibnamefont {Vishwanath}}, \bibinfo {author} {\bibfnamefont {S.}~\bibnamefont {Trebst}}, \ and\ \bibinfo {author} {\bibfnamefont {R.}~\bibnamefont {Verresen}},\ }\href {\doibase 10.1103/PhysRevLett.131.200201} {\bibfield  {journal} {\bibinfo  {journal} {Phys. Rev. Lett.}\ }\textbf {\bibinfo {volume} {131}},\ \bibinfo {pages} {200201} (\bibinfo {year} {2023})}\BibitemShut {NoStop}%
\bibitem [{\citenamefont {Zhu}\ \emph {et~al.}(2024)\citenamefont {Zhu}, \citenamefont {Tantivasadakarn},\ and\ \citenamefont {Trebst}}]{Zhu2023}%
  \BibitemOpen
  \bibfield  {author} {\bibinfo {author} {\bibfnamefont {G.-Y.}\ \bibnamefont {Zhu}}, \bibinfo {author} {\bibfnamefont {N.}~\bibnamefont {Tantivasadakarn}}, \ and\ \bibinfo {author} {\bibfnamefont {S.}~\bibnamefont {Trebst}},\ }\href {\doibase 10.1103/PhysRevResearch.6.L042063} {\bibfield  {journal} {\bibinfo  {journal} {Phys. Rev. Res.}\ }\textbf {\bibinfo {volume} {6}},\ \bibinfo {pages} {L042063} (\bibinfo {year} {2024})}\BibitemShut {NoStop}%
\bibitem [{\citenamefont {Ippoliti}\ \emph {et~al.}(2021)\citenamefont {Ippoliti}, \citenamefont {Gullans}, \citenamefont {Gopalakrishnan}, \citenamefont {Huse},\ and\ \citenamefont {Khemani}}]{ippoliti2020}%
  \BibitemOpen
  \bibfield  {author} {\bibinfo {author} {\bibfnamefont {M.}~\bibnamefont {Ippoliti}}, \bibinfo {author} {\bibfnamefont {M.~J.}\ \bibnamefont {Gullans}}, \bibinfo {author} {\bibfnamefont {S.}~\bibnamefont {Gopalakrishnan}}, \bibinfo {author} {\bibfnamefont {D.~A.}\ \bibnamefont {Huse}}, \ and\ \bibinfo {author} {\bibfnamefont {V.}~\bibnamefont {Khemani}},\ }\href {\doibase 10.1103/PhysRevX.11.011030} {\bibfield  {journal} {\bibinfo  {journal} {Phys. Rev. X}\ }\textbf {\bibinfo {volume} {11}},\ \bibinfo {pages} {011030} (\bibinfo {year} {2021})}\BibitemShut {NoStop}%
\bibitem [{\citenamefont {Klocke}\ and\ \citenamefont {Buchhold}(2022)}]{Klocke2022}%
  \BibitemOpen
  \bibfield  {author} {\bibinfo {author} {\bibfnamefont {K.}~\bibnamefont {Klocke}}\ and\ \bibinfo {author} {\bibfnamefont {M.}~\bibnamefont {Buchhold}},\ }\href {\doibase 10.1103/PhysRevB.106.104307} {\bibfield  {journal} {\bibinfo  {journal} {Phys. Rev. B}\ }\textbf {\bibinfo {volume} {106}},\ \bibinfo {pages} {104307} (\bibinfo {year} {2022})}\BibitemShut {NoStop}%
\bibitem [{\citenamefont {Klocke}\ and\ \citenamefont {Buchhold}(2023)}]{Klocke2023}%
  \BibitemOpen
  \bibfield  {author} {\bibinfo {author} {\bibfnamefont {K.}~\bibnamefont {Klocke}}\ and\ \bibinfo {author} {\bibfnamefont {M.}~\bibnamefont {Buchhold}},\ }\href {\doibase 10.1103/PhysRevX.13.041028} {\bibfield  {journal} {\bibinfo  {journal} {Phys. Rev. X}\ }\textbf {\bibinfo {volume} {13}},\ \bibinfo {pages} {041028} (\bibinfo {year} {2023})}\BibitemShut {NoStop}%
\bibitem [{\citenamefont {Lavasani}\ \emph {et~al.}(2023)\citenamefont {Lavasani}, \citenamefont {Luo},\ and\ \citenamefont {Vijay}}]{SagarKitaev}%
  \BibitemOpen
  \bibfield  {author} {\bibinfo {author} {\bibfnamefont {A.}~\bibnamefont {Lavasani}}, \bibinfo {author} {\bibfnamefont {Z.-X.}\ \bibnamefont {Luo}}, \ and\ \bibinfo {author} {\bibfnamefont {S.}~\bibnamefont {Vijay}},\ }\href {\doibase 10.1103/PhysRevB.108.115135} {\bibfield  {journal} {\bibinfo  {journal} {Phys. Rev. B}\ }\textbf {\bibinfo {volume} {108}},\ \bibinfo {pages} {115135} (\bibinfo {year} {2023})}\BibitemShut {NoStop}%
\bibitem [{\citenamefont {Morral-Yepes}\ \emph {et~al.}(2023)\citenamefont {Morral-Yepes}, \citenamefont {Pollmann},\ and\ \citenamefont {Lovas}}]{Morral2022}%
  \BibitemOpen
  \bibfield  {author} {\bibinfo {author} {\bibfnamefont {R.}~\bibnamefont {Morral-Yepes}}, \bibinfo {author} {\bibfnamefont {F.}~\bibnamefont {Pollmann}}, \ and\ \bibinfo {author} {\bibfnamefont {I.}~\bibnamefont {Lovas}},\ }\href {\doibase 10.1103/PhysRevB.108.224304} {\bibfield  {journal} {\bibinfo  {journal} {Phys. Rev. B}\ }\textbf {\bibinfo {volume} {108}},\ \bibinfo {pages} {224304} (\bibinfo {year} {2023})}\BibitemShut {NoStop}%
\bibitem [{\citenamefont {Lira-Solanilla}\ \emph {et~al.}(2025)\citenamefont {Lira-Solanilla}, \citenamefont {Turkeshi},\ and\ \citenamefont {Pappalardi}}]{lirasolanilla2025}%
  \BibitemOpen
  \bibfield  {author} {\bibinfo {author} {\bibfnamefont {A.}~\bibnamefont {Lira-Solanilla}}, \bibinfo {author} {\bibfnamefont {X.}~\bibnamefont {Turkeshi}}, \ and\ \bibinfo {author} {\bibfnamefont {S.}~\bibnamefont {Pappalardi}},\ }\href {\doibase 10.1103/fl34-h1p1} {\bibfield  {journal} {\bibinfo  {journal} {Phys. Rev. Lett.}\ }\textbf {\bibinfo {volume} {135}},\ \bibinfo {pages} {080401} (\bibinfo {year} {2025})}\BibitemShut {NoStop}%
\bibitem [{\citenamefont {Cao}\ \emph {et~al.}(2019)\citenamefont {Cao}, \citenamefont {Tilloy},\ and\ \citenamefont {Luca}}]{Cao2019}%
  \BibitemOpen
  \bibfield  {author} {\bibinfo {author} {\bibfnamefont {X.}~\bibnamefont {Cao}}, \bibinfo {author} {\bibfnamefont {A.}~\bibnamefont {Tilloy}}, \ and\ \bibinfo {author} {\bibfnamefont {A.~D.}\ \bibnamefont {Luca}},\ }\href {\doibase 10.21468/SciPostPhys.7.2.024} {\bibfield  {journal} {\bibinfo  {journal} {SciPost Phys.}\ }\textbf {\bibinfo {volume} {7}},\ \bibinfo {pages} {24} (\bibinfo {year} {2019})}\BibitemShut {NoStop}%
\bibitem [{\citenamefont {Alberton}\ \emph {et~al.}(2021)\citenamefont {Alberton}, \citenamefont {Buchhold},\ and\ \citenamefont {Diehl}}]{alberton2021enttrans}%
  \BibitemOpen
  \bibfield  {author} {\bibinfo {author} {\bibfnamefont {O.}~\bibnamefont {Alberton}}, \bibinfo {author} {\bibfnamefont {M.}~\bibnamefont {Buchhold}}, \ and\ \bibinfo {author} {\bibfnamefont {S.}~\bibnamefont {Diehl}},\ }\href {\doibase 10.1103/PhysRevLett.126.170602} {\bibfield  {journal} {\bibinfo  {journal} {Phys. Rev. Lett.}\ }\textbf {\bibinfo {volume} {126}},\ \bibinfo {pages} {170602} (\bibinfo {year} {2021})}\BibitemShut {NoStop}%
\bibitem [{\citenamefont {Bernard}\ \emph {et~al.}(2018)\citenamefont {Bernard}, \citenamefont {Jin},\ and\ \citenamefont {Shpielberg}}]{Bernard_2018}%
  \BibitemOpen
  \bibfield  {author} {\bibinfo {author} {\bibfnamefont {D.}~\bibnamefont {Bernard}}, \bibinfo {author} {\bibfnamefont {T.}~\bibnamefont {Jin}}, \ and\ \bibinfo {author} {\bibfnamefont {O.}~\bibnamefont {Shpielberg}},\ }\href {\doibase 10.1209/0295-5075/121/60006} {\bibfield  {journal} {\bibinfo  {journal} {Europhysics Letters}\ }\textbf {\bibinfo {volume} {121}},\ \bibinfo {pages} {60006} (\bibinfo {year} {2018})}\BibitemShut {NoStop}%
\bibitem [{\citenamefont {Buchhold}\ \emph {et~al.}(2021)\citenamefont {Buchhold}, \citenamefont {Minoguchi}, \citenamefont {Altland},\ and\ \citenamefont {Diehl}}]{buchhold2021effective}%
  \BibitemOpen
  \bibfield  {author} {\bibinfo {author} {\bibfnamefont {M.}~\bibnamefont {Buchhold}}, \bibinfo {author} {\bibfnamefont {Y.}~\bibnamefont {Minoguchi}}, \bibinfo {author} {\bibfnamefont {A.}~\bibnamefont {Altland}}, \ and\ \bibinfo {author} {\bibfnamefont {S.}~\bibnamefont {Diehl}},\ }\href {\doibase 10.1103/PhysRevX.11.041004} {\bibfield  {journal} {\bibinfo  {journal} {Phys. Rev. X}\ }\textbf {\bibinfo {volume} {11}},\ \bibinfo {pages} {041004} (\bibinfo {year} {2021})}\BibitemShut {NoStop}%
\bibitem [{\citenamefont {Poboiko}\ \emph {et~al.}(2023)\citenamefont {Poboiko}, \citenamefont {P\"opperl}, \citenamefont {Gornyi},\ and\ \citenamefont {Mirlin}}]{poboiko}%
  \BibitemOpen
  \bibfield  {author} {\bibinfo {author} {\bibfnamefont {I.}~\bibnamefont {Poboiko}}, \bibinfo {author} {\bibfnamefont {P.}~\bibnamefont {P\"opperl}}, \bibinfo {author} {\bibfnamefont {I.~V.}\ \bibnamefont {Gornyi}}, \ and\ \bibinfo {author} {\bibfnamefont {A.~D.}\ \bibnamefont {Mirlin}},\ }\href {\doibase 10.1103/PhysRevX.13.041046} {\bibfield  {journal} {\bibinfo  {journal} {Phys. Rev. X}\ }\textbf {\bibinfo {volume} {13}},\ \bibinfo {pages} {041046} (\bibinfo {year} {2023})}\BibitemShut {NoStop}%
\bibitem [{\citenamefont {Poboiko}\ \emph {et~al.}(2024)\citenamefont {Poboiko}, \citenamefont {Gornyi},\ and\ \citenamefont {Mirlin}}]{poboiko2023measurementinduced}%
  \BibitemOpen
  \bibfield  {author} {\bibinfo {author} {\bibfnamefont {I.}~\bibnamefont {Poboiko}}, \bibinfo {author} {\bibfnamefont {I.~V.}\ \bibnamefont {Gornyi}}, \ and\ \bibinfo {author} {\bibfnamefont {A.~D.}\ \bibnamefont {Mirlin}},\ }\href {\doibase 10.1103/PhysRevLett.132.110403} {\bibfield  {journal} {\bibinfo  {journal} {Phys. Rev. Lett.}\ }\textbf {\bibinfo {volume} {132}},\ \bibinfo {pages} {110403} (\bibinfo {year} {2024})}\BibitemShut {NoStop}%
\bibitem [{\citenamefont {Fava}\ \emph {et~al.}(2023)\citenamefont {Fava}, \citenamefont {Piroli}, \citenamefont {Swann}, \citenamefont {Bernard},\ and\ \citenamefont {Nahum}}]{Fava2023}%
  \BibitemOpen
  \bibfield  {author} {\bibinfo {author} {\bibfnamefont {M.}~\bibnamefont {Fava}}, \bibinfo {author} {\bibfnamefont {L.}~\bibnamefont {Piroli}}, \bibinfo {author} {\bibfnamefont {T.}~\bibnamefont {Swann}}, \bibinfo {author} {\bibfnamefont {D.}~\bibnamefont {Bernard}}, \ and\ \bibinfo {author} {\bibfnamefont {A.}~\bibnamefont {Nahum}},\ }\href {\doibase 10.1103/PhysRevX.13.041045} {\bibfield  {journal} {\bibinfo  {journal} {Phys. Rev. X}\ }\textbf {\bibinfo {volume} {13}},\ \bibinfo {pages} {041045} (\bibinfo {year} {2023})}\BibitemShut {NoStop}%
\bibitem [{\citenamefont {Fava}\ \emph {et~al.}(2024)\citenamefont {Fava}, \citenamefont {Piroli}, \citenamefont {Bernard},\ and\ \citenamefont {Nahum}}]{Fava2024}%
  \BibitemOpen
  \bibfield  {author} {\bibinfo {author} {\bibfnamefont {M.}~\bibnamefont {Fava}}, \bibinfo {author} {\bibfnamefont {L.}~\bibnamefont {Piroli}}, \bibinfo {author} {\bibfnamefont {D.}~\bibnamefont {Bernard}}, \ and\ \bibinfo {author} {\bibfnamefont {A.}~\bibnamefont {Nahum}},\ }\href {\doibase 10.1103/PhysRevResearch.6.043246} {\bibfield  {journal} {\bibinfo  {journal} {Phys. Rev. Res.}\ }\textbf {\bibinfo {volume} {6}},\ \bibinfo {pages} {043246} (\bibinfo {year} {2024})}\BibitemShut {NoStop}%
\bibitem [{\citenamefont {Chahine}\ and\ \citenamefont {Buchhold}(2024)}]{chahine2024}%
  \BibitemOpen
  \bibfield  {author} {\bibinfo {author} {\bibfnamefont {K.}~\bibnamefont {Chahine}}\ and\ \bibinfo {author} {\bibfnamefont {M.}~\bibnamefont {Buchhold}},\ }\href {\doibase 10.1103/PhysRevB.110.054313} {\bibfield  {journal} {\bibinfo  {journal} {Phys. Rev. B}\ }\textbf {\bibinfo {volume} {110}},\ \bibinfo {pages} {054313} (\bibinfo {year} {2024})}\BibitemShut {NoStop}%
\bibitem [{\citenamefont {Klocke}\ \emph {et~al.}(2025)\citenamefont {Klocke}, \citenamefont {Simm}, \citenamefont {Zhu}, \citenamefont {Trebst},\ and\ \citenamefont {Buchhold}}]{klocke2025}%
  \BibitemOpen
  \bibfield  {author} {\bibinfo {author} {\bibfnamefont {K.}~\bibnamefont {Klocke}}, \bibinfo {author} {\bibfnamefont {D.}~\bibnamefont {Simm}}, \bibinfo {author} {\bibfnamefont {G.-Y.}\ \bibnamefont {Zhu}}, \bibinfo {author} {\bibfnamefont {S.}~\bibnamefont {Trebst}}, \ and\ \bibinfo {author} {\bibfnamefont {M.}~\bibnamefont {Buchhold}},\ }\href {\doibase 10.1103/PhysRevB.111.224301} {\bibfield  {journal} {\bibinfo  {journal} {Phys. Rev. B}\ }\textbf {\bibinfo {volume} {111}},\ \bibinfo {pages} {224301} (\bibinfo {year} {2025})}\BibitemShut {NoStop}%
\bibitem [{\citenamefont {Ladewig}\ \emph {et~al.}(2022)\citenamefont {Ladewig}, \citenamefont {Diehl},\ and\ \citenamefont {Buchhold}}]{Ladewig2022}%
  \BibitemOpen
  \bibfield  {author} {\bibinfo {author} {\bibfnamefont {B.}~\bibnamefont {Ladewig}}, \bibinfo {author} {\bibfnamefont {S.}~\bibnamefont {Diehl}}, \ and\ \bibinfo {author} {\bibfnamefont {M.}~\bibnamefont {Buchhold}},\ }\href {\doibase 10.1103/PhysRevResearch.4.033001} {\bibfield  {journal} {\bibinfo  {journal} {Phys. Rev. Res.}\ }\textbf {\bibinfo {volume} {4}},\ \bibinfo {pages} {033001} (\bibinfo {year} {2022})}\BibitemShut {NoStop}%
\bibitem [{\citenamefont {Minato}\ \emph {et~al.}(2022)\citenamefont {Minato}, \citenamefont {Sugimoto}, \citenamefont {Kuwahara},\ and\ \citenamefont {Saito}}]{minato2021fate}%
  \BibitemOpen
  \bibfield  {author} {\bibinfo {author} {\bibfnamefont {T.}~\bibnamefont {Minato}}, \bibinfo {author} {\bibfnamefont {K.}~\bibnamefont {Sugimoto}}, \bibinfo {author} {\bibfnamefont {T.}~\bibnamefont {Kuwahara}}, \ and\ \bibinfo {author} {\bibfnamefont {K.}~\bibnamefont {Saito}},\ }\href {\doibase 10.1103/PhysRevLett.128.010603} {\bibfield  {journal} {\bibinfo  {journal} {Phys. Rev. Lett.}\ }\textbf {\bibinfo {volume} {128}},\ \bibinfo {pages} {010603} (\bibinfo {year} {2022})}\BibitemShut {NoStop}%
\bibitem [{\citenamefont {Szyniszewski}\ \emph {et~al.}(2023)\citenamefont {Szyniszewski}, \citenamefont {Lunt},\ and\ \citenamefont {Pal}}]{szyniszewski2023disordered}%
  \BibitemOpen
  \bibfield  {author} {\bibinfo {author} {\bibfnamefont {M.}~\bibnamefont {Szyniszewski}}, \bibinfo {author} {\bibfnamefont {O.}~\bibnamefont {Lunt}}, \ and\ \bibinfo {author} {\bibfnamefont {A.}~\bibnamefont {Pal}},\ }\href {\doibase 10.1103/PhysRevB.108.165126} {\bibfield  {journal} {\bibinfo  {journal} {Phys. Rev. B}\ }\textbf {\bibinfo {volume} {108}},\ \bibinfo {pages} {165126} (\bibinfo {year} {2023})}\BibitemShut {NoStop}%
\bibitem [{\citenamefont {Kells}\ \emph {et~al.}(2023)\citenamefont {Kells}, \citenamefont {Meidan},\ and\ \citenamefont {Romito}}]{kells2023}%
  \BibitemOpen
  \bibfield  {author} {\bibinfo {author} {\bibfnamefont {G.}~\bibnamefont {Kells}}, \bibinfo {author} {\bibfnamefont {D.}~\bibnamefont {Meidan}}, \ and\ \bibinfo {author} {\bibfnamefont {A.}~\bibnamefont {Romito}},\ }\href {\doibase 10.21468/SciPostPhys.14.3.031} {\bibfield  {journal} {\bibinfo  {journal} {SciPost Phys.}\ }\textbf {\bibinfo {volume} {14}},\ \bibinfo {pages} {031} (\bibinfo {year} {2023})}\BibitemShut {NoStop}%
\bibitem [{\citenamefont {Jin}\ and\ \citenamefont {Martin}(2024)}]{jin2023measurementinduced}%
  \BibitemOpen
  \bibfield  {author} {\bibinfo {author} {\bibfnamefont {T.}~\bibnamefont {Jin}}\ and\ \bibinfo {author} {\bibfnamefont {D.~G.}\ \bibnamefont {Martin}},\ }\href {\doibase 10.1103/PhysRevB.110.L060202} {\bibfield  {journal} {\bibinfo  {journal} {Phys. Rev. B}\ }\textbf {\bibinfo {volume} {110}},\ \bibinfo {pages} {L060202} (\bibinfo {year} {2024})}\BibitemShut {NoStop}%
\bibitem [{\citenamefont {{Poboiko}}\ \emph {et~al.}(2025)\citenamefont {{Poboiko}}, \citenamefont {{Gornyi}},\ and\ \citenamefont {{Mirlin}}}]{poboiko2025}%
  \BibitemOpen
  \bibfield  {author} {\bibinfo {author} {\bibfnamefont {I.}~\bibnamefont {{Poboiko}}}, \bibinfo {author} {\bibfnamefont {I.~V.}\ \bibnamefont {{Gornyi}}}, \ and\ \bibinfo {author} {\bibfnamefont {A.~D.}\ \bibnamefont {{Mirlin}}},\ }\href {\doibase 10.48550/arXiv.2507.11312} {\  (\bibinfo {year} {2025}),\ 10.48550/arXiv.2507.11312},\ \Eprint {http://arxiv.org/abs/2507.11312} {arXiv:2507.11312 [quant-ph]} \BibitemShut {NoStop}%
\bibitem [{\citenamefont {Poboiko}\ \emph {et~al.}(2025)\citenamefont {Poboiko}, \citenamefont {P\"opperl}, \citenamefont {Gornyi},\ and\ \citenamefont {Mirlin}}]{poboiko2025interacting}%
  \BibitemOpen
  \bibfield  {author} {\bibinfo {author} {\bibfnamefont {I.}~\bibnamefont {Poboiko}}, \bibinfo {author} {\bibfnamefont {P.}~\bibnamefont {P\"opperl}}, \bibinfo {author} {\bibfnamefont {I.~V.}\ \bibnamefont {Gornyi}}, \ and\ \bibinfo {author} {\bibfnamefont {A.~D.}\ \bibnamefont {Mirlin}},\ }\href {\doibase 10.1103/PhysRevB.111.024204} {\bibfield  {journal} {\bibinfo  {journal} {Phys. Rev. B}\ }\textbf {\bibinfo {volume} {111}},\ \bibinfo {pages} {024204} (\bibinfo {year} {2025})}\BibitemShut {NoStop}%
\bibitem [{\citenamefont {{Poboiko}}\ \emph {et~al.}(2025)\citenamefont {{Poboiko}}, \citenamefont {{Szyniszewski}}, \citenamefont {{Turner}}, \citenamefont {{Gornyi}}, \citenamefont {{Mirlin}},\ and\ \citenamefont {{Pal}}}]{poboiko2025levyflights}%
  \BibitemOpen
  \bibfield  {author} {\bibinfo {author} {\bibfnamefont {I.}~\bibnamefont {{Poboiko}}}, \bibinfo {author} {\bibfnamefont {M.}~\bibnamefont {{Szyniszewski}}}, \bibinfo {author} {\bibfnamefont {C.~J.}\ \bibnamefont {{Turner}}}, \bibinfo {author} {\bibfnamefont {I.~V.}\ \bibnamefont {{Gornyi}}}, \bibinfo {author} {\bibfnamefont {A.~D.}\ \bibnamefont {{Mirlin}}}, \ and\ \bibinfo {author} {\bibfnamefont {A.}~\bibnamefont {{Pal}}},\ }\href {\doibase 10.48550/arXiv.2501.12903} {\bibfield  {journal} {\bibinfo  {journal} {arXiv e-prints}\ } (\bibinfo {year} {2025}),\ 10.48550/arXiv.2501.12903}\BibitemShut {NoStop}%
\bibitem [{\citenamefont {Doggen}\ \emph {et~al.}(2022)\citenamefont {Doggen}, \citenamefont {Gefen}, \citenamefont {Gornyi}, \citenamefont {Mirlin},\ and\ \citenamefont {Polyakov}}]{doggen2021generalized}%
  \BibitemOpen
  \bibfield  {author} {\bibinfo {author} {\bibfnamefont {E.~V.~H.}\ \bibnamefont {Doggen}}, \bibinfo {author} {\bibfnamefont {Y.}~\bibnamefont {Gefen}}, \bibinfo {author} {\bibfnamefont {I.~V.}\ \bibnamefont {Gornyi}}, \bibinfo {author} {\bibfnamefont {A.~D.}\ \bibnamefont {Mirlin}}, \ and\ \bibinfo {author} {\bibfnamefont {D.~G.}\ \bibnamefont {Polyakov}},\ }\href {\doibase 10.1103/PhysRevResearch.4.023146} {\bibfield  {journal} {\bibinfo  {journal} {Phys. Rev. Res.}\ }\textbf {\bibinfo {volume} {4}},\ \bibinfo {pages} {023146} (\bibinfo {year} {2022})}\BibitemShut {NoStop}%
\bibitem [{\citenamefont {Tsitsishvili}\ \emph {et~al.}(2024)\citenamefont {Tsitsishvili}, \citenamefont {Poletti}, \citenamefont {Dalmonte},\ and\ \citenamefont {Chiriacò}}]{tsitsishvili2023measurement}%
  \BibitemOpen
  \bibfield  {author} {\bibinfo {author} {\bibfnamefont {M.}~\bibnamefont {Tsitsishvili}}, \bibinfo {author} {\bibfnamefont {D.}~\bibnamefont {Poletti}}, \bibinfo {author} {\bibfnamefont {M.}~\bibnamefont {Dalmonte}}, \ and\ \bibinfo {author} {\bibfnamefont {G.}~\bibnamefont {Chiriacò}},\ }\href {\doibase 10.21468/SciPostPhysCore.7.1.011} {\bibfield  {journal} {\bibinfo  {journal} {SciPost Phys. Core}\ }\textbf {\bibinfo {volume} {7}},\ \bibinfo {pages} {011} (\bibinfo {year} {2024})}\BibitemShut {NoStop}%
\bibitem [{\citenamefont {Turkeshi}\ \emph {et~al.}(2022)\citenamefont {Turkeshi}, \citenamefont {Dalmonte}, \citenamefont {Fazio},\ and\ \citenamefont {Schir\`o}}]{turkeshi2022entanglement}%
  \BibitemOpen
  \bibfield  {author} {\bibinfo {author} {\bibfnamefont {X.}~\bibnamefont {Turkeshi}}, \bibinfo {author} {\bibfnamefont {M.}~\bibnamefont {Dalmonte}}, \bibinfo {author} {\bibfnamefont {R.}~\bibnamefont {Fazio}}, \ and\ \bibinfo {author} {\bibfnamefont {M.}~\bibnamefont {Schir\`o}},\ }\href {\doibase 10.1103/PhysRevB.105.L241114} {\bibfield  {journal} {\bibinfo  {journal} {Phys. Rev. B}\ }\textbf {\bibinfo {volume} {105}},\ \bibinfo {pages} {L241114} (\bibinfo {year} {2022})}\BibitemShut {NoStop}%
\bibitem [{\citenamefont {Piccitto}\ \emph {et~al.}(2022)\citenamefont {Piccitto}, \citenamefont {Russomanno},\ and\ \citenamefont {Rossini}}]{piccitto2022}%
  \BibitemOpen
  \bibfield  {author} {\bibinfo {author} {\bibfnamefont {G.}~\bibnamefont {Piccitto}}, \bibinfo {author} {\bibfnamefont {A.}~\bibnamefont {Russomanno}}, \ and\ \bibinfo {author} {\bibfnamefont {D.}~\bibnamefont {Rossini}},\ }\href {\doibase 10.1103/PhysRevB.105.064305} {\bibfield  {journal} {\bibinfo  {journal} {Phys. Rev. B}\ }\textbf {\bibinfo {volume} {105}},\ \bibinfo {pages} {064305} (\bibinfo {year} {2022})}\BibitemShut {NoStop}%
\bibitem [{\citenamefont {M\"uller}\ \emph {et~al.}(2022)\citenamefont {M\"uller}, \citenamefont {Diehl},\ and\ \citenamefont {Buchhold}}]{Mueller2022}%
  \BibitemOpen
  \bibfield  {author} {\bibinfo {author} {\bibfnamefont {T.}~\bibnamefont {M\"uller}}, \bibinfo {author} {\bibfnamefont {S.}~\bibnamefont {Diehl}}, \ and\ \bibinfo {author} {\bibfnamefont {M.}~\bibnamefont {Buchhold}},\ }\href {\doibase 10.1103/PhysRevLett.128.010605} {\bibfield  {journal} {\bibinfo  {journal} {Phys. Rev. Lett.}\ }\textbf {\bibinfo {volume} {128}},\ \bibinfo {pages} {010605} (\bibinfo {year} {2022})}\BibitemShut {NoStop}%
\bibitem [{\citenamefont {Oshima}\ and\ \citenamefont {Fuji}(2023)}]{oshima2023}%
  \BibitemOpen
  \bibfield  {author} {\bibinfo {author} {\bibfnamefont {H.}~\bibnamefont {Oshima}}\ and\ \bibinfo {author} {\bibfnamefont {Y.}~\bibnamefont {Fuji}},\ }\href {\doibase 10.1103/PhysRevB.107.014308} {\bibfield  {journal} {\bibinfo  {journal} {Phys. Rev. B}\ }\textbf {\bibinfo {volume} {107}},\ \bibinfo {pages} {014308} (\bibinfo {year} {2023})}\BibitemShut {NoStop}%
\bibitem [{\citenamefont {Paviglianiti}\ and\ \citenamefont {Silva}(2023)}]{paviglianiti2023multipartite}%
  \BibitemOpen
  \bibfield  {author} {\bibinfo {author} {\bibfnamefont {A.}~\bibnamefont {Paviglianiti}}\ and\ \bibinfo {author} {\bibfnamefont {A.}~\bibnamefont {Silva}},\ }\href {\doibase 10.1103/PhysRevB.108.184302} {\bibfield  {journal} {\bibinfo  {journal} {Phys. Rev. B}\ }\textbf {\bibinfo {volume} {108}},\ \bibinfo {pages} {184302} (\bibinfo {year} {2023})}\BibitemShut {NoStop}%
\bibitem [{\citenamefont {Minoguchi}\ \emph {et~al.}(2022)\citenamefont {Minoguchi}, \citenamefont {Rabl},\ and\ \citenamefont {Buchhold}}]{minoguchi2021continuous}%
  \BibitemOpen
  \bibfield  {author} {\bibinfo {author} {\bibfnamefont {Y.}~\bibnamefont {Minoguchi}}, \bibinfo {author} {\bibfnamefont {P.}~\bibnamefont {Rabl}}, \ and\ \bibinfo {author} {\bibfnamefont {M.}~\bibnamefont {Buchhold}},\ }\href {\doibase 10.21468/SciPostPhys.12.1.009} {\bibfield  {journal} {\bibinfo  {journal} {SciPost Phys.}\ }\textbf {\bibinfo {volume} {12}},\ \bibinfo {pages} {009} (\bibinfo {year} {2022})}\BibitemShut {NoStop}%
\bibitem [{\citenamefont {Tirrito}\ \emph {et~al.}(2023)\citenamefont {Tirrito}, \citenamefont {Santini}, \citenamefont {Fazio},\ and\ \citenamefont {Collura}}]{tirrito2023}%
  \BibitemOpen
  \bibfield  {author} {\bibinfo {author} {\bibfnamefont {E.}~\bibnamefont {Tirrito}}, \bibinfo {author} {\bibfnamefont {A.}~\bibnamefont {Santini}}, \bibinfo {author} {\bibfnamefont {R.}~\bibnamefont {Fazio}}, \ and\ \bibinfo {author} {\bibfnamefont {M.}~\bibnamefont {Collura}},\ }\href {\doibase 10.21468/SciPostPhys.15.3.096} {\bibfield  {journal} {\bibinfo  {journal} {SciPost Phys.}\ }\textbf {\bibinfo {volume} {15}},\ \bibinfo {pages} {096} (\bibinfo {year} {2023})}\BibitemShut {NoStop}%
\bibitem [{\citenamefont {Xing}\ \emph {et~al.}(2024)\citenamefont {Xing}, \citenamefont {Turkeshi}, \citenamefont {Schir\'o}, \citenamefont {Fazio},\ and\ \citenamefont {Poletti}}]{xing2023interactions}%
  \BibitemOpen
  \bibfield  {author} {\bibinfo {author} {\bibfnamefont {B.}~\bibnamefont {Xing}}, \bibinfo {author} {\bibfnamefont {X.}~\bibnamefont {Turkeshi}}, \bibinfo {author} {\bibfnamefont {M.}~\bibnamefont {Schir\'o}}, \bibinfo {author} {\bibfnamefont {R.}~\bibnamefont {Fazio}}, \ and\ \bibinfo {author} {\bibfnamefont {D.}~\bibnamefont {Poletti}},\ }\href {\doibase 10.1103/PhysRevB.109.L060302} {\bibfield  {journal} {\bibinfo  {journal} {Phys. Rev. B}\ }\textbf {\bibinfo {volume} {109}},\ \bibinfo {pages} {L060302} (\bibinfo {year} {2024})}\BibitemShut {NoStop}%
\bibitem [{\citenamefont {Turkeshi}\ and\ \citenamefont {Schir\'o}(2023)}]{turkeshi2023entanglement}%
  \BibitemOpen
  \bibfield  {author} {\bibinfo {author} {\bibfnamefont {X.}~\bibnamefont {Turkeshi}}\ and\ \bibinfo {author} {\bibfnamefont {M.}~\bibnamefont {Schir\'o}},\ }\href {\doibase 10.1103/PhysRevB.107.L020403} {\bibfield  {journal} {\bibinfo  {journal} {Phys. Rev. B}\ }\textbf {\bibinfo {volume} {107}},\ \bibinfo {pages} {L020403} (\bibinfo {year} {2023})}\BibitemShut {NoStop}%
\bibitem [{\citenamefont {Fuji}\ and\ \citenamefont {Ashida}(2020)}]{fuji2020}%
  \BibitemOpen
  \bibfield  {author} {\bibinfo {author} {\bibfnamefont {Y.}~\bibnamefont {Fuji}}\ and\ \bibinfo {author} {\bibfnamefont {Y.}~\bibnamefont {Ashida}},\ }\href {\doibase 10.1103/PhysRevB.102.054302} {\bibfield  {journal} {\bibinfo  {journal} {Phys. Rev. B}\ }\textbf {\bibinfo {volume} {102}},\ \bibinfo {pages} {054302} (\bibinfo {year} {2020})}\BibitemShut {NoStop}%
\bibitem [{\citenamefont {Altland}\ \emph {et~al.}(2022)\citenamefont {Altland}, \citenamefont {Buchhold}, \citenamefont {Diehl},\ and\ \citenamefont {Micklitz}}]{Altland2022}%
  \BibitemOpen
  \bibfield  {author} {\bibinfo {author} {\bibfnamefont {A.}~\bibnamefont {Altland}}, \bibinfo {author} {\bibfnamefont {M.}~\bibnamefont {Buchhold}}, \bibinfo {author} {\bibfnamefont {S.}~\bibnamefont {Diehl}}, \ and\ \bibinfo {author} {\bibfnamefont {T.}~\bibnamefont {Micklitz}},\ }\href {\doibase 10.1103/PhysRevResearch.4.L022066} {\bibfield  {journal} {\bibinfo  {journal} {Phys. Rev. Res.}\ }\textbf {\bibinfo {volume} {4}},\ \bibinfo {pages} {L022066} (\bibinfo {year} {2022})}\BibitemShut {NoStop}%
\bibitem [{\citenamefont {Turkeshi}\ \emph {et~al.}(2024)\citenamefont {Turkeshi}, \citenamefont {Piroli},\ and\ \citenamefont {Schir\`o}}]{turkeshi2023density}%
  \BibitemOpen
  \bibfield  {author} {\bibinfo {author} {\bibfnamefont {X.}~\bibnamefont {Turkeshi}}, \bibinfo {author} {\bibfnamefont {L.}~\bibnamefont {Piroli}}, \ and\ \bibinfo {author} {\bibfnamefont {M.}~\bibnamefont {Schir\`o}},\ }\href {\doibase 10.1103/PhysRevB.109.144306} {\bibfield  {journal} {\bibinfo  {journal} {Phys. Rev. B}\ }\textbf {\bibinfo {volume} {109}},\ \bibinfo {pages} {144306} (\bibinfo {year} {2024})}\BibitemShut {NoStop}%
\bibitem [{\citenamefont {Yang}\ \emph {et~al.}(2023)\citenamefont {Yang}, \citenamefont {Zuo},\ and\ \citenamefont {Liu}}]{Yang2023Keldysh}%
  \BibitemOpen
  \bibfield  {author} {\bibinfo {author} {\bibfnamefont {Q.}~\bibnamefont {Yang}}, \bibinfo {author} {\bibfnamefont {Y.}~\bibnamefont {Zuo}}, \ and\ \bibinfo {author} {\bibfnamefont {D.~E.}\ \bibnamefont {Liu}},\ }\href {\doibase 10.1103/PhysRevResearch.5.033174} {\bibfield  {journal} {\bibinfo  {journal} {Phys. Rev. Res.}\ }\textbf {\bibinfo {volume} {5}},\ \bibinfo {pages} {033174} (\bibinfo {year} {2023})}\BibitemShut {NoStop}%
\bibitem [{\citenamefont {{Fan}}\ \emph {et~al.}(2025)\citenamefont {{Fan}}, \citenamefont {{Yin}},\ and\ \citenamefont {{Garc{\'\i}a-Garc{\'\i}a}}}]{fan2025}%
  \BibitemOpen
  \bibfield  {author} {\bibinfo {author} {\bibfnamefont {B.}~\bibnamefont {{Fan}}}, \bibinfo {author} {\bibfnamefont {C.}~\bibnamefont {{Yin}}}, \ and\ \bibinfo {author} {\bibfnamefont {A.~M.}\ \bibnamefont {{Garc{\'\i}a-Garc{\'\i}a}}},\ }\href {\doibase 10.48550/arXiv.2508.18468} {\  (\bibinfo {year} {2025}),\ 10.48550/arXiv.2508.18468},\ \Eprint {http://arxiv.org/abs/2508.18468} {arXiv:2508.18468 [quant-ph]} \BibitemShut {NoStop}%
\bibitem [{\citenamefont {Brody}\ \emph {et~al.}(1981)\citenamefont {Brody}, \citenamefont {Flores}, \citenamefont {French}, \citenamefont {Mello}, \citenamefont {Pandey},\ and\ \citenamefont {Wong}}]{brody1981}%
  \BibitemOpen
  \bibfield  {author} {\bibinfo {author} {\bibfnamefont {T.~A.}\ \bibnamefont {Brody}}, \bibinfo {author} {\bibfnamefont {J.}~\bibnamefont {Flores}}, \bibinfo {author} {\bibfnamefont {J.~B.}\ \bibnamefont {French}}, \bibinfo {author} {\bibfnamefont {P.~A.}\ \bibnamefont {Mello}}, \bibinfo {author} {\bibfnamefont {A.}~\bibnamefont {Pandey}}, \ and\ \bibinfo {author} {\bibfnamefont {S.~S.~M.}\ \bibnamefont {Wong}},\ }\href {\doibase 10.1103/RevModPhys.53.385} {\bibfield  {journal} {\bibinfo  {journal} {Rev. Mod. Phys.}\ }\textbf {\bibinfo {volume} {53}},\ \bibinfo {pages} {385} (\bibinfo {year} {1981})}\BibitemShut {NoStop}%
\bibitem [{\citenamefont {Metha}(1967)}]{mehta1967}%
  \BibitemOpen
  \bibfield  {author} {\bibinfo {author} {\bibfnamefont {M.}~\bibnamefont {Metha}},\ }\href {\doibase https://doi.org/10.1016/B978-1-4832-3258-4.50001-X} {\emph {\bibinfo {title} {Random Matrices and the Statistical Theory of Energy Levels}}}\ (\bibinfo  {publisher} {Academic Press},\ \bibinfo {year} {1967})\BibitemShut {NoStop}%
\bibitem [{\citenamefont {Malakar}\ \emph {et~al.}(2024)\citenamefont {Malakar}, \citenamefont {Brenes}, \citenamefont {Segal},\ and\ \citenamefont {Silva}}]{Malakar2024}%
  \BibitemOpen
  \bibfield  {author} {\bibinfo {author} {\bibfnamefont {M.}~\bibnamefont {Malakar}}, \bibinfo {author} {\bibfnamefont {M.}~\bibnamefont {Brenes}}, \bibinfo {author} {\bibfnamefont {D.}~\bibnamefont {Segal}}, \ and\ \bibinfo {author} {\bibfnamefont {A.}~\bibnamefont {Silva}},\ }\href {\doibase 10.1103/PhysRevB.110.134316} {\bibfield  {journal} {\bibinfo  {journal} {Phys. Rev. B}\ }\textbf {\bibinfo {volume} {110}},\ \bibinfo {pages} {134316} (\bibinfo {year} {2024})}\BibitemShut {NoStop}%
\bibitem [{\citenamefont {Zerba}\ and\ \citenamefont {Silva}(2023)}]{zerba2023}%
  \BibitemOpen
  \bibfield  {author} {\bibinfo {author} {\bibfnamefont {C.}~\bibnamefont {Zerba}}\ and\ \bibinfo {author} {\bibfnamefont {A.}~\bibnamefont {Silva}},\ }\href {\doibase 10.21468/SciPostPhysCore.6.3.051} {\bibfield  {journal} {\bibinfo  {journal} {SciPost Phys. Core}\ }\textbf {\bibinfo {volume} {6}},\ \bibinfo {pages} {051} (\bibinfo {year} {2023})}\BibitemShut {NoStop}%
\bibitem [{\citenamefont {Nahum}\ and\ \citenamefont {Skinner}(2020)}]{Nahum20a}%
  \BibitemOpen
  \bibfield  {author} {\bibinfo {author} {\bibfnamefont {A.}~\bibnamefont {Nahum}}\ and\ \bibinfo {author} {\bibfnamefont {B.}~\bibnamefont {Skinner}},\ }\href {\doibase 10.1103/PhysRevResearch.2.023288} {\bibfield  {journal} {\bibinfo  {journal} {Phys. Rev. Research}\ }\textbf {\bibinfo {volume} {2}},\ \bibinfo {pages} {023288} (\bibinfo {year} {2020})}\BibitemShut {NoStop}%
\bibitem [{\citenamefont {Merritt}\ and\ \citenamefont {Fidkowski}(2023)}]{merritt2023}%
  \BibitemOpen
  \bibfield  {author} {\bibinfo {author} {\bibfnamefont {J.}~\bibnamefont {Merritt}}\ and\ \bibinfo {author} {\bibfnamefont {L.}~\bibnamefont {Fidkowski}},\ }\href {\doibase 10.1103/PhysRevB.107.064303} {\bibfield  {journal} {\bibinfo  {journal} {Phys. Rev. B}\ }\textbf {\bibinfo {volume} {107}},\ \bibinfo {pages} {064303} (\bibinfo {year} {2023})}\BibitemShut {NoStop}%
\bibitem [{\citenamefont {Sang}\ and\ \citenamefont {Hsieh}(021a)}]{sang2021}%
  \BibitemOpen
  \bibfield  {author} {\bibinfo {author} {\bibfnamefont {S.}~\bibnamefont {Sang}}\ and\ \bibinfo {author} {\bibfnamefont {T.~H.}\ \bibnamefont {Hsieh}},\ }\href {\doibase 10.1103/PhysRevResearch.3.023200} {\bibfield  {journal} {\bibinfo  {journal} {Phys. Rev. Res.}\ }\textbf {\bibinfo {volume} {3}},\ \bibinfo {pages} {023200} (\bibinfo {year} {2021a})}\BibitemShut {NoStop}%
\bibitem [{\citenamefont {Sang}\ \emph {et~al.}(2021)\citenamefont {Sang}, \citenamefont {Li}, \citenamefont {Zhou}, \citenamefont {Chen}, \citenamefont {Hsieh},\ and\ \citenamefont {Fisher}}]{sang2021b}%
  \BibitemOpen
  \bibfield  {author} {\bibinfo {author} {\bibfnamefont {S.}~\bibnamefont {Sang}}, \bibinfo {author} {\bibfnamefont {Y.}~\bibnamefont {Li}}, \bibinfo {author} {\bibfnamefont {T.}~\bibnamefont {Zhou}}, \bibinfo {author} {\bibfnamefont {X.}~\bibnamefont {Chen}}, \bibinfo {author} {\bibfnamefont {T.~H.}\ \bibnamefont {Hsieh}}, \ and\ \bibinfo {author} {\bibfnamefont {M.~P.}\ \bibnamefont {Fisher}},\ }\href {\doibase 10.1103/PRXQuantum.2.030313} {\bibfield  {journal} {\bibinfo  {journal} {PRX Quantum}\ }\textbf {\bibinfo {volume} {2}},\ \bibinfo {pages} {030313} (\bibinfo {year} {2021})}\BibitemShut {NoStop}%
\bibitem [{\citenamefont {Walls}\ and\ \citenamefont {Milburn}(2008)}]{book_QO_Walls}%
  \BibitemOpen
  \bibfield  {author} {\bibinfo {author} {\bibfnamefont {D.~F.}\ \bibnamefont {Walls}}\ and\ \bibinfo {author} {\bibfnamefont {G.~J.}\ \bibnamefont {Milburn}},\ }\href {\doibase 10.1007/978-3-540-28574-8} {\emph {\bibinfo {title} {Quantum Optics}}}\ (\bibinfo  {publisher} {Springer-Verlag Berlin Heidelberg},\ \bibinfo {year} {2008})\BibitemShut {NoStop}%
\bibitem [{\citenamefont {Wiseman}\ and\ \citenamefont {Milburn}(2009)}]{book_QControl_Wiseman}%
  \BibitemOpen
  \bibfield  {author} {\bibinfo {author} {\bibfnamefont {H.~M.}\ \bibnamefont {Wiseman}}\ and\ \bibinfo {author} {\bibfnamefont {G.~J.}\ \bibnamefont {Milburn}},\ }\href {\doibase 10.1017/CBO9780511813948} {\emph {\bibinfo {title} {Quantum Measurement and Control}}}\ (\bibinfo  {publisher} {Cambridge University Press},\ \bibinfo {year} {2009})\BibitemShut {NoStop}%
\bibitem [{\citenamefont {Bianchi}\ \emph {et~al.}(2021)\citenamefont {Bianchi}, \citenamefont {Hackl},\ and\ \citenamefont {Kieburg}}]{bianchi2021}%
  \BibitemOpen
  \bibfield  {author} {\bibinfo {author} {\bibfnamefont {E.}~\bibnamefont {Bianchi}}, \bibinfo {author} {\bibfnamefont {L.}~\bibnamefont {Hackl}}, \ and\ \bibinfo {author} {\bibfnamefont {M.}~\bibnamefont {Kieburg}},\ }\href {\doibase 10.1103/PhysRevB.103.L241118} {\bibfield  {journal} {\bibinfo  {journal} {Phys. Rev. B}\ }\textbf {\bibinfo {volume} {103}},\ \bibinfo {pages} {L241118} (\bibinfo {year} {2021})}\BibitemShut {NoStop}%
\bibitem [{\citenamefont {Alba}\ and\ \citenamefont {Calabrese}(2018)}]{Alba2018}%
  \BibitemOpen
  \bibfield  {author} {\bibinfo {author} {\bibfnamefont {V.}~\bibnamefont {Alba}}\ and\ \bibinfo {author} {\bibfnamefont {P.}~\bibnamefont {Calabrese}},\ }\href {\doibase 10.21468/SciPostPhys.4.3.017} {\bibfield  {journal} {\bibinfo  {journal} {SciPost Phys.}\ }\textbf {\bibinfo {volume} {4}},\ \bibinfo {pages} {17} (\bibinfo {year} {2018})}\BibitemShut {NoStop}%
\bibitem [{\citenamefont {Calabrese}\ and\ \citenamefont {Cardy}(2005)}]{Calabrese_2005}%
  \BibitemOpen
  \bibfield  {author} {\bibinfo {author} {\bibfnamefont {P.}~\bibnamefont {Calabrese}}\ and\ \bibinfo {author} {\bibfnamefont {J.}~\bibnamefont {Cardy}},\ }\href {\doibase 10.1088/1742-5468/2005/04/p04010} {\bibfield  {journal} {\bibinfo  {journal} {Journal of Statistical Mechanics: Theory and Experiment}\ }\textbf {\bibinfo {volume} {2005}},\ \bibinfo {pages} {P04010} (\bibinfo {year} {2005})}\BibitemShut {NoStop}%
\bibitem [{\citenamefont {Ardonne}\ \emph {et~al.}(2004)\citenamefont {Ardonne}, \citenamefont {Fendley},\ and\ \citenamefont {Fradkin}}]{ardonne2004}%
  \BibitemOpen
  \bibfield  {author} {\bibinfo {author} {\bibfnamefont {E.}~\bibnamefont {Ardonne}}, \bibinfo {author} {\bibfnamefont {P.}~\bibnamefont {Fendley}}, \ and\ \bibinfo {author} {\bibfnamefont {E.}~\bibnamefont {Fradkin}},\ }\href {\doibase https://doi.org/10.1016/j.aop.2004.01.004} {\bibfield  {journal} {\bibinfo  {journal} {Annals of Physics}\ }\textbf {\bibinfo {volume} {310}},\ \bibinfo {pages} {493} (\bibinfo {year} {2004})}\BibitemShut {NoStop}%
\bibitem [{\citenamefont {Inglis}\ and\ \citenamefont {Melko}(2013)}]{inglis2013}%
  \BibitemOpen
  \bibfield  {author} {\bibinfo {author} {\bibfnamefont {S.}~\bibnamefont {Inglis}}\ and\ \bibinfo {author} {\bibfnamefont {R.~G.}\ \bibnamefont {Melko}},\ }\href {\doibase 10.1088/1367-2630/15/7/073048} {\bibfield  {journal} {\bibinfo  {journal} {New Journal of Physics}\ }\textbf {\bibinfo {volume} {15}},\ \bibinfo {pages} {073048} (\bibinfo {year} {2013})}\BibitemShut {NoStop}%
\bibitem [{\citenamefont {Chen}\ \emph {et~al.}(2015)\citenamefont {Chen}, \citenamefont {Cho}, \citenamefont {Faulkner},\ and\ \citenamefont {Fradkin}}]{chen2015}%
  \BibitemOpen
  \bibfield  {author} {\bibinfo {author} {\bibfnamefont {X.}~\bibnamefont {Chen}}, \bibinfo {author} {\bibfnamefont {G.~Y.}\ \bibnamefont {Cho}}, \bibinfo {author} {\bibfnamefont {T.}~\bibnamefont {Faulkner}}, \ and\ \bibinfo {author} {\bibfnamefont {E.}~\bibnamefont {Fradkin}},\ }\href {\doibase 10.1088/1742-5468/2015/02/P02010} {\bibfield  {journal} {\bibinfo  {journal} {Journal of Statistical Mechanics: Theory and Experiment}\ }\textbf {\bibinfo {volume} {2015}},\ \bibinfo {pages} {P02010} (\bibinfo {year} {2015})}\BibitemShut {NoStop}%
\bibitem [{\citenamefont {Fradkin}(2013)}]{Fradkin2013}%
  \BibitemOpen
  \bibfield  {author} {\bibinfo {author} {\bibfnamefont {E.}~\bibnamefont {Fradkin}},\ }\href@noop {} {\emph {\bibinfo {title} {Field Theories of Condensed Matter Physics}}},\ \bibinfo {edition} {2nd}\ ed.\ (\bibinfo  {publisher} {Cambridge University Press},\ \bibinfo {year} {2013})\BibitemShut {NoStop}%
\bibitem [{Note1()}]{Note1}%
  \BibitemOpen
  \bibinfo {note} {This can be easily verified by confirming Wick's theorem for $\rho _A$.}\BibitemShut {Stop}%
\bibitem [{Note2()}]{Note2}%
  \BibitemOpen
  \bibinfo {note} {The original wave functions $\psi ^s_{\ell ,t}$ get renormalized at every time step. This changes the basis at each time step and provides an ambiguous set of wave functions, hindering numerical probes applied to them~\cite {Szyniszewski2024}.}\BibitemShut {Stop}%
\bibitem [{\citenamefont {Zharekeshev}\ and\ \citenamefont {Kramer}(1997)}]{Zharekeshev1997}%
  \BibitemOpen
  \bibfield  {author} {\bibinfo {author} {\bibfnamefont {I.~K.}\ \bibnamefont {Zharekeshev}}\ and\ \bibinfo {author} {\bibfnamefont {B.}~\bibnamefont {Kramer}},\ }\href {\doibase 10.1103/PhysRevLett.79.717} {\bibfield  {journal} {\bibinfo  {journal} {Phys. Rev. Lett.}\ }\textbf {\bibinfo {volume} {79}},\ \bibinfo {pages} {717} (\bibinfo {year} {1997})}\BibitemShut {NoStop}%
\bibitem [{\citenamefont {Serbyn}\ and\ \citenamefont {Moore}(2016)}]{Serbyn2016}%
  \BibitemOpen
  \bibfield  {author} {\bibinfo {author} {\bibfnamefont {M.}~\bibnamefont {Serbyn}}\ and\ \bibinfo {author} {\bibfnamefont {J.~E.}\ \bibnamefont {Moore}},\ }\href {\doibase 10.1103/PhysRevB.93.041424} {\bibfield  {journal} {\bibinfo  {journal} {Phys. Rev. B}\ }\textbf {\bibinfo {volume} {93}},\ \bibinfo {pages} {041424} (\bibinfo {year} {2016})}\BibitemShut {NoStop}%
\bibitem [{\citenamefont {Sierant}\ and\ \citenamefont {Zakrzewski}(2019)}]{Sierant2019}%
  \BibitemOpen
  \bibfield  {author} {\bibinfo {author} {\bibfnamefont {P.}~\bibnamefont {Sierant}}\ and\ \bibinfo {author} {\bibfnamefont {J.}~\bibnamefont {Zakrzewski}},\ }\href {\doibase 10.1103/PhysRevB.99.104205} {\bibfield  {journal} {\bibinfo  {journal} {Phys. Rev. B}\ }\textbf {\bibinfo {volume} {99}},\ \bibinfo {pages} {104205} (\bibinfo {year} {2019})}\BibitemShut {NoStop}%
\bibitem [{\citenamefont {Pal}\ and\ \citenamefont {Huse}(2010)}]{Pal2010}%
  \BibitemOpen
  \bibfield  {author} {\bibinfo {author} {\bibfnamefont {A.}~\bibnamefont {Pal}}\ and\ \bibinfo {author} {\bibfnamefont {D.~A.}\ \bibnamefont {Huse}},\ }\href {\doibase 10.1103/PhysRevB.82.174411} {\bibfield  {journal} {\bibinfo  {journal} {Phys. Rev. B}\ }\textbf {\bibinfo {volume} {82}},\ \bibinfo {pages} {174411} (\bibinfo {year} {2010})}\BibitemShut {NoStop}%
\bibitem [{\citenamefont {\ifmmode~\check{S}\else \v{S}\fi{}untajs}\ \emph {et~al.}(2020)\citenamefont {\ifmmode~\check{S}\else \v{S}\fi{}untajs}, \citenamefont {Bon\ifmmode~\check{c}\else \v{c}\fi{}a}, \citenamefont {Prosen},\ and\ \citenamefont {Vidmar}}]{suntajs2020}%
  \BibitemOpen
  \bibfield  {author} {\bibinfo {author} {\bibfnamefont {J.}~\bibnamefont {\ifmmode~\check{S}\else \v{S}\fi{}untajs}}, \bibinfo {author} {\bibfnamefont {J.}~\bibnamefont {Bon\ifmmode~\check{c}\else \v{c}\fi{}a}}, \bibinfo {author} {\bibfnamefont {T.~c.~v.}\ \bibnamefont {Prosen}}, \ and\ \bibinfo {author} {\bibfnamefont {L.}~\bibnamefont {Vidmar}},\ }\href {\doibase 10.1103/PhysRevE.102.062144} {\bibfield  {journal} {\bibinfo  {journal} {Phys. Rev. E}\ }\textbf {\bibinfo {volume} {102}},\ \bibinfo {pages} {062144} (\bibinfo {year} {2020})}\BibitemShut {NoStop}%
\bibitem [{\citenamefont {Berry}\ and\ \citenamefont {Tabor}(1977)}]{berry1977}%
  \BibitemOpen
  \bibfield  {author} {\bibinfo {author} {\bibfnamefont {M.~V.}\ \bibnamefont {Berry}}\ and\ \bibinfo {author} {\bibfnamefont {M.}~\bibnamefont {Tabor}},\ }\href {\doibase 10.1098/rspa.1977.0140} {\bibfield  {journal} {\bibinfo  {journal} {Proceedings of the Royal Society of London. A. Mathematical and Physical Sciences}\ }\textbf {\bibinfo {volume} {356}},\ \bibinfo {pages} {375} (\bibinfo {year} {1977})},\ \Eprint {http://arxiv.org/abs/https://royalsocietypublishing.org/doi/pdf/10.1098/rspa.1977.0140} {https://royalsocietypublishing.org/doi/pdf/10.1098/rspa.1977.0140} \BibitemShut {NoStop}%
\bibitem [{\citenamefont {Bohigas}\ \emph {et~al.}(1984)\citenamefont {Bohigas}, \citenamefont {Giannoni},\ and\ \citenamefont {Schmit}}]{bohigas1984}%
  \BibitemOpen
  \bibfield  {author} {\bibinfo {author} {\bibfnamefont {O.}~\bibnamefont {Bohigas}}, \bibinfo {author} {\bibfnamefont {M.~J.}\ \bibnamefont {Giannoni}}, \ and\ \bibinfo {author} {\bibfnamefont {C.}~\bibnamefont {Schmit}},\ }\href {\doibase 10.1103/PhysRevLett.52.1} {\bibfield  {journal} {\bibinfo  {journal} {Phys. Rev. Lett.}\ }\textbf {\bibinfo {volume} {52}},\ \bibinfo {pages} {1} (\bibinfo {year} {1984})}\BibitemShut {NoStop}%
\bibitem [{\citenamefont {Atas}\ \emph {et~al.}(2013)\citenamefont {Atas}, \citenamefont {Bogomolny}, \citenamefont {Giraud},\ and\ \citenamefont {Roux}}]{atas2013}%
  \BibitemOpen
  \bibfield  {author} {\bibinfo {author} {\bibfnamefont {Y.~Y.}\ \bibnamefont {Atas}}, \bibinfo {author} {\bibfnamefont {E.}~\bibnamefont {Bogomolny}}, \bibinfo {author} {\bibfnamefont {O.}~\bibnamefont {Giraud}}, \ and\ \bibinfo {author} {\bibfnamefont {G.}~\bibnamefont {Roux}},\ }\href {\doibase 10.1103/PhysRevLett.110.084101} {\bibfield  {journal} {\bibinfo  {journal} {Phys. Rev. Lett.}\ }\textbf {\bibinfo {volume} {110}},\ \bibinfo {pages} {084101} (\bibinfo {year} {2013})}\BibitemShut {NoStop}%
\bibitem [{\citenamefont {Kullback}\ and\ \citenamefont {Leibler}(1951)}]{Kullback1951}%
  \BibitemOpen
  \bibfield  {author} {\bibinfo {author} {\bibfnamefont {S.}~\bibnamefont {Kullback}}\ and\ \bibinfo {author} {\bibfnamefont {R.~A.}\ \bibnamefont {Leibler}},\ }\href {\doibase 10.1214/aoms/1177729694} {\bibfield  {journal} {\bibinfo  {journal} {The Annals of Mathematical Statistics}\ }\textbf {\bibinfo {volume} {22}},\ \bibinfo {pages} {79 } (\bibinfo {year} {1951})}\BibitemShut {NoStop}%
\bibitem [{\citenamefont {Luitz}\ \emph {et~al.}(2015)\citenamefont {Luitz}, \citenamefont {Laflorencie},\ and\ \citenamefont {Alet}}]{Luitz2015}%
  \BibitemOpen
  \bibfield  {author} {\bibinfo {author} {\bibfnamefont {D.~J.}\ \bibnamefont {Luitz}}, \bibinfo {author} {\bibfnamefont {N.}~\bibnamefont {Laflorencie}}, \ and\ \bibinfo {author} {\bibfnamefont {F.}~\bibnamefont {Alet}},\ }\href {\doibase 10.1103/PhysRevB.91.081103} {\bibfield  {journal} {\bibinfo  {journal} {Phys. Rev. B}\ }\textbf {\bibinfo {volume} {91}},\ \bibinfo {pages} {081103} (\bibinfo {year} {2015})}\BibitemShut {NoStop}%
\bibitem [{\citenamefont {Colbois}\ \emph {et~al.}(2024)\citenamefont {Colbois}, \citenamefont {Alet},\ and\ \citenamefont {Laflorencie}}]{Colbois2024}%
  \BibitemOpen
  \bibfield  {author} {\bibinfo {author} {\bibfnamefont {J.}~\bibnamefont {Colbois}}, \bibinfo {author} {\bibfnamefont {F.}~\bibnamefont {Alet}}, \ and\ \bibinfo {author} {\bibfnamefont {N.}~\bibnamefont {Laflorencie}},\ }\href {\doibase 10.1103/PhysRevLett.133.116502} {\bibfield  {journal} {\bibinfo  {journal} {Phys. Rev. Lett.}\ }\textbf {\bibinfo {volume} {133}},\ \bibinfo {pages} {116502} (\bibinfo {year} {2024})}\BibitemShut {NoStop}%
\bibitem [{\citenamefont {Khaymovich}\ \emph {et~al.}(2020)\citenamefont {Khaymovich}, \citenamefont {Kravtsov}, \citenamefont {Altshuler},\ and\ \citenamefont {Ioffe}}]{Khaymovich2020}%
  \BibitemOpen
  \bibfield  {author} {\bibinfo {author} {\bibfnamefont {I.~M.}\ \bibnamefont {Khaymovich}}, \bibinfo {author} {\bibfnamefont {V.~E.}\ \bibnamefont {Kravtsov}}, \bibinfo {author} {\bibfnamefont {B.~L.}\ \bibnamefont {Altshuler}}, \ and\ \bibinfo {author} {\bibfnamefont {L.~B.}\ \bibnamefont {Ioffe}},\ }\href {\doibase 10.1103/PhysRevResearch.2.043346} {\bibfield  {journal} {\bibinfo  {journal} {Phys. Rev. Res.}\ }\textbf {\bibinfo {volume} {2}},\ \bibinfo {pages} {043346} (\bibinfo {year} {2020})}\BibitemShut {NoStop}%
\bibitem [{\citenamefont {Kravtsov}\ \emph {et~al.}(2020)\citenamefont {Kravtsov}, \citenamefont {Khaymovich}, \citenamefont {Altshuler},\ and\ \citenamefont {Ioffe}}]{kravtsov2020}%
  \BibitemOpen
  \bibfield  {author} {\bibinfo {author} {\bibfnamefont {V.~E.}\ \bibnamefont {Kravtsov}}, \bibinfo {author} {\bibfnamefont {I.~M.}\ \bibnamefont {Khaymovich}}, \bibinfo {author} {\bibfnamefont {B.~L.}\ \bibnamefont {Altshuler}}, \ and\ \bibinfo {author} {\bibfnamefont {L.~B.}\ \bibnamefont {Ioffe}},\ }\href {https://arxiv.org/abs/2002.02979} {\enquote {\bibinfo {title} {Localization transition on the random regular graph as an unstable tricritical point in a log-normal rosenzweig-porter random matrix ensemble},}\ } (\bibinfo {year} {2020}),\ \Eprint {http://arxiv.org/abs/2002.02979} {arXiv:2002.02979 [cond-mat.dis-nn]} \BibitemShut {NoStop}%
\bibitem [{\citenamefont {Pino}\ \emph {et~al.}(2019)\citenamefont {Pino}, \citenamefont {Tabanera},\ and\ \citenamefont {Serna}}]{Pino2019}%
  \BibitemOpen
  \bibfield  {author} {\bibinfo {author} {\bibfnamefont {M.}~\bibnamefont {Pino}}, \bibinfo {author} {\bibfnamefont {J.}~\bibnamefont {Tabanera}}, \ and\ \bibinfo {author} {\bibfnamefont {P.}~\bibnamefont {Serna}},\ }\href {\doibase 10.1088/1751-8121/ab4b76} {\bibfield  {journal} {\bibinfo  {journal} {Journal of Physics A: Mathematical and Theoretical}\ }\textbf {\bibinfo {volume} {52}},\ \bibinfo {pages} {475101} (\bibinfo {year} {2019})}\BibitemShut {NoStop}%
\bibitem [{\citenamefont {Edwards}\ and\ \citenamefont {Thouless}(1972)}]{edwards1972}%
  \BibitemOpen
  \bibfield  {author} {\bibinfo {author} {\bibfnamefont {J.~T.}\ \bibnamefont {Edwards}}\ and\ \bibinfo {author} {\bibfnamefont {D.~J.}\ \bibnamefont {Thouless}},\ }\href {\doibase 10.1088/0022-3719/5/8/007} {\bibfield  {journal} {\bibinfo  {journal} {Journal of Physics C: Solid State Physics}\ }\textbf {\bibinfo {volume} {5}},\ \bibinfo {pages} {807} (\bibinfo {year} {1972})}\BibitemShut {NoStop}%
\bibitem [{\citenamefont {Hopjan}\ and\ \citenamefont {Vidmar}(2023)}]{Hopjan2023}%
  \BibitemOpen
  \bibfield  {author} {\bibinfo {author} {\bibfnamefont {M.}~\bibnamefont {Hopjan}}\ and\ \bibinfo {author} {\bibfnamefont {L.}~\bibnamefont {Vidmar}},\ }\href {\doibase 10.1103/PhysRevResearch.5.043301} {\bibfield  {journal} {\bibinfo  {journal} {Phys. Rev. Res.}\ }\textbf {\bibinfo {volume} {5}},\ \bibinfo {pages} {043301} (\bibinfo {year} {2023})}\BibitemShut {NoStop}%
\bibitem [{\citenamefont {Kos}\ \emph {et~al.}(2018)\citenamefont {Kos}, \citenamefont {Ljubotina},\ and\ \citenamefont {Prosen}}]{Kos2018}%
  \BibitemOpen
  \bibfield  {author} {\bibinfo {author} {\bibfnamefont {P.}~\bibnamefont {Kos}}, \bibinfo {author} {\bibfnamefont {M.}~\bibnamefont {Ljubotina}}, \ and\ \bibinfo {author} {\bibfnamefont {T.~c.~v.}\ \bibnamefont {Prosen}},\ }\href {\doibase 10.1103/PhysRevX.8.021062} {\bibfield  {journal} {\bibinfo  {journal} {Phys. Rev. X}\ }\textbf {\bibinfo {volume} {8}},\ \bibinfo {pages} {021062} (\bibinfo {year} {2018})}\BibitemShut {NoStop}%
\bibitem [{Note3()}]{Note3}%
  \BibitemOpen
  \bibinfo {note} {In systems for which the Hilbert space is not exponentially growing with system size, the Heisenberg time also scales polynomially. Crucially, it still grows faster than $T_\protect \text {Th}$ so that the ratio $\tau _\protect \text {Th}=T_\protect \text {Th}/T_\protect \text {H}$ still vanishes. For instance, in the $3D$ Anderson model, the Thouless time in the diffusive regime is found to scale quadratically in $L$, while the Heisenberg time scales as $T_\protect \text {H}\sim L^3$~\cite {sierant2020thouless}.}\BibitemShut {Stop}%
\bibitem [{\citenamefont {Rosenzweig}\ and\ \citenamefont {Porter}(1960)}]{Rosenzweig1960}%
  \BibitemOpen
  \bibfield  {author} {\bibinfo {author} {\bibfnamefont {N.}~\bibnamefont {Rosenzweig}}\ and\ \bibinfo {author} {\bibfnamefont {C.~E.}\ \bibnamefont {Porter}},\ }\href {\doibase 10.1103/PhysRev.120.1698} {\bibfield  {journal} {\bibinfo  {journal} {Phys. Rev.}\ }\textbf {\bibinfo {volume} {120}},\ \bibinfo {pages} {1698} (\bibinfo {year} {1960})}\BibitemShut {NoStop}%
\bibitem [{\citenamefont {Kravtsov}\ \emph {et~al.}(2015)\citenamefont {Kravtsov}, \citenamefont {Khaymovich}, \citenamefont {Cuevas},\ and\ \citenamefont {Amini}}]{Kravtsov2015rp}%
  \BibitemOpen
  \bibfield  {author} {\bibinfo {author} {\bibfnamefont {V.~E.}\ \bibnamefont {Kravtsov}}, \bibinfo {author} {\bibfnamefont {I.~M.}\ \bibnamefont {Khaymovich}}, \bibinfo {author} {\bibfnamefont {E.}~\bibnamefont {Cuevas}}, \ and\ \bibinfo {author} {\bibfnamefont {M.}~\bibnamefont {Amini}},\ }\href {\doibase 10.1088/1367-2630/17/12/122002} {\bibfield  {journal} {\bibinfo  {journal} {New Journal of Physics}\ }\textbf {\bibinfo {volume} {17}},\ \bibinfo {pages} {122002} (\bibinfo {year} {2015})}\BibitemShut {NoStop}%
\bibitem [{Note4()}]{Note4}%
  \BibitemOpen
  \bibinfo {note} {Similar behavior was observed for the monitored Kitaev model in $D=2$~\cite {klocke2025}.}\BibitemShut {Stop}%
\bibitem [{\citenamefont {Starchl}\ \emph {et~al.}(2025)\citenamefont {Starchl}, \citenamefont {Fischer},\ and\ \citenamefont {Sieberer}}]{Starchl2025}%
  \BibitemOpen
  \bibfield  {author} {\bibinfo {author} {\bibfnamefont {E.}~\bibnamefont {Starchl}}, \bibinfo {author} {\bibfnamefont {M.~H.}\ \bibnamefont {Fischer}}, \ and\ \bibinfo {author} {\bibfnamefont {L.~M.}\ \bibnamefont {Sieberer}},\ }\href {\doibase 10.1103/jppz-vdgn} {\bibfield  {journal} {\bibinfo  {journal} {PRX Quantum}\ }\textbf {\bibinfo {volume} {6}},\ \bibinfo {pages} {030302} (\bibinfo {year} {2025})}\BibitemShut {NoStop}%
\bibitem [{\citenamefont {De~Luca}\ \emph {et~al.}(2014)\citenamefont {De~Luca}, \citenamefont {Altshuler}, \citenamefont {Kravtsov},\ and\ \citenamefont {Scardicchio}}]{deluca2014}%
  \BibitemOpen
  \bibfield  {author} {\bibinfo {author} {\bibfnamefont {A.}~\bibnamefont {De~Luca}}, \bibinfo {author} {\bibfnamefont {B.~L.}\ \bibnamefont {Altshuler}}, \bibinfo {author} {\bibfnamefont {V.~E.}\ \bibnamefont {Kravtsov}}, \ and\ \bibinfo {author} {\bibfnamefont {A.}~\bibnamefont {Scardicchio}},\ }\href {\doibase 10.1103/PhysRevLett.113.046806} {\bibfield  {journal} {\bibinfo  {journal} {Phys. Rev. Lett.}\ }\textbf {\bibinfo {volume} {113}},\ \bibinfo {pages} {046806} (\bibinfo {year} {2014})}\BibitemShut {NoStop}%
\bibitem [{\citenamefont {Sierant}\ \emph {et~al.}(2023)\citenamefont {Sierant}, \citenamefont {Lewenstein},\ and\ \citenamefont {Scardicchio}}]{Sierant2023}%
  \BibitemOpen
  \bibfield  {author} {\bibinfo {author} {\bibfnamefont {P.}~\bibnamefont {Sierant}}, \bibinfo {author} {\bibfnamefont {M.}~\bibnamefont {Lewenstein}}, \ and\ \bibinfo {author} {\bibfnamefont {A.}~\bibnamefont {Scardicchio}},\ }\href {\doibase 10.21468/SciPostPhys.15.2.045} {\bibfield  {journal} {\bibinfo  {journal} {SciPost Phys.}\ }\textbf {\bibinfo {volume} {15}},\ \bibinfo {pages} {045} (\bibinfo {year} {2023})}\BibitemShut {NoStop}%
\bibitem [{\citenamefont {Biroli}\ \emph {et~al.}(2022)\citenamefont {Biroli}, \citenamefont {Hartmann},\ and\ \citenamefont {Tarzia}}]{biroli2022}%
  \BibitemOpen
  \bibfield  {author} {\bibinfo {author} {\bibfnamefont {G.}~\bibnamefont {Biroli}}, \bibinfo {author} {\bibfnamefont {A.~K.}\ \bibnamefont {Hartmann}}, \ and\ \bibinfo {author} {\bibfnamefont {M.}~\bibnamefont {Tarzia}},\ }\href {\doibase 10.1103/PhysRevB.105.094202} {\bibfield  {journal} {\bibinfo  {journal} {Phys. Rev. B}\ }\textbf {\bibinfo {volume} {105}},\ \bibinfo {pages} {094202} (\bibinfo {year} {2022})}\BibitemShut {NoStop}%
\bibitem [{\citenamefont {Bezanson}\ \emph {et~al.}(2017)\citenamefont {Bezanson}, \citenamefont {Edelman}, \citenamefont {Karpinski},\ and\ \citenamefont {Shah}}]{bezanson17}%
  \BibitemOpen
  \bibfield  {author} {\bibinfo {author} {\bibfnamefont {J.}~\bibnamefont {Bezanson}}, \bibinfo {author} {\bibfnamefont {A.}~\bibnamefont {Edelman}}, \bibinfo {author} {\bibfnamefont {S.}~\bibnamefont {Karpinski}}, \ and\ \bibinfo {author} {\bibfnamefont {V.~B.}\ \bibnamefont {Shah}},\ }\href {https://doi.org/10.1137/141000671} {\bibfield  {journal} {\bibinfo  {journal} {SIAM review}\ }\textbf {\bibinfo {volume} {59}},\ \bibinfo {pages} {65} (\bibinfo {year} {2017})}\BibitemShut {NoStop}%
\bibitem [{\citenamefont {Szyniszewski}(2024)}]{Szyniszewski2024}%
  \BibitemOpen
  \bibfield  {author} {\bibinfo {author} {\bibfnamefont {M.}~\bibnamefont {Szyniszewski}},\ }\href {\doibase 10.1103/PhysRevB.110.024303} {\bibfield  {journal} {\bibinfo  {journal} {Phys. Rev. B}\ }\textbf {\bibinfo {volume} {110}},\ \bibinfo {pages} {024303} (\bibinfo {year} {2024})}\BibitemShut {NoStop}%
\bibitem [{\citenamefont {Sierant}\ \emph {et~al.}(2020)\citenamefont {Sierant}, \citenamefont {Delande},\ and\ \citenamefont {Zakrzewski}}]{sierant2020thouless}%
  \BibitemOpen
  \bibfield  {author} {\bibinfo {author} {\bibfnamefont {P.}~\bibnamefont {Sierant}}, \bibinfo {author} {\bibfnamefont {D.}~\bibnamefont {Delande}}, \ and\ \bibinfo {author} {\bibfnamefont {J.}~\bibnamefont {Zakrzewski}},\ }\href {\doibase 10.1103/PhysRevLett.124.186601} {\bibfield  {journal} {\bibinfo  {journal} {Phys. Rev. Lett.}\ }\textbf {\bibinfo {volume} {124}},\ \bibinfo {pages} {186601} (\bibinfo {year} {2020})}\BibitemShut {NoStop}%
\bibitem [{\citenamefont {Wolf}\ \emph {et~al.}(2008)\citenamefont {Wolf}, \citenamefont {Verstraete}, \citenamefont {Hastings},\ and\ \citenamefont {Cirac}}]{wolf2008area}%
  \BibitemOpen
  \bibfield  {author} {\bibinfo {author} {\bibfnamefont {M.~M.}\ \bibnamefont {Wolf}}, \bibinfo {author} {\bibfnamefont {F.}~\bibnamefont {Verstraete}}, \bibinfo {author} {\bibfnamefont {M.~B.}\ \bibnamefont {Hastings}}, \ and\ \bibinfo {author} {\bibfnamefont {J.~I.}\ \bibnamefont {Cirac}},\ }\href@noop {} {\bibfield  {journal} {\bibinfo  {journal} {Physical review letters}\ }\textbf {\bibinfo {volume} {100}},\ \bibinfo {pages} {070502} (\bibinfo {year} {2008})}\BibitemShut {NoStop}%
\bibitem [{\citenamefont {Klich}\ and\ \citenamefont {Levitov}(2009)}]{KlichLevitov}%
  \BibitemOpen
  \bibfield  {author} {\bibinfo {author} {\bibfnamefont {I.}~\bibnamefont {Klich}}\ and\ \bibinfo {author} {\bibfnamefont {L.}~\bibnamefont {Levitov}},\ }\href {\doibase 10.1103/PhysRevLett.102.100502} {\bibfield  {journal} {\bibinfo  {journal} {Phys. Rev. Lett.}\ }\textbf {\bibinfo {volume} {102}},\ \bibinfo {pages} {100502} (\bibinfo {year} {2009})}\BibitemShut {NoStop}%
\bibitem [{\citenamefont {Song}\ \emph {et~al.}(2012)\citenamefont {Song}, \citenamefont {Rachel}, \citenamefont {Flindt}, \citenamefont {Klich}, \citenamefont {Laflorencie},\ and\ \citenamefont {Le~Hur}}]{KlichEntanglement}%
  \BibitemOpen
  \bibfield  {author} {\bibinfo {author} {\bibfnamefont {H.~F.}\ \bibnamefont {Song}}, \bibinfo {author} {\bibfnamefont {S.}~\bibnamefont {Rachel}}, \bibinfo {author} {\bibfnamefont {C.}~\bibnamefont {Flindt}}, \bibinfo {author} {\bibfnamefont {I.}~\bibnamefont {Klich}}, \bibinfo {author} {\bibfnamefont {N.}~\bibnamefont {Laflorencie}}, \ and\ \bibinfo {author} {\bibfnamefont {K.}~\bibnamefont {Le~Hur}},\ }\href {\doibase 10.1103/PhysRevB.85.035409} {\bibfield  {journal} {\bibinfo  {journal} {Phys. Rev. B}\ }\textbf {\bibinfo {volume} {85}},\ \bibinfo {pages} {035409} (\bibinfo {year} {2012})}\BibitemShut {NoStop}%
\end{thebibliography}%
\begin{appendix}
\section{Entanglement prefactor and mutual information as probes of the non-trivial fixed points}\label{appendixA}
\begin{figure}[th!]
\includegraphics[width=\linewidth]{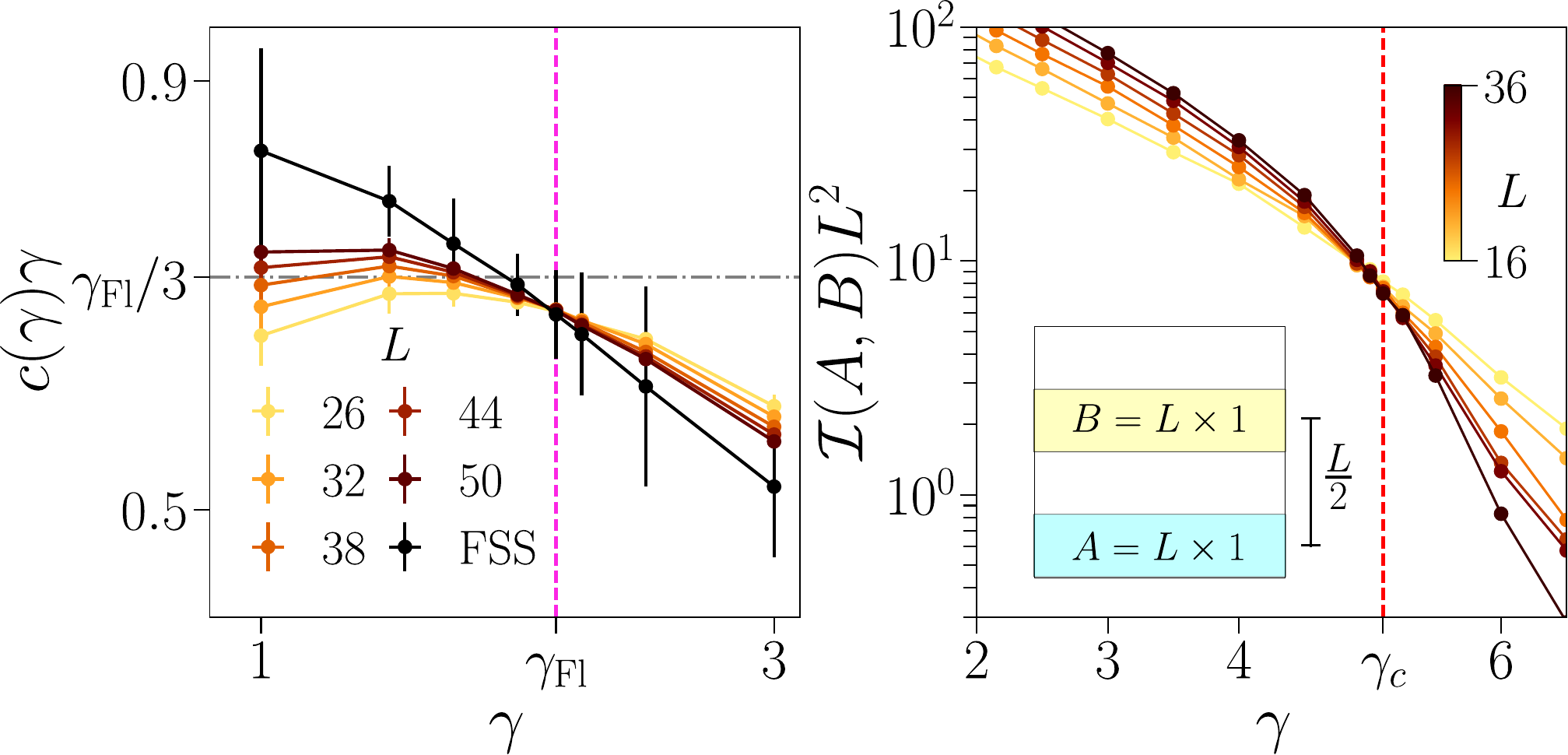}
\caption{\label{fig:MI}\textbf{Two fixed points of monitored fermions in 2D.} (a) Prefactor $c(\gamma)$ of the half-system entanglement entropy $S(A=L\times L/2)=c(\gamma)L\ln L+b(\gamma)L$ as a function of the measurement rate $\gamma$ in the log-law phase. A sharp crossing point reveals the Fermi liquid fixed point. The value of the prefactor is consistent with $c(\gamma_\text{Fl})=1/3$, agreeing with the subsystem scaling and entanglement dynamics data. (b) The mutual information $\mathcal{I}(A,B)$ between disjoint subsystems $A, B$ reveals the MIPT. Here, $L=16, 20, 24, 28, 32, 36$ and the subsystems are strips of size $L\times1$ separated by a distance $L/2$.   }
\end{figure}
In this section, we provide further signatures of the non-trivial fixed points $\gamma_\text{Fl}$ and $\gamma_c$. The Fermi liquid fixed point $\gamma_\text{Fl}$ is revealed by studying the prefactor of the half-system entanglement entropy as a function of the linear system size $L$, see fig. \ref{fig:MI}(a). We extract the prefactor as a function of monitoring rate $\gamma$ and linear system size $L$ as follows. For a fixed monitoring rate $\gamma<\gamma_c$, we assume that the entanglement entropy is $S(A)=c(\gamma)L\ln L+b(\gamma)L$. For any given $L$, we perform a linear fit of $S(A)/\tilde L$ vs $\ln \tilde L$ for all even $\tilde L$ values in $\tilde L\in[2,L]$ and with $L$ values up to $L=50$. We further perform an extrapolation to $L\to+\infty$ of the prefactor as a function of $1/L$. In particular, we have performed a linear fit $c(\gamma)$ = $m/L+c(L\to+\infty)$ to extract $c(L\to+\infty)$. Finite size scaling of the prefactor $c(\gamma)$ reveals a sharp crossing point at $\gamma=\gamma_\text{Fl}$. Furthermore, its value at that point is consistent with $1/3$, thus agreeing with the subsystem scaling and entanglement dynamics data discussed in the main text.

The MIPT is instead revealed by the mutual information between disjoint subsystems, see fig. \ref{fig:MI}(b). This quantity provides an upper bound for the correlations between disjoint subsystems $A, B$~\cite{wolf2008area}. For free fermions it is determined by particle number fluctuations between $A$ and $B$~\cite{KlichLevitov,KlichEntanglement}. In particular, we consider subsystems $A, B$ of equal geometry and size $L\times1$, separated by a distance $L/2$. Two different scaling behaviours are expected on either side of the entanglement transition\cite{poboiko2023measurementinduced}. Through a finite size scaling analysis of mutual information, we indeed find a crossing at $\gamma\approx\gamma_c$. However, we find that the entanglement Hamiltonian toolkit described in the main text is much more accurate in characterizing the critical point and exponent.
\end{appendix}

\end{document}